\documentstyle[supertabular,epsf,rotate]{mn}
\newcommand{\lta}{\stackrel{<}{_{\sim}}}
\newcommand{\gta}{\stackrel{>}{_{\sim}}}
\newcommand{\ltsimeq}{\raisebox{-0.6ex}{$\,\stackrel 
        {\raisebox{-.2ex}{$\textstyle <$}}{\sim}\,$}} 
\newcommand{\gtsimeq}{\raisebox{-0.6ex}{$\,\stackrel 
        {\raisebox{-.2ex}{$\textstyle >$}}{\sim}\,$}} 

\input epsf.sty

\def\lesssim{\mathrel{\hbox{\rlap{\hbox{\lower2pt\hbox{$\sim$}}}\raise2pt\hbox{$<$}}}}
\def\grtsim{\mathrel{\hbox{\rlap{\hbox{\lower2pt\hbox{$\sim$}}}\raise2pt\hbox{$>$}}}}

\newcommand{\ltsim}{\mbox{{\raisebox{-0.4ex}{$\stackrel{<}{{\scriptstyle\sim}}$}}}}

\begin{document}

\title[The TOOT00 survey I: the data]
{The {\sc TexOx}-1000 redshift survey of radio sources I: the TOOT00
region}

\author[Vardoulaki et al.]{
Eleni Vardoulaki$^{1}$\footnotemark, Steve Rawlings$^{1}$, Gary J. Hill$^{2}$, Tom Mauch$^{1}$,   
\\
\\
{\LARGE\rm 
Katherine J. Inskip$^{3}$, Julia Riley$^{4}$, Kate Brand$^{1}$, Steve Croft$^{5}$, }
\\
\\
{\LARGE\rm 
Chris J. Willott$^{6}$}
\\
\\
$^{1}$ 
Astrophysics, Denys Wilkinson Building, Keble Road, Oxford, OX1 3RH, UK\\
$^{2}$ 
University of Texas at Austin, 1 University Station, C1402, Austin, TX 78712, 
USA\\
$^{3}$ 
Max-Planck-Institut fur Astronomie, Konigstuhl 17, D-69117 Heidelberg, Germany\\
$^{4}$ 
Cavendish Astrophysics, Department of Physics, Madingley Road, Cambridge, 
CB3 OHE, UK\\
$^{5}$ 
University of California, Berkeley, Dept. of Astronomy, 601 Campbell Hall $\#$3411, Berkeley, CA 94720, USA\\
$^{6}$ 
Herzberg Institute of Astrophysics, National Research Council, 5071 West 
Saanich Rd, Victoria, BC V9E 2E7, Canada\\
}

\maketitle

\begin{abstract}
\noindent
We present optical spectroscopy, near-infrared (mostly $K-$band) and radio 
(151-MHz and 1.4-GHz) imaging of the 
first complete region (TOOT00) of the TexOx-1000 (TOOT) redshift survey of 
radio sources. The 0.0015-sr ($\sim$ 5 deg$^{2}$) TOOT00 region is selected 
from pointed 
observations of the Cambridge Low-Frequency Survey Telescope at 151 MHz at a 
flux density limit of $\simeq$ 100 mJy, $\sim$ 5-times fainter than the 7C 
Redshift Survey (7CRS), and contains 47 radio sources. We have obtained 40 
spectroscopic redshifts ($\sim$ 85\% completeness). Adding redshifts 
estimated for the 7 other cases yields a median redshift $z_{\rm med} \sim$ 
1.25. We find a significant population of objects with FRI-like radio 
structures at radio luminosities above both the low-redshift FRI/II break and 
the break in the radio luminosity function. The redshift distribution and 
sub-populations of TOOT00 are broadly consistent with extrapolations from the 
7CRS/6CE/3CRR datasets underlying the SKADS Simulated Skies Semi-Empirical 
Extragalactic Database, S$^{3}$-SEX.
\end{abstract}


\footnotetext{Email: eleniv@astro.ox.ac.uk}

\section{Introduction}
\label{sec:intro}

\addtocounter{figure}{0}

\begin{figure*}
\begin{center}
\setlength{\unitlength}{1mm}
\begin{picture}(150,110)
\put(155,-10){\includegraphics{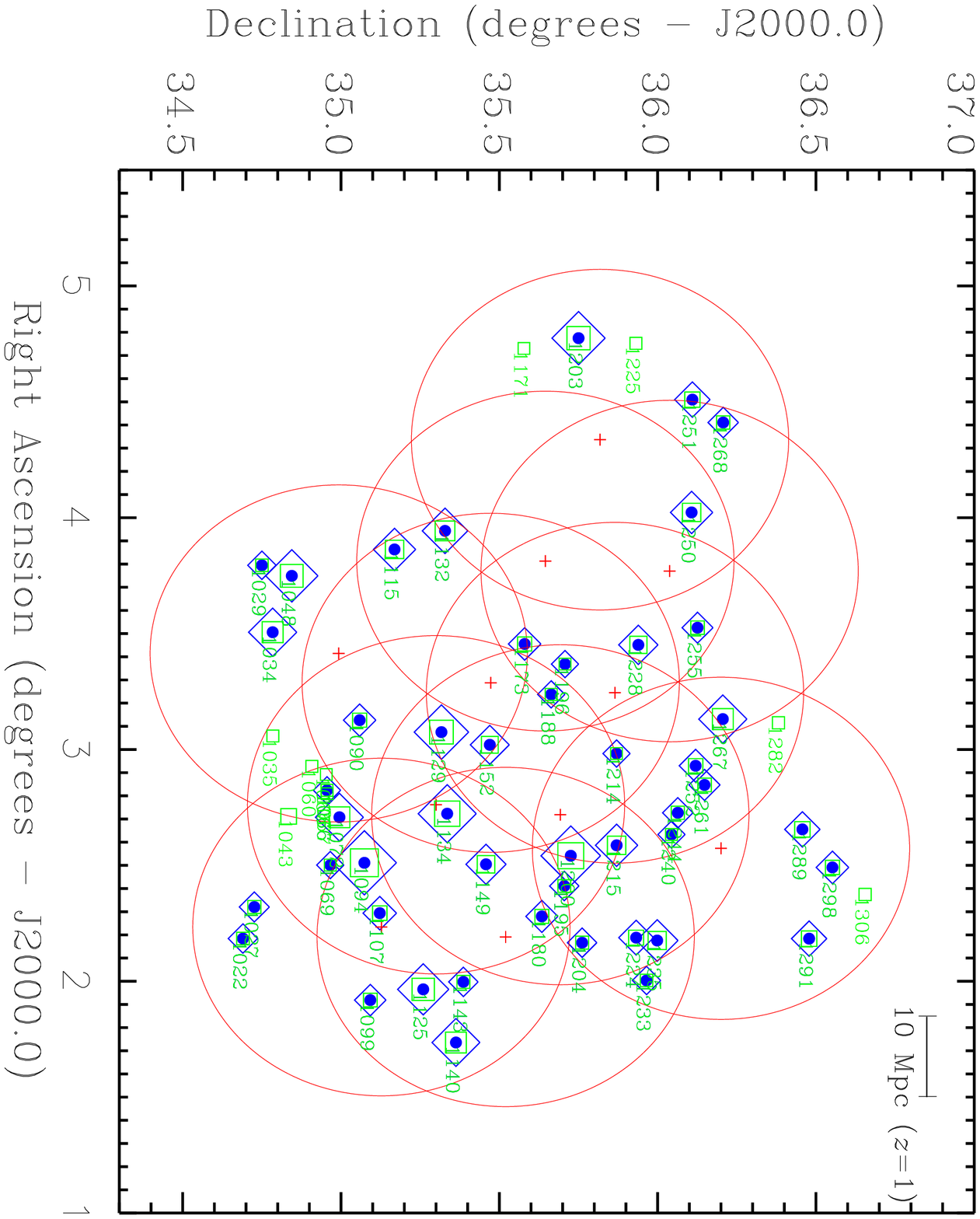}}
\end{picture}
\end{center}
\vspace{0.05in}
{\caption[Sky are of the TOOT00 region]{\label{fig:radec} Sky area of the 
TOOT00 region with the bar 
showing the angle corresponding to a proper distance of 10 Mpc observed at 
redshift $z$ = 1. Symbols: blue diamonds 
show the 47 TOOT00 radio sources, where their size is proportional to the 
base-10 logarithm of the 1.4-GHz flux density of each object; objects detected 
in the 74-MHz VLSS survey (Cohen et al.\ 2007) are indicated with a solid blue 
circle. Green boxes represent the 56 objects of the 7C sample in the same sky 
area with $S_{151 \rm MHz} >$ 95 mJy; their size is proportional to the base-10 
logarithm of the 151-MHz flux density of each object. The 11 A-array 1.4-GHz 
VLA pointings are shown as red crosses `+', where the circles around them have 
a radius of 35.7 arcmin in order to include all the TOOT00 objects, and 
corresponding roughly to the 32-arcmin-diameter primary beam of the VLA. The 
9 `off-edge' objects described in Section~\ref{sec:sample} are: 
TOOT00\_1306 at 00 11 50.33 +36 39 19.34 with $S_{151 \rm MHz}$ = 0.11 Jy, 
TOOT00\_1043 at 00 13 12.77 +34 50 04.78 with $S_{151 \rm MHz}$ = 0.21 Jy, 
TOOT00\_1067 at 00 13 42.51 +34 57 22.04 with $S_{151 \rm MHz}$ = 0.10 Jy, 
TOOT00\_1065 at 00 14 06.90 +34 57 12.28 with $S_{151 \rm MHz}$ = 0.097 Jy,  
TOTO00\_1060 at 00 14 16.62 +34 54 28.83 with $S_{151 \rm MHz}$ = 0.10 Jy, 
TOOT00\_1035 at 00 14 53.67 +34 47 04.59 with $S_{151 \rm MHz}$ = 0.10 Jy, 
TOOT00\_1282 at 00 15 28.61 +36 22 53.84 with $S_{151 \rm MHz}$ = 0.11 Jy, 
TOOT00\_1171 at 00 23 15.84 +35 34 39.66 with $S_{151 \rm MHz}$ = 0.098 Jy, 
TOOT00\_1225 at 00 23 28.63 +35 55 52.86 with $S_{151 \rm MHz}$ = 0.10 Jy.
}}
\end{figure*}

Radio observations of Active Galactic Nuclei (AGN) reveal some of the 
complexity of the AGN phenomenon. A central black hole accretes matter through 
an accretion 
disk, which along the plane of the disk is hidden by a dusty and gas-rich 
torus. Radio emission emerges from the immediate vicinity of the black hole 
through the synchrotron process (e.g. from the base of jets) and potentially 
other processes like optically thin bremsstrahlung from a slow, dense disk 
wind (Blundell \& Kuncic 2007). Powerful jets can propagate to large scales 
and radio telescopes enable the study of these `radio-loud' objects in detail, 
which are believed to live from $\lta 10^{4}$ yr for small GHz-peaked sources 
to $\sim 10^{7-8}$ yr for larger steep-spectrum sources. Best et al.\ (2005) 
found that 25\% of the most massive galaxies show radio-loud AGN activity, and 
argued that such AGN activity is constantly re-triggered. Radio sources seem to 
have a `duty-cycle', i.e. a recurrent radio-loud AGN activity, which 
is believed to play an important role in balancing the energy losses from 
hot gas that surrounds massive elliptical galaxies (Best et al.\ 2006). This 
is done by re-heating the cooling gas inside the galaxy halo, 
in a way that provides a self-regulating feedback mechanism capable of 
controlling the rate of growth of galaxies (Tabor \& Binney 1993). The 
`duty-cycle' or, in other 
words the lifetime of radio sources, differs according to radio-source type.

Fanaroff \& Riley (1974) were the first to divide radio sources according to 
radio structure and radio luminosity: FRIs are bright close to the core 
whereas FRIIs are edge-brightened radio sources. The radio luminosity divide 
is at $\log_{10}(L_{178 \rm MHz}/ \rm W Hz^{-1} sr^{-1}) \approx$ 25 
($H_{0}=50~ {\rm km~s^{-1}Mpc^{-1}}$, $\Omega_{\rm M}=1$ and 
$\Omega_{\Lambda}=0$)\footnote{This corresponds to 
$\log_{10}( L_{151 \rm MHz}/ \rm W Hz^{-1} sr^{-1}) \approx$ 25 in our 
adopted cosmology ($H_{0}=70~ {\rm km~s^{-1}Mpc^{-1}}$, 
$\Omega_{\rm M}=0.3$ and $\Omega_{\Lambda}=0.7$ for an $\alpha \simeq$ 0.8 radio 
source) at the typical redshift ($z \simeq$ 0.1) of objects defining the 
FRI/FRII division of the Fanaroff \& Riley (1974) 
study.}; above this value lie the FRIIs, and below that the FRIs. Kaiser \& 
Best (2007a, 2007b) suggest that all sources start out with an FRII structure, 
and the final form they take depends on the host galaxy. On the other hand, 
Parma et al.\ (2002) studied a sample of FRI and FRII radio sources with a 
double-double structure and argued that it is unlikely that many FRIIs evolve 
to FRIs.

Bird et al.\ (2008) argue that the 
average FRII-type (Fanaroff \& Riley 1974) radio-source lifetimes are 
$\sim 1.5\times 10^{7}$ yr with a duty cycle\footnote{We define duty cycle as 
the dimensionless ratio of time active to the sum of time active and dormant.} 
$\sim$ 0.02 
(for $\sim 8 \times 10^{8}$ yr time off). 
Parma et al.\ (2002) claim that the ages of FRIs are $\sim$ 5-10 times larger 
than the ages of FRII radio sources, consistent with the Best et al. (2005) 
inference that massive galaxies are active in either an FRI or FRII phase 
roughly 25\% of the time. 

The difference in FRI and FRII radio-source lifetimes may be linked to 
studies of HI gas as probe of the host environment of these objects. 
Emonts et al.\ (2008) studied a sample of nearby FRI radio sources up to 
redshift of $z \sim$ 0.04 and a sample of FRIIs with $z \ltsim$ 0.06, and found 
that different types of radio sources contain different large-scale 
HI properties; the HI detection rate in FRIIs is significantly higher than 
the detection rate for the FRI radio galaxies. These findings suggest 
a different triggering history for different types of radio sources: FRIIs are 
likely triggered by galaxy mergers and collisions as the large-scale HI is 
often distributed in tail- or bridge-like structures; FRI radio 
sources are likely fed in other ways, such as steady accretion from a hot IGM.

Radio sources can be observed throughout different cosmic epochs. Studies of 
radio galaxies and radio-loud quasars (RLQ) can determine the evolution of the 
radio-source population by defining the Radio Luminosity Function (RLF; e.g. 
Dunlop \& Peacock 1990; Rawlings, Eales \& Lacy 2001; Willott et 
al. 2001), i.e. the number density of sources per co-moving volume. The sharp 
decline in the (co-moving) space density of radio sources from epochs 
corresponding to $z \sim 2.5$ to those corresponding to $z \sim 0$ was 
inferred from early counts of radio sources by Longair (1966). By combining
constraints from radio source counts with surveys of bright 
high-radio-frequency-selected samples, using spectroscopic and estimated 
redshifts, Dunlop \& Peacock (1990) were the first to map out the cosmic 
evolution of this population in any detail. They claimed to detect a 
measurable drop, or redshift-cutoff, in the space density of sources at 
$z \gtsimeq 2.5$. Returning to this problem with virtually-complete 
spectroscopic redshift information on low-radio-frequency-selected samples, 
Willott et al.\ (2001) measured a RLF in rough agreement with that of 
Dunlop \& Peacock, but differing in important aspects such as a factor 
$\sim 2$ lower normalisation at $z \sim 2.5$. Such seemingly subtle 
differences, attributable largely to biases in the redshift estimates used by 
Dunlop \& Peacock, removed any convincing evidence for the redshift cut-off, a 
finding in agreement with work focused on the most luminous steep-spectrum 
radio sources (Jarvis et al. 2001).

It seems clear, therefore, that spectroscopic redshift surveys of radio 
sources are of the utmost importance in making robust measurements of the 
cosmic evolution of the RLF at high redshift. It is also clear that, 
as emphasised by Willott et al. (2001), the next generation of radio source 
redshift surveys should be carefully targeted on those objects which make the 
dominant contribution to the radio luminosity density $\Pi$\footnote{We define 
the radio luminosity density by $\Pi = \int \int L_{\nu} 
[{\rm d} \rho / {\rm d} \log_{10} L_{\nu}] {\rm d}
(\log_{10} L_{\nu}) {\rm d} \nu$, where 
${\rm d} \rho / {\rm d} \log_{10} L_{\nu}$ is the 
co-moving space density (in units of $\rm Mpc^{-3}$) of radio sources.}. 
Evaluated as a function of cosmic time $t$, $\Pi(t)$ encodes the information 
needed to estimate crucial physical quantities like the quantity of heat, or 
entropy, radio sources inject into the inter-galactic medium (Yamada \& Fujita 
2001; Gopal-Krishna \& Wiita 2001; Rawlings 2002). Current estimates of 
$\Pi(t)$, even at epochs corresponding to the supposedly well-studied redshift 
range $2 \ltsimeq z \ltsimeq 3$, are highly uncertain because they rely on 
extrapolating from measurements of the most luminous radio sources down a very 
steep RLF. Robust luminosity density measurements require that space densities 
are measured for objects much closer to the break in the RLF. 

The RLF is often modelled as a broken power law. The break luminosity is close 
to, but above, the dividing line between the two FR structural 
classes for the large-scale radio structure of these objects. 
Willott et al.\ (1998) found that the break in the RLF for radio-loud quasars 
is at $\log_{10}(L_{151 MHz}/\rm W Hz^{-1} sr^{-1})$ $\ltsim$ 27\footnote{This 
corresponds to $\log_{10}( L_{151 \rm MHz}/\rm W Hz^{-1} sr^{-1}) \approx$ 26 in 
our adopted cosmology ($H_{0}=70~ {\rm km~s^{-1}Mpc^{-1}}$, 
$\Omega_{\rm M}=0.3$ and $\Omega_{\Lambda}=0.7$ for an $\alpha \simeq$ 0.8 radio 
source) at the typical redshift ($z \simeq$ 1) of the 3CRR/6CR/7CRS datasets.} 
($H_{0}=50~ {\rm km~s^{-1}Mpc^{-1}}$, $\Omega_{\rm M}=1$ and 
$\Omega_{\Lambda}=0$), while there is no evidence for a decline in the 
co-moving space density of RLQs at higher redshift than the peak of the RLF 
at $z \sim$ 2. This is important for radio galaxies, since from unified 
AGN schemes RLQs and radio galaxies are the same objects where the apparent 
difference in their properties is a result of the orientation of the radio 
jets (e.g. Antonucci 1993).

Surveys such as the 3CRR, 6CE and 7CRS (Laing, Riley \& Longair 1983; 
Rawlings, Eales \& Lacy 2001 and Willott et al. 2003, respectively) were 
designed in order to study complete, down to a certain flux density limit, 
spectroscopic samples of radio sources. The goal was to study the RLF. 
However, at $z >$ 2 these surveys are not 
deep enough to make a good estimate of the density of radio sources 
close to the RLF break. In order to target the radio-source population 
responsible for the bulk of the cosmic heating, one needs to study a deeper 
low-radio-frequency-selected 
sample around redshift $z \sim 2$. Willott et al. (2001) suggested a 
new study that would target a large sample ($\sim$ 1000) of radio sources 
selected from the 7CRS survey at a flux density limit of 100 mJy at 151 MHz. 

The {\sc TexOx}-1000 (TOOT; Hill \& Rawlings 2003) survey is the realisation 
of a hypothetical survey proposed by Willott et al.\ (2001). The primary aim 
of the TOOT survey is to measure spectroscopic redshifts for $\approx 1000$ 
radio sources from a low-frequency-selected sample reaching a flux-density 
limit $S_{151} \sim 0.1 ~ \rm Jy$ over several separate sky patches totalling 
$\sim 0.03 ~ \rm sr$ (see Tables 1 \& 2 of Hill \& Rawlings 2003, for details 
on the TOOT regions and status of the survey). Another objective of the TOOT 
survey was to target radio sources at redshift $z \sim 2$ where the density of 
the radio source population is at its maximum (e.g. Rawlings 2003).

In this paper we study the 00$^{\rm h}$ targeted sky area of the TexOx-1000, 
hereafter TOOT00, which includes 47 radio sources with a flux density limit of 
$\simeq$ 100 mJy at 151 MHz. In Section~\ref{sec:sample} we present the 151 
MHz and 1.4 GHz radio imaging and optical spectroscopic and near-infrared 
(mostly $K-$band) imaging data on the TOOT00 sample. 
In Section~\ref{sec:analysis} we present a basic analysis on the TOOT00 
sample: a 
comparison of spectroscopic and photometric redshifts in 
Section~\ref{sec:optclass}; an analysis of radio spectral indices in 
Section~\ref{sec:alpha}; an investigation of relationships between radio 
structure, radio luminosity and redshift in Section~\ref{sec:radclass}; and a 
discussion of the quasar fraction in Section~\ref{sec:qfraction}. 
In Section~\ref{sec:distribution} we present the redshift distribution 
of the TOOT00 sample (Figure~\ref{n_z_frclass}) and compare it to simulations 
of the radio sky from the SKADS Simulated Skies Semi-Empirical Extragalactic 
Database (S$^{3}$-SEX; Wilman et al.\ 2008). In Section~\ref{sec:conclusions} 
we make some concluding remarks. In the Appendix A we present notes 
on each object, photometry in the optical and near-IR, as well as a figure 
with near-IR/radio overlays, optical spectroscopy and Spectral Energy 
Distributions (SEDs) for the 47 TOOT00 objects.

We use J2000.0 positions and the convention for all spectral 
indices, $\alpha$, that flux density $S_{\nu} \propto \nu^{-\alpha}$,
where $\nu$ is the observing frequency. 
We assume throughout a low-density, $\Lambda$-dominated Universe in which
$H_{0}=70~ {\rm km~s^{-1}Mpc^{-1}}$, $\Omega_{\rm M}=0.3$ and $\Omega_
{\Lambda}=0.7$.

\section{Data and observations}
\label{sec:sample}

The 7C radio source 
survey was performed using the Cambridge Low Frequency Synthesis Telescope 
(CLFST) operating at 151 MHz with a resolution of 70$\times$70 cosec(Dec) 
arcsec$^{2}$. General principles of its operation are described by Rees 
(1990). The instrument was used to survey about one hundred fields north of 
declination 30 degrees and well away from the galactic plane. The rms noise 
and hence the limiting flux density of the survey varies both within any one 
field and between different fields owing to beam distortion, primary beam and 
ionospheric effects. Some data on individual fields and groups of fields have 
been published in various papers (see Riley, Waldram \& Riley 1999 and 
references therein). The construction of a final unified catalogue is 
described in Hales et al. (2007) and is available via the 7C 
website\footnote{$http://www.mrao.cam.ac.uk/surveys/7C$}. The 
final 7C catalogue lists 43683 sources over an area of about 1.7 sr.

The TOOT00 radio-source sample was selected from the 7C survey to include 
sources with peak flux densities above $\simeq$ 100 mJy at 151 MHz; the final 
lower limit was 95 mJy. The area studied is illustrated in 
Figure~\ref{fig:radec}; 56 sources lie in this area. All of the TOOT00 objects 
were observed with the VLA, in A and B configurations at 1.4 GHz, where 11 
pointings were used to cover the TOOT00 area (Figure~\ref{fig:radec}). The 
A-array observations were performed on 12 August 1999 and the B-array on 25 
June 2002. To avoid excessive bandwidth smearing these observations were made 
using seven 3.125-MHz channels centred on 1.370 GHz plus seven 3.125-MHz 
channels centred on 1.444 GHz. Following standard reduction procedures in 
\textsc{AIPS} we synthesised maps of all our objects using both separate and 
combined A- and B-array visibilities. In this paper we present B-array images 
of the TOOT00 radio sources (Figure~\ref{seds} in Appendix A), unless the 
A-array maps reveal important information about the radio structure.

In the final sample selection all sources with flux 
densities $S_{151 \rm MHz} >$ 220 mJy were included, as well as a sub-set of 
sources with 95 mJy $< S_{151 \rm MHz} \leq$ 220 mJy, close enough to the VLA 
pointing centres to have good maps in the VLA A-array observations. The 9 
sources excluded are characterised as `off-edge' objects and are typically 
$\sim$ 30 arcmin away from the centre of the positions where the VLA was 
pointed. These are faint enough that no reasonable map could be made. Note 
that this introduces a selection bias against fainter radio sources in the 
TOOT00 survey. The final TOOT00 sample includes 47 radio sources. 

The sky area of TOOT00 was measured using a Monte-Carlo approach: random 
numbers were generated, uniform in RA and with a sin $\theta$ distribution 
in Dec, within a rectangle enclosing the all 56 original TOOT00 sources. We 
tested whether the randomly selected points lay within 35.7 arcmin (see 
Figure~\ref{fig:radec}) of any of the 11 VLA pointing positions. The ratio of 
points passing this test to the total number of random points (0.68) was 
multiplied by the sky area of the rectangle. The result is 0.0015 sr or 
$\sim$ 5.0 deg$^{2}$.

Table 1 gives the basic radio data on the 47 TOOT00 radio sources. The flux 
densities used subsequently are shown in bold, and were chosen according to 
the following criteria: (i) if the integrated flux density has a significantly 
larger value than the peak flux density and is not the result of confusion, we 
use the former and (ii) in the case of a compact radio source we always use the 
peak flux density. 
The radio spectral indices $\alpha_{151{\rm MHz}}^{1.4{\rm GHz}}$ were 
calculated using the 1.4 GHz (NRAO/VLA Sky Survey - NVSS; Condon et al.\ 1998) 
and 151 MHz (7C) flux densities for the TOOT00 sources. We also obtained 74 
MHz flux densities for 12 out of 47 TOOT00 radio sources from the VLA 
Low-frequency Sky Survey (VLSS; Cohen et al.\ 2007). The VLSS has a resolution 
of 80 arcsec (FWHM), a noise level of $\approx$ 0.1 Jy beam$^{-1}$ 
and a typical catalogue point-source detection limit of 0.7 Jy beam$^{-1}$; 
the 74 MHz maps are available for searching for fainter sources. The angular 
sizes of the TOOT00 sources were measured from the 
VLA (A or B configuration) images. We use our VLA data and the Owen \& 
Laing (1989) radio-source classification to place our objects in four 
categories: Classical Double (CD), Fat Double (FD), Twin Jet (TJ) and Compact 
(COM) radio sources. CDs correspond to FRIIs where there is at least one 
bright hotspot at the ends of the structure in both high- and low-resolution 
radio maps; FDs represent FRI/II division objects, where the radio structure 
is more diffuse on the high resolution radio maps than in the case of CDs, and 
no compact hotspot-like features are visible in the A-array radio map; 
TJs are FRI type radio sources which are brightest near the centre of the 
sources; COM radio sources don't show any evidence of extended radio emission 
on $\gta$ 5-arcsec scales.

Optical spectroscopic observations were undertaken with the ISIS spectrometer 
on the William Herschel Telescope (WHT\footnote{The WHT is operated by the 
Isaac Newton Group of Telescopes (ING), at the Roque de Los Muchachos 
Observatory in La Palma, Spain.}) in August 2000, 2004, 2006 and 2009 
(see Table 2); TOOT00\_1134 was observed with the Marcario 
Low Resolution Spectrograph (LRS) on the Hobby-Eberly Telescope 
(HET\footnote{The Marcario Low Resolution Spectrograph is a joint project of 
the Hobby - Eberly Telescope partnership and the Instituto de Astronomia 
de la Universidad Nacional Autonoma de Mexico. The Hobby - Eberly 
Telescope is operated by McDonald Observatory on behalf of 
The University of Texas at Austin, the Pennsylvania State 
University, Stanford University, Ludwig-Maximilians-Universitaet 
Muenchen, and Georg-August-Universitaet Goettingen.}; 
Hill et al.\ 1998). All ISIS observations used the R158R and R158B gratings 
with the beam split using the 5400-$\rm \AA$ dichroic; the R158B 
grating was used for the blue arm, providing 1.62 $\rm \AA$/pixel and 
0.2 arcsec/pixel scales with the EEV12 detector. 
For the August 2000 observing 
run the R158R grating was used with the TEK4 detector on 
the red arm, which gave 2.90 $\rm \AA$/pixel and 0.36 arcsec/pixel scales. 
For the August 2004 and August 2006 observing runs the red MARCONI detector 
was used, which provides similar intrinsic 
spatial and spectral scales to the blue 
EEV12 detector. For the August 2009 observing run the EEV12 was used 
alongside the REDPLUS detector: after $3 \times 2$ (spectral $\times$ spatial) 
binning in the blue, and $3 \times 1$ binning in the red, this gave 
gave $\approx$ $5 \rm \AA$/pixel and $\approx 0.44/0.22$ (blue/red) 
arcsec/pixel scales. 
The LRS was used with a 2-arcsec-wide slit and 300 l/mm grism 
giving a resolution of 16.7 $\rm \AA$ and coverage of 4150 to 10100 $\rm \AA$. 
Observations and data analysis followed the standard 
methods described by Rawlings, Eales \& Lacy (2001). We offset from nearby 
bright stars to either optical or radio positions, choosing the position angle 
(PA) of the slit either on the basis of encompassing companion 
optical objects or to line up with the radio axis. All the targets were 
observed below an air mass of 1.2, and all but two were observed below an air 
mass of 1.1. All the data presented here were taken in photometric 
conditions with the seeing close to 1 arcsec. In most cases 
spectral features, emission lines and/or spectral breaks were detected, 
and the reduced 1D spectra are presented in Figure~\ref{seds}. For cosmetic 
purposes, most of the spectra have been smoothed by replacing the value of 
each spectral bin by the average of the values in the bin and its two 
neighbours. All measurements from the spectra were made prior to this process, 
and the results are tabulated in Table 4.

Multi-colour 
optical images from the INT Wide Field Camera data were obtained as part of the 
Oxford-Dartmouth 30-Degrees (ODT) survey (see MacDonald et al. 2004). 
The ODT survey is a multi-band imaging survey, with limiting depths 
(5$\sigma$, 2$''$ aperture Vega magnitudes) of $U$ = 25.3, $B$ = 26.2, 
$V$ = 25.7, $R$ = 25.4, $i'$ = 24.6, $Z$ = 21.9 and $K$ = 18.5. 
In the case where ODT data were unavailable, estimates of 
the photometry were made from the optical spectrum of the object when available.

Near-IR imaging in the $K-$band was obtained at the United Kingdom Infrared 
Telescope UKIRT\footnote{The United Kingdom Infrared Telescope is 
operated by the Joint Astronomy Centre on behalf of the Science and Technology 
Facilities Council of the U.K.} by using either UIST (Howatt et al.\ 2004) in 
January 2003 or UFTI (the UKIRT Fast--Track Imager; Roche et al. 2002) in 
July 2000, December 2000, January 2001 and 2002 and September 2004. Further 
infrared observations of a $z \sim$ 1 TOOT00 sub-sample were carried out 
at the UKIRT in 2003 with UIST and in 2004 with UFTI with typical exposures of 
$\sim$ 9-18 min; we term this the `Inskip' sub-sample. UIST is a 
1--5 $\mu$m imager-spectrometer with a 1024$\times$1024 InSb array. 
The instrument is designed to switch quickly and accurately between imaging and 
spectroscopic modes. We used the imaging mode, where two plate scales are 
available, 0.12 arcsec / pixel or 0.06 arcsec / pixel, giving fields of 
view of 2$\times$2 or 1$\times$1 arcmin$^{2}$. UFTI is a 1--2.5 $\mu$m camera 
with a $1024 \times 1024$ HgCdTe array and a plate scale of 0.091 arcsec / 
pixel, which gives a field of view of 92 arcsec. Each object in the `Inskip' 
sub-sample was observed in the $J$, $H$ and $K$ infrared wavebands. The 
$K$-band observations were generally of $\sim 2$ hours duration in 
total. $J$-band observations were typically 18-63 minutes, and $H$-band 
observations typically 18-72 minutes (depending on source redshift). 
Table A1 (see Appendix A) shows the photometric data used to 
calculate photometric redshifts, and Table A2 shows complete photometry 
on each object. 

All near-IR observations used a nine-point jitter pattern, with offsets of
roughly 10 arcsec between each 1 minute exposure. The observational data 
were dark subtracted, and masked for bad pixels. A first-pass sky 
flat-field image was created from groups of roughly 60--100 consecutive 
frames of data, which were combined and median filtered. This accounted 
for the majority of the pixel-to-pixel variations of the chip, but some 
larger-scale illumination gradients remained, due to the changing 
position of the telescope through the course of the night. Smaller 
blocks of 9--18 first-pass flat-fielded images were similarly 
combined to create residual sky flat-field images, allowing us to remove 
the remaining variations. The first-pass flat-fielded data were used to 
generate masks for any bright objects in the images, and a repeat of the 
whole process allowed the data to be cleanly flat-fielded without any 
contamination from stars or galaxies.

The flat-fielded data for each source were sky-subtracted and combined 
using the IRAF package DIMSUM, creating a final mosaiced image of 
approximately $115 \times 115$ arcsec$^2$\footnote{The highest 
signal-to-noise level is restricted to the central $70 \times 70$ 
arcsec$^2$.}, which was flux calibrated using observations of UKIRT faint 
standard stars. Astrometry corrections were performed to the finalised 
near-IR images using the package \textsc{karma}. 
Photometry was carried out within apertures ranging from 
3 to 9 arcsec in diameter, using the IRAF routine \textsc{apphot} and a 
single sky annulus. All sources have been corrected for galactic 
extinction using the $E(B-V)$ for the Milky Way from the NASA 
Extragalactic Database (NED) and the parametrised galactic extinction law 
of Howarth (1983).

Optical and near-infrared photometric data were used to create SEDs for our 
objects as shown in Figure~\ref{seds} (in Appendix A). Seven of our objects 
do not as yet have spectroscopically-confirmed redshifts. Table 3 summarises 
the results on the spectroscopic and photometric redshifts, as well as the $K$ 
magnitude of each object and whether or not the IDs look resolved in the 
$K-$band. It also gives the Optical (see Section~\ref{sec:optclass}) and 
Radio-Optical (see Section~\ref{sec:distribution}) classification of the TOOT00 
radio sources. 

Table 4 presents the measured observed FWHM (in km/s) and EW (in $\rm \AA$) of 
each line, the radio luminosity at 151 MHz $L_{151 \rm MHz}$ and 
the flux of the [OII]$_{\lambda3727}$ emission line, since it is the 
most frequently observed line in the optical in quasars and radio galaxies. 
In the case where the [OII] line was not present, we measured another narrow 
line (e.g.\ [OIII]$_{\lambda5007}$), or any narrow-line profile of the 
Ly$\alpha$, [CIV]$_{\lambda1549}$, [CIII] or MgII emission lines.  
For objects without emission lines in the optical spectrum, a rough estimate 
of a limit on the [OII] line flux was made as follows: i) for galaxies, i.e. 
in the case the continuum is clearly strong and stellar, we took a limit of 
$\sim$ 10 $\rm \AA$ which corresponds roughly to lines clearly detectable in 
a typical spectrum of 
$z \sim$ 0.5 ellipticals, observed with moderate S/N ratio; ii) for fainter 
objects in which the origin of the continuum is unclear, we supply as a limit 
the flux of the brightest, but probably spurious, line-like feature in the 
spectrum adjacent to the estimated location (given the photometric redshift) 
of the brightest expected narrow line.

\clearpage
\addtocounter{table}{0}

\scriptsize
\begin{table*}
\begin{center}
 {\caption[Radio properties of the 47 TOOT00 radio sources]
{\label{tab:sample_toot00} Radio properties of the 47 TOOT00 radio 
sources; values in bold are those used in the analysis. {\bf Columns 1, 2 \& 
3} give the name of the object and its 151-MHz radio position from the 
7C survey (J2000.0). {\bf Column 4} gives the angular size of the radio 
source (A- or B-array in line with Figure~\ref{seds}; the characters 
`os' denote that the measurement was made from the core to the extent of 
an one-sided structure). {\bf Columns 5 \& 6} give the 
peak flux density and integrated flux density per beam at 151 MHz 
respectively. {\bf Column 7} gives the signal-to-noise (S/N) ratio at 151 
MHz. {\bf Column 8} gives the flux density at 1.4 GHz as 
was measured from the B-array VLA maps using \textsc{IMSTAT} task in 
\textsc{AIPS}; the symbol `+' 
indicates that the flux density is the sum of two components. 
{\bf Columns 9 \& 10} give the flux density at 1.4 GHz taken from the NVSS 
survey; the symbol `+' indicates that the flux of each component was 
measured separately and then added together. {\bf Columns 11 \& 12} give the 
peak flux density and integrated flux density per beam at 74 MHz from the VLSS 
survey; objects that have flux densities below $\simeq$ 0.3 Jy are marked as 
2$\sigma$ limits. Flux 
densities measured directly from the VLSS maps using the \textsc{IMSTAT} task 
in \textsc{AIPS} are marked with an `m'. {\bf Columns 13 \& 14} give 
the spectral index of each object calculated from 1.4 GHz to 151 MHz and 
151 MHz to 74 MHz respectively. Fluxes in bold have been used to calculate the 
spectral index, making use of the NVSS flux density for each one of the radio 
sources. 
{\bf Column 15} gives the radio classification (R Cl) of each object (Owen \& 
Laing, 1989): CD stands for classical double, TJ for twin-jet, FD for fat 
double and COM for a compact radio source.
}}
\begin{picture}(50,180)
\put(-280,-470){\includegraphics{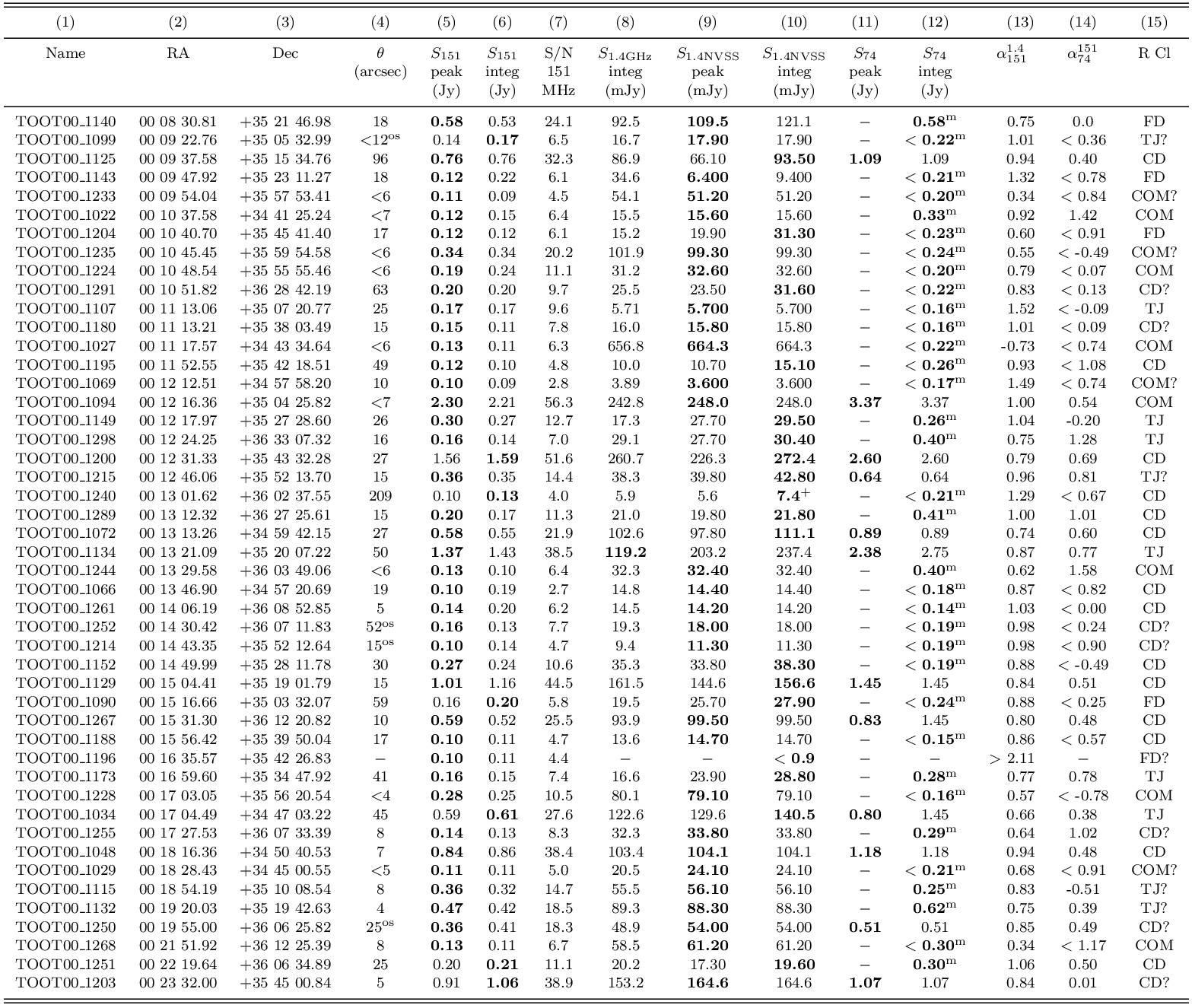}}
\end{picture}
 \end{center}
 \end{table*}

\normalsize
\clearpage

\addtocounter{table}{0}

\scriptsize
\begin{table*}
\begin{center}
 {\caption[Table~\ref{tab:spectralog}]{
Observing log for the optical spectroscopy: In {\bf Column 1} we give the name 
of the object. {\bf Column 2} gives the observing date. {\bf Column 3} gives 
the coordinates where the slit was centred. {\bf Column 4} gives the 
exposure time (in seconds) in the blue `B' and the red `R' part of the WHT ISIS 
spectrum; `LRS' denotes spectroscopy taken with the HET. {\bf Columns 5 \& 6} 
give the width and the position angle of the long axis of the slit respectively.
}}
\begin{picture}(50,180)
\put(-250,-535){\includegraphics{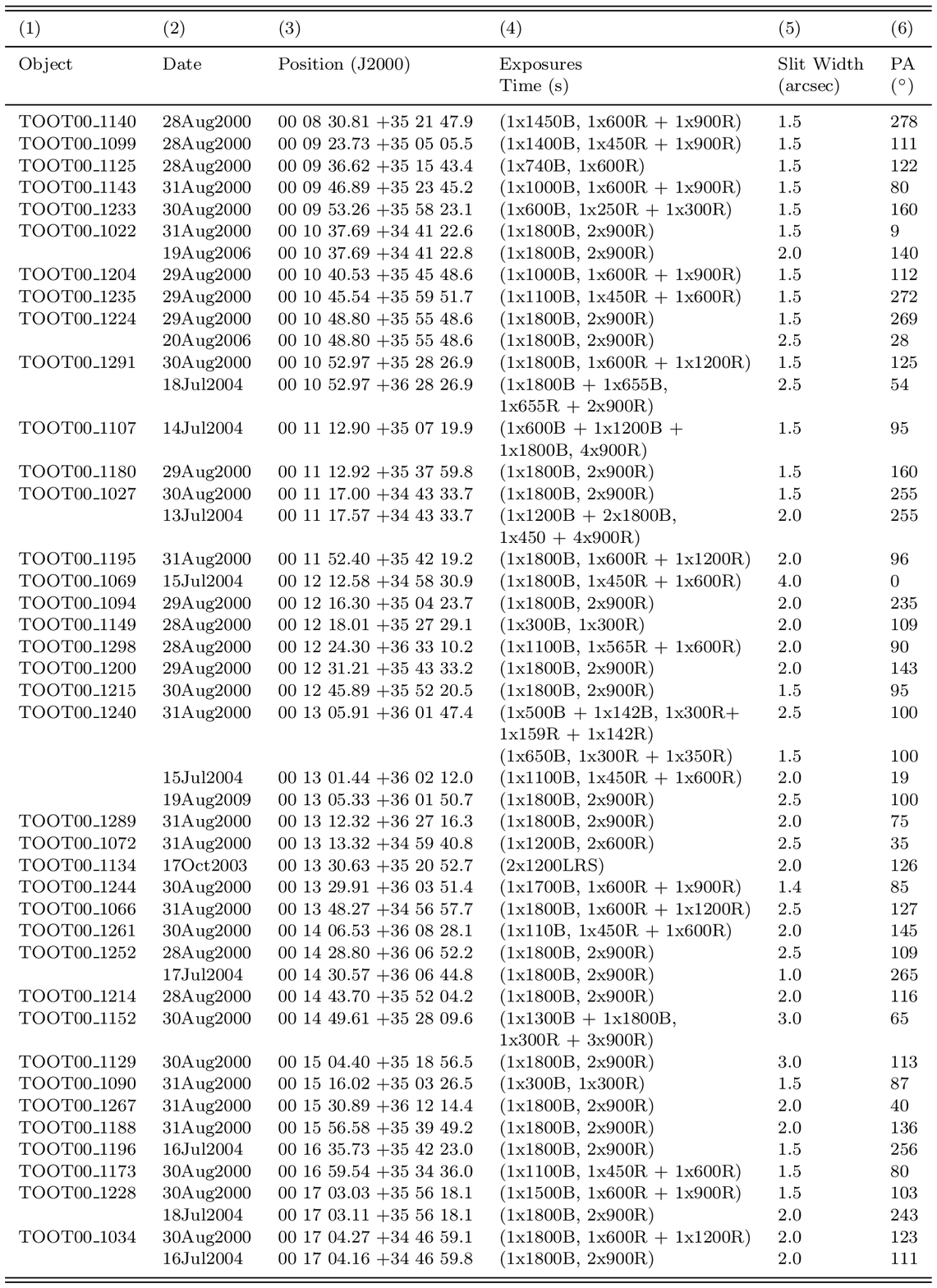}}
\end{picture}
\end{center}
 \end{table*}
\normalsize
\clearpage
\addtocounter{table}{-1}

\scriptsize
\begin{table*}
\begin{center}
{\caption[Table~\ref{tab:spectralog}]{
(continued)
}}
\begin{picture}(50,180)
\put(-250,-535){\includegraphics{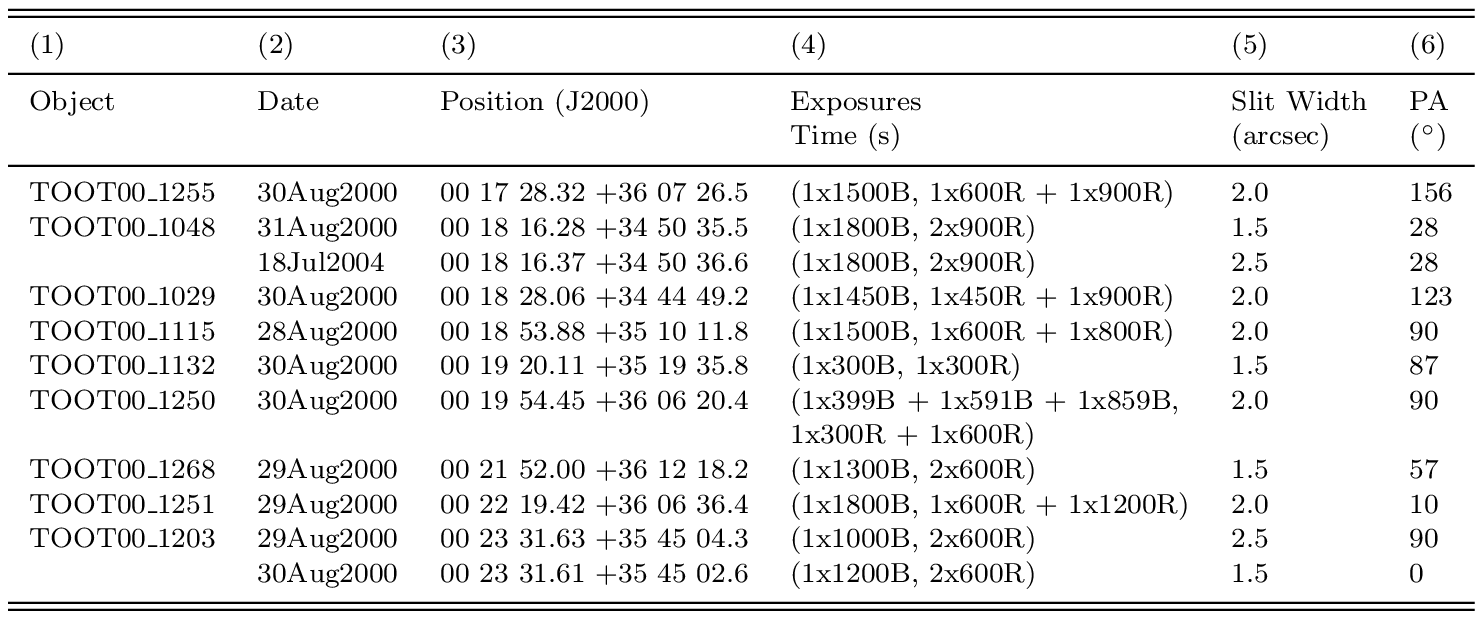}}
\end{picture}
\end{center}
 \end{table*}
\normalsize
\clearpage
\addtocounter{table}{0}

\scriptsize
\begin{table*}
\begin{center}
 {\caption[Spectroscopic and photometric redshifts]{\label{tab:class}
Spectroscopic and photometric redshifts: 
{\bf Column 1} gives the name of the object and {\bf Columns 2 \& 3} give 
the RA and Dec of the near-IR $K$-band identification of each radio source 
respectively. {\bf Column 4} gives the optical classification (OpCl) of the 
radio source, where `G' stands for a galaxy and `Q' for a quasar; `a' denotes a 
broad-absorption-line QSO (BALQSO). {\bf Column 5} presents the radio-optical 
classification (R-OpCl): FRIIs correspond to CD objects; FRIs to FD and TJ; 
Q-F corresponds to quasars with flat radio spectral index (`F' denotes flat 
radio spectral index $\alpha^{1.4 \rm GHz}_{151 \rm MHz} \lta$ 0.5; a `?' denotes 
$\alpha$ close to 0.5); CSS-Q corresponds to quasars with compact radio 
structure; FRII-Q denotes a quasar that has FRII radio structure; CSS denotes 
a compact steep spectrum radio source. {\bf Column 6} gives the spectroscopic 
redshift as measured from the spectra (Figure~\ref{seds}); a `?' denotes 
uncertainty or a featureless continuum. {\bf Column 7} gives the 
photometric redshift calculated from HyperZ using BC templates (see 
Section~\ref{sec:optclass}); redshifts 
corresponding to secondary best-fit templates are marked with `2nd'. Values 
in brackets are deemed uncertain because there is no clear step change in 
flux associated with the 4000-$\rm \AA$ break in the SED (see 
Figure~\ref{seds}). {\bf Column 8} gives redshifts estimated from the $K-z$ 
relation of Willott et al.\ (2003) as described in Section~\ref{sec:optclass}. 
{\bf Column 9} gives the 
$K$ magnitude of each object in the Vega system, measured with 4-arcsec (or 
3-arcsec in case of confusion) diameter aperture from the centre of its 
infrared position, denoted as `(4)' (or `(3)'); in the absence of such a 
measurement we use the one measured from the ODT survey (MacDonald et al.\ 
2004), marked with an (O). The symbol `p' is used in the case where the object 
is a point source at $K$; applicable for magnitudes $K <$ 18 as a 
$\sim$10$\sigma$ signal-to-noise ratio is necessary to clarify whether or not 
an object is a point 
source at $K$. The symbol `$*$' denotes data from the `Inskip' near-IR sample 
(see Section~\ref{sec:sample}). 
}}
\begin{picture}(50,160)
\put(-240,-512){\includegraphics{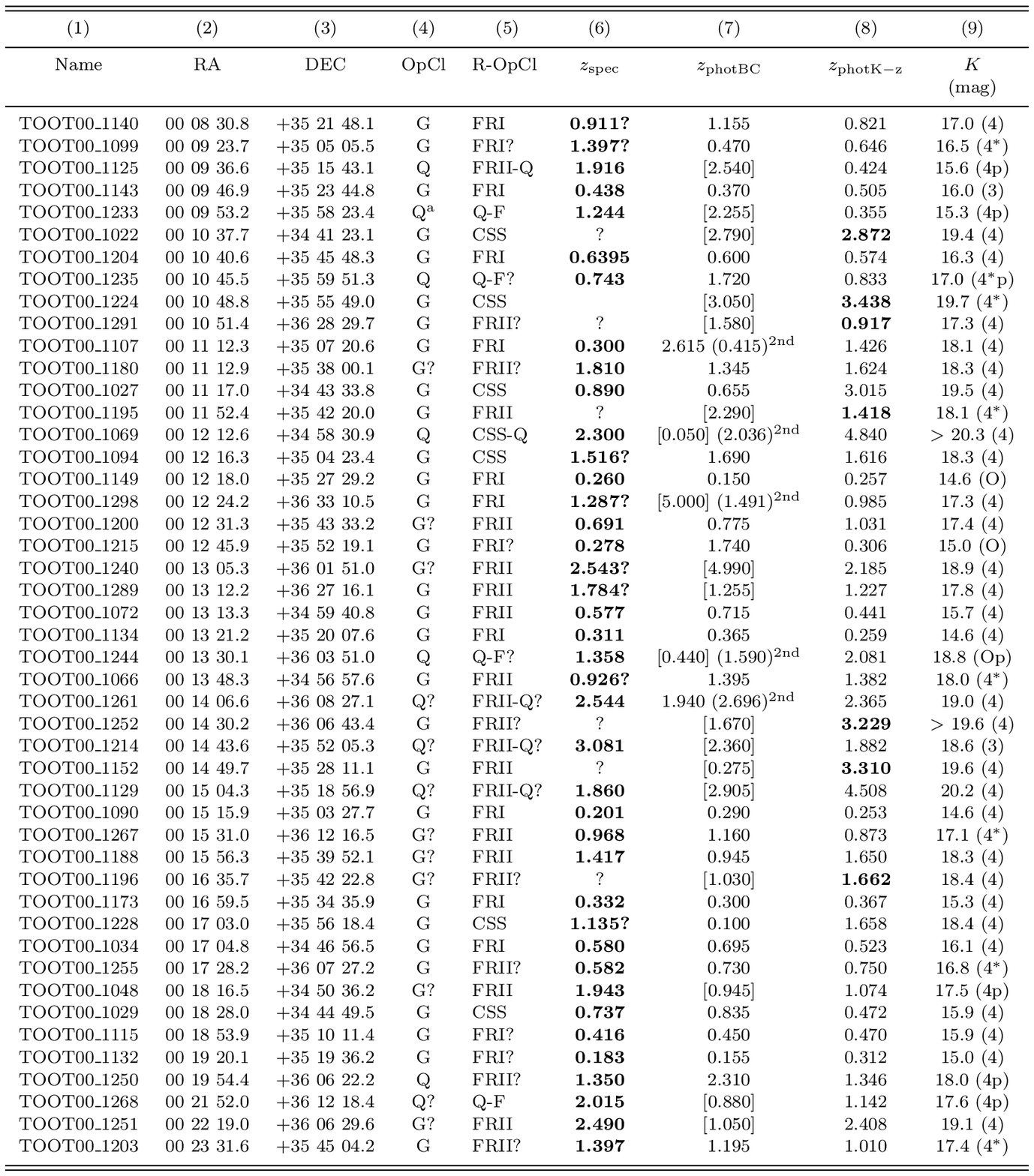}}
\end{picture}
\end{center}
 \end{table*}

\normalsize

\clearpage
\addtocounter{table}{0}

\scriptsize
\begin{table*}
\begin{center}
 {\caption[Emission line properties]{\label{tab:line} Emission line 
properties: {\bf Column 1} gives the name of the object. {\bf Columns 2 \& 3} 
give the measured (observed) values of the equivalent width (EW in $\rm \AA$) 
and the FWHM (in km/s) of the most prominent emission line in the optical 
spectrum. {\bf Column 4} gives the flux of the [OII]$_{3727}$ emission 
line, $S_{line}$ in units of $10^{-19}$ $\rm W m^{-2}$, as measured from the 
spectrum of each object; in the case where the [OII] line is not present we 
use the brightest available line as shown; `n' denotes measurement of a narrow 
line profile at FWHM using a Gaussian fit; a `---' is used when no spectroscopy 
is available. {\bf Column 5} gives the emission line used in 
the calculations. {\bf Column 6} gives the base-10 logarithm of the rest-frame 
luminosity at 151 MHz, which is used in Figures~\ref{l151a},~\ref{l151_z_ts} 
\&~\ref{l151d}.
}}
\begin{picture}(50,180)
\put(-240,-535){\includegraphics{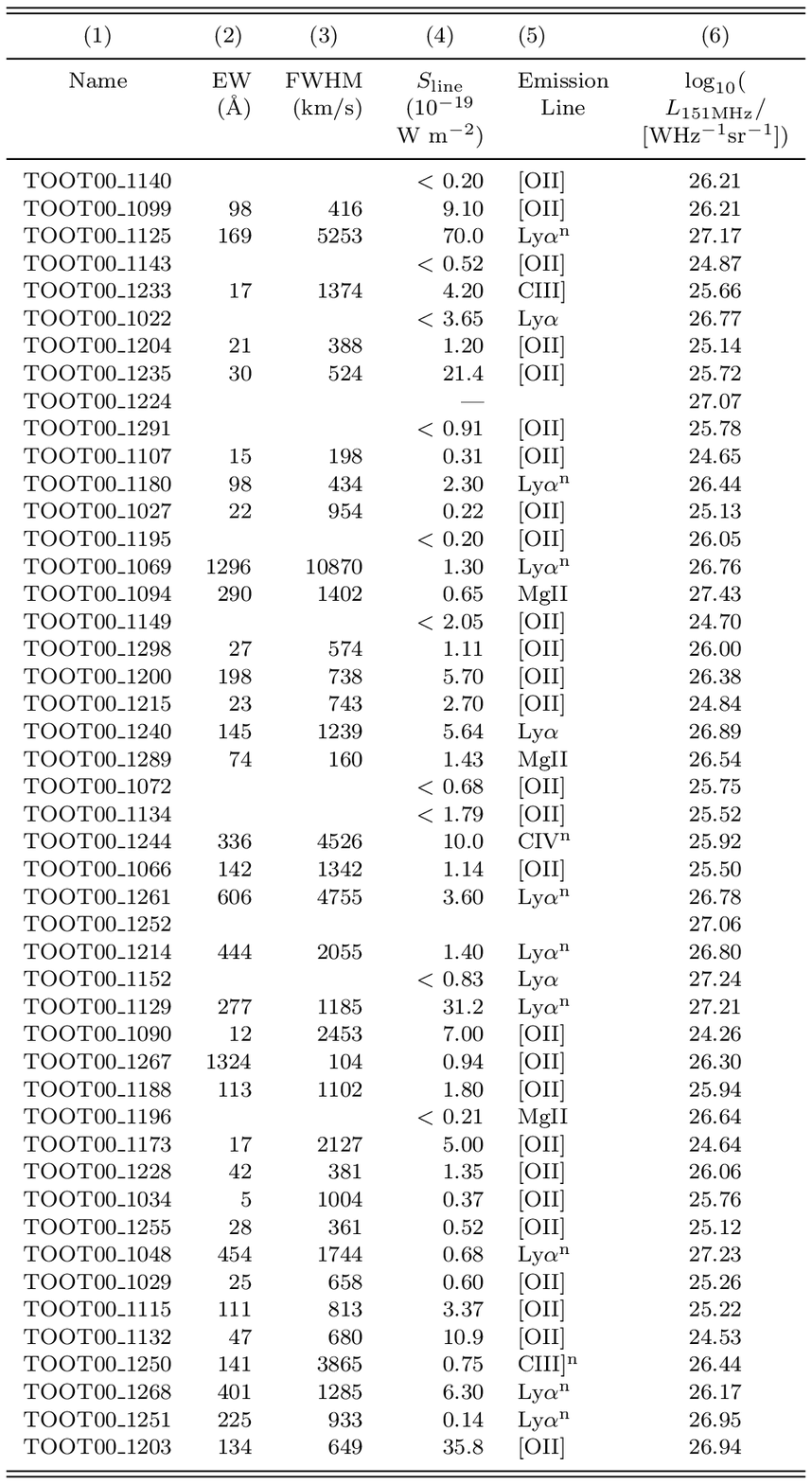}}
\end{picture}
\end{center}
 \end{table*}

\normalsize

\clearpage

\newpage

\begin{figure*}
\begin{center}
\setlength{\unitlength}{1mm}
\begin{picture}(150,70)
\put(75,-15){\includegraphics{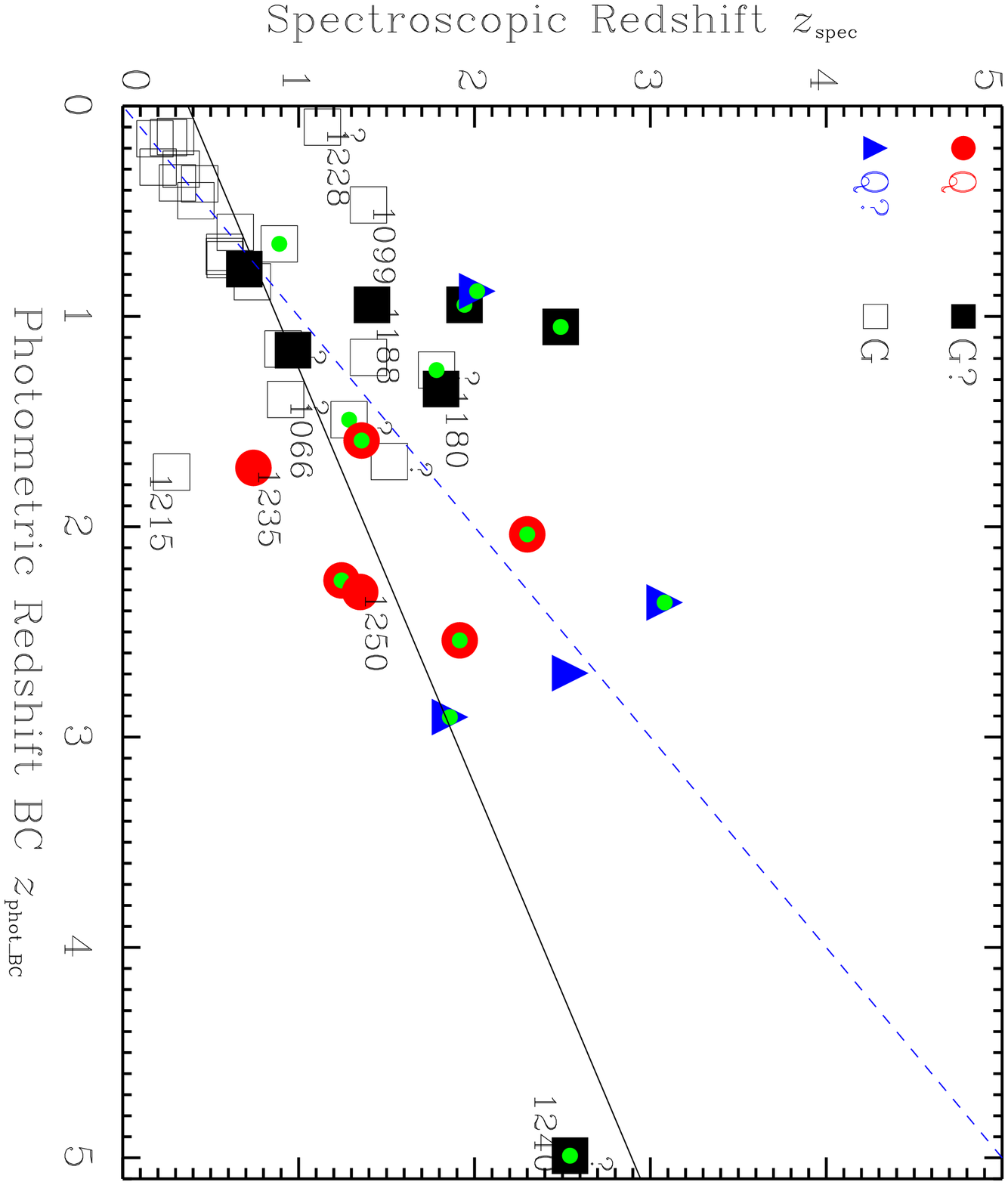}}
\put(165,-15){\includegraphics{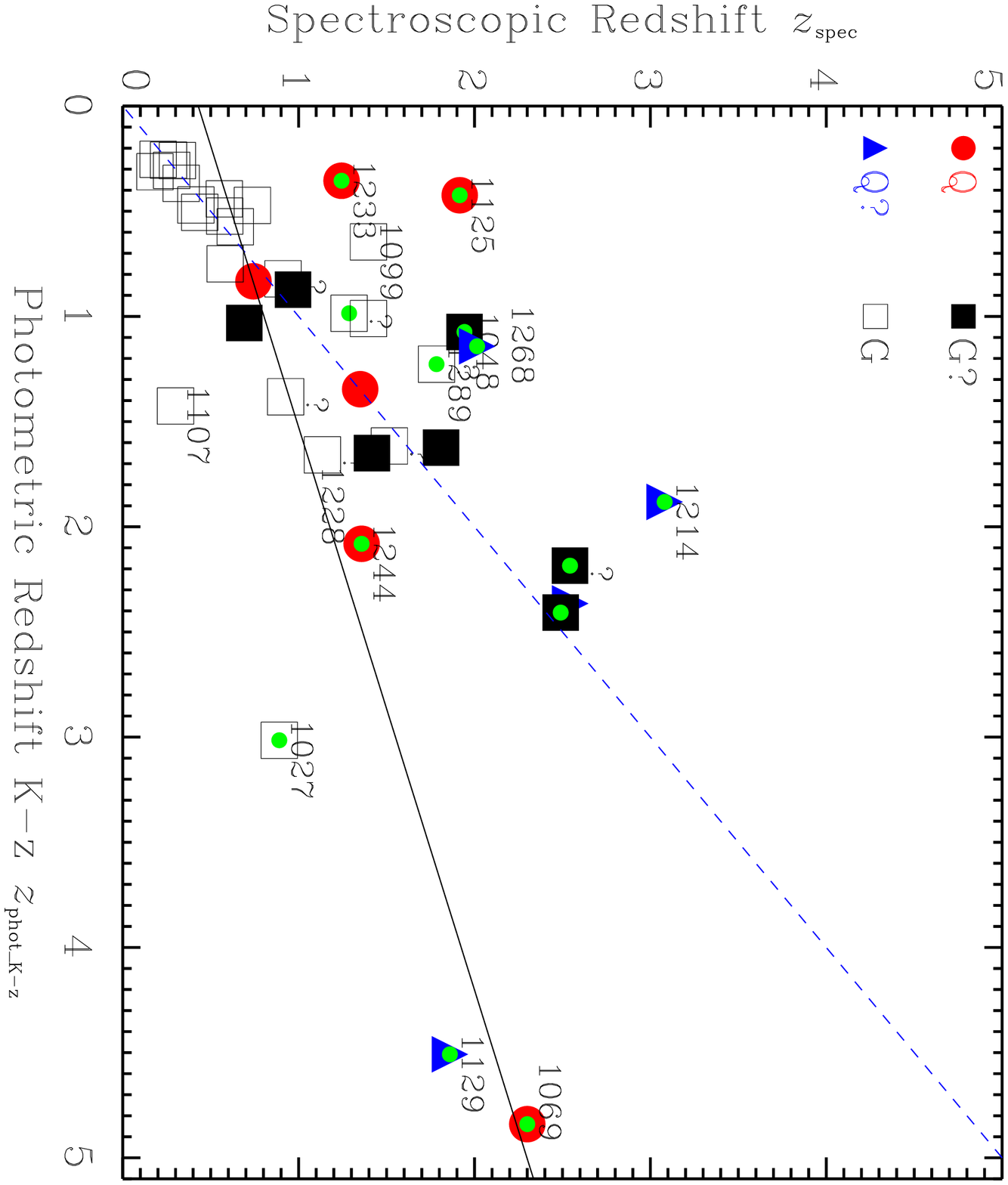}}
\end{picture}
\end{center}
\vspace{.4in}
{\caption[Spectroscopic vs photometric redshift]{\label{zspzph}
{\bf a) Left}: Spectroscopic versus photometric redshift calculated from 
HyperZ. Symbols indicate optical classification as explained in 
Section~\ref{sec:optclass}: filled circles for quasars `Q', blue filled 
triangles for possible quasars `Q?', black filled squares for possible 
galaxies `G?', and black squares for secure galaxies `G'. The black solid line 
is the best fit line for G objects in the $z_{\rm spec}$ vs $z_{\rm phot BC}$ 
correlation, and was calculated using the \textsc{LINFIT} function in 
\textsc{IDL}, where the objects without spectroscopic redshifts were not 
included in the calculation: 
$z_{\rm spec} = 0.50\times z_{\rm phot BC} + 0.37$; the linear Pearson 
correlation is 0.548 and the Spearman correlation is 0.568 with a 99.53\% 
probability for a correlation.  Objects with uncertain photometric redshift are 
marked with a filled green circle; the character `?' denotes objects with 
an uncertain spectroscopic redshift. Sources that fall off the dashed blue 
$z_{\rm spec} = z_{\rm phot}$ 
line, and do not have an uncertain photometric redshift, are marked with their 
name. The 7 objects without spectroscopic redshift have been omitted.
{\bf b) Right}: Spectroscopic versus photometric redshift calculated using 
the $K_{\rm W}-z$ relation. Symbols are the same as in figure {\bf a} on the 
left. The black solid line is the 
best fit line for G objects in the $z_{\rm spec}$ vs $z_{\rm phot K-z}$ 
correlation, and was calculated as stated above: 
$z_{\rm spec} = 0.37\times z_{\rm phot K-z} + 0.43$, where the linear Pearson 
correlation is  0.514 and the Spearman correlation is 0.762 with a 99.99\% 
probability for a correlation. The character `?' denotes objects with an 
uncertain spectroscopic redshift; objects with uncertain photometric redshift 
from the HyperZ code are marked with a filled green circle. Sources with 
$|z_{\rm spec} - z_{\rm phot}| > 0.5$ are marked with their name. Again, 
the 7 objects without spectroscopic redshift have been omitted. The 
generalised Spearman correlation coefficient is calculated using survival 
analysis statistics ASURV (Lavalley et al. 1992).
}}
\end{figure*}

\section{Analysis}
\label{sec:analysis}

\subsection{Optical classification and redshifts}
\label{sec:optclass}

The optical spectroscopy of the 47 TOOT00 radio sources provided 40 out 
of 47 redshifts. For the other 7 objects there was either a lack of emission 
lines, or the spectrum was totally blank. Thus the TOOT00 radio-source 
sample is 85\% spectroscopically complete. Using the results of the 
spectroscopy, and in combination with individual characteristics for each of 
them, we classify the 47 TOOT00 radio sources according to the 
following optical criteria:\\
i) {\bf Q}: Objects with definite broad lines in their optical spectrum 
(6/47); all of them have blue SEDs and all apart from one (TOOT00\_1069) are 
point sources at $K$.\\
ii) {\bf Q?}: Objects without definite broad lines in the optical spectrum, 
but that are either point sources at $K$, or there is a clear hint of broad 
lines in the optical spectrum, or both (4/47); all objects have a blue SED.\\
iii) {\bf G?}: Evidence from the optical spectrum that at least one 
high-excitation (see e.g. Jackson \& Rawlings 1997) narrow emission line 
exists (8/47).\\
iv) {\bf G}: All the objects that don't fall in the previous categories 
(29/47); all cases are resolved at $K$.\\

One of the TOOT00 `Q' objects, TOOT00\_1233, is possibly a 
broad-absorption-line QSO (BALQSO; see notes on objects in Appendix A), as 
shown in Figure~\ref{seds}.

We calculate photometric redshifts $z_{\rm phot BC}$ with the HyperZ 
photometric code (Bolzonella et al.\ 2000) using photometry from $U-$band 
($>$ 3600 $\rm \AA$) to the $K-$band (2.2 $\mu \rm m$) and Bruzual-Charlot 
(Bruzual \& Charlot 2003; BC) templates. In nearly all cases we let reddening 
(Calzetti et al. 2000) be a free parameter (see notes on objects in 
Appendix A for a few exceptions), but the derived values of A$_{\rm V}$ were 
typically small. Results are shown in Table 3. In many cases we 
put the value obtained in square brackets, as the probability density function 
for the redshift is not sharply peaked, typically because the SED is not 
galaxy-like or is too sparsely sampled above the putative 4000 $\rm \AA$ break. 

In Figure~\ref{zspzph}a we present a comparison of spectroscopic and 
photometric redshifts. The Q and Q? type objects fall off the correlation 
line, which is unsurprising because all but two of them have a highly 
uncertain photometric redshift due, presumably, to significant contributions to 
their SEDs from non-stellar light causing the galaxy HyperZ templates to be 
inappropriate; the exception is TOOT00\_1244 for which the spectroscopic and 
photometric $z_{\rm phot BC}$ have similar values. The `G' type objects 
with high $z_{\rm spec}/z_{\rm phot BC}$ typically fall off the correlation, 
because they have a highly uncertain $z_{\rm phot BC}$; the exceptions being 
TOOT00\_1099 and TOOT00\_1228, although the latter has an uncertain 
spectroscopic redshift. In TOOT00\_1215 the disagreement between spectroscopic 
and photometric HyperZ redshift probably reflects the fact that more 
photometric points are needed for an 
accurate fit of the BC template. In Figure~\ref{zspzph}b we see that quasars 
tend to have lower photometric redshifts than their spectroscopic ones, with 
the exception of TOOT00\_1069, presumably because their $K$-band light is 
contaminated by non-stellar emission. Objects TOOT00\_1027, TOOT00\_1107, 
TOOT00\_1129, TOOT00\_1069 are all seemingly intrinsically faint objects in 
the near-IR $K$-band, which yields the high $z_{\rm phot K-z}$ value (see 
Willott et al. 2003).

Due to a $\sim$85\% (40 out of 47 objects) spectroscopic completeness we only 
need to use photometric redshifts for 7 out of the 47 objects. 
For these 7 objects we use a photometric redshift calculated from the $K-z$ 
relation of Willott et al. (2003; hereafter $K_{\rm W}-z$ relation). The 
$K_{\rm W}-z$ fits the 3CRR, 6CE and 7CRS data sets, where the $K$ magnitude 
is aperture and emission line corrected:
\begin{equation}
K_W = 17.37 + 4.53\times(\log_{10} z) - 0.31\times(\log_{10} z)^2,
\end{equation}
where $K$ is the $K-$band magnitude and $z$ the redshift. 

When using the $K_{\rm W}-z$ relation we kept in mind that the $K$ magnitude 
of our sample is not emission-line or aperture corrected. This means that the 
$K_{\rm W}-z$ relation may give a slightly incorrect estimate 
for the photometric redshift of our objects. We believe, though, that these 
errors are negligible for our sample. According to Willott et al. (1999), one 
expects the line flux to scale with radio flux density at $K$-band, so the 
line contamination of the $K$-band magnitude should be small in TOOT. 
Additionally, we chose not to perform aperture 
or emission line corrections in our $K$ magnitude for the following 
reasons: (i) above redshift $z > $ 0.6 aperture corrections are $\sim$ 0.05 
mag (Eales et al.\ 1997); (ii) most of the objects have $K$-band magnitude 
measured in the same aperture, apart from some exceptions due to confusion 
by a nearby object (see Table 1 and Appendix A) and any corrections would be 
highly dependent on the assumed light spatial profile which is uncertain. 
In our analysis, whenever we use the redshift calculated from the 
$K_{\rm W}-z$, it is only for objects at high redshift $z >$ 0.9, where line 
and aperture corrections, although uncertain, are probably small.

\subsection{Radio classification, spectral indices and sizes}
\label{sec:alpha}

Figure~\ref{seds} (see Appendix A) presents near-IR/radio overlays for all 
the TOOT00 objects, where either the A- or B-array VLA maps were used. Another 
classification we use is based on their radio structure, as shown in Table 1. 
In the same figure we also present optical spectra and SEDs that were 
constructed from the photometric data (Table A1 in Appendix A), where the 
best fit BC templates are shown. In the cases where optical photometry was 
unavailable from the ODT, the optical magnitude was estimated directly from 
the spectrum; a mean value of the continuum was used to estimate the flux 
density in each optical band where the fluxes measured were approximately 
corrected for systematic errors due to light losses from the slit. This value 
is approximately 25\% for a 1.5-arcsec wide slit; we chose to use a value 
calculated for each object of $\sim$ 10\% to 40\% dependent on the slit width 
(see Table 2). The errors on the magnitudes (see Table A1 in Appendix A) are 
taken to be equal to the percentage errors estimated for light losses from the 
slit.

\begin{figure}
\begin{center}
\setlength{\unitlength}{1mm}
\begin{picture}(150,50)
\put(81,-20){\includegraphics{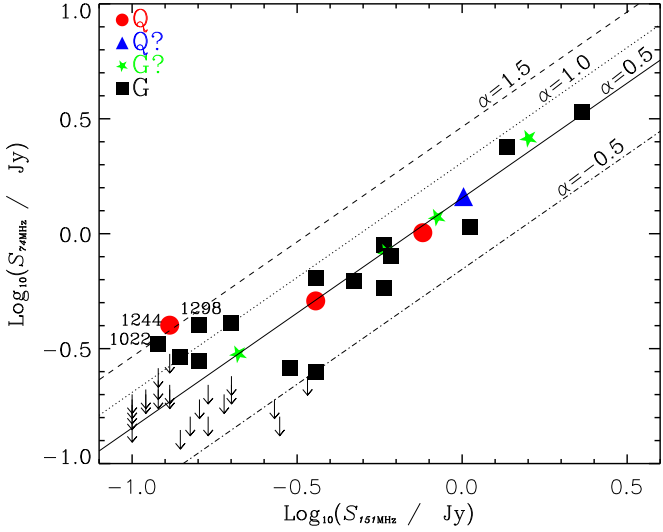}}
\end{picture}
\end{center}
\vspace{.5in}
{\caption[$S_{74 \rm MHz}$ vs $S_{151 \rm MHz}$]{\label{s74s151}
Flux density at 74 MHz $S_{74 \rm MHz}$ versus flux density at 151 MHz 
$S_{151 \rm MHz}$. Symbols: red circles denote quasars `Q', blue triangles 
denote possible quasars `Q?', green stars stand for possible galaxies `G?', 
and black squares for secure galaxies `G'. 12 objects were detected by the 
VLSS and plotted with their symbol. Another 11 have flux densities at 74 MHz 
measured directly from the VLSS map (see Table 1). Thus, 23 out of 47 TOOT00 
objects are detected both at 74 and 151 MHz. The other 24 objects, 
that have no measurement at 74 MHz, are indicated as 2$\sigma$ upper limits 
and plotted with black arrows.  The solid black line is for $\alpha$ = 0.5, the 
dotted black line for $\alpha$ = 1.0, the dashed black line for $\alpha$ = 1.5 
and the dotted-dashed line for $\alpha$ = -0.5.
}}
\end{figure}

In Figure~\ref{s74s151} we present the $S_{74 \rm MHz}$ versus $S_{151 \rm MHz}$ 
diagram. The VLSS catalogue provides flux densities for 12 (out of 47) TOOT00 
radio sources. Direct measurements from the 74-MHz maps were made for a 
further 11 TOOT00 objects that were not included in the VLSS catalogue, but 
appeared to be detected in the 74 MHz maps; the measurements 
were performed using the {\textsc IMSTAT} package in {\textsc AIPS}. 
For the other 24 TOOT00 objects, a 2$\sigma$ detection was adopted as an 
upper limit (see Table 1). The median spectral index calculated between 74 and 
151 MHz including the limits is $\alpha^{151 \rm MHz}_{74 \rm MHz} \sim$ 0.5. 
This confirms, as expected, that the average low-frequency radio spectra 
approach the $\alpha \sim$ 0.5 value expected in emission from un-aged 
electron populations at the low rest-frame frequency of $\sim$ 200 MHz (using 
the median redshift $z \sim$ 1.25 of the sample; see 
Section~\ref{sec:distribution}). The fact that we have almost the same 
amount of limits above and below $\alpha$ = 0.5 is re-assuring. 
The overplotted lines in Figure~\ref{s74s151} indicate that the TOOT00 radio 
sources not detected at 74 MHz are mainly faint enough sources at 151 MHz that 
with any reasonable radio spectral index they would fall close to the 
limit of the VLSS survey. In Figure~\ref{s74s151}, we label three cases 
(TOOT00\_1022, TOOT00\_1244 \& TOOT00\_1298) which, in Table 1, 
have $\alpha^{151 \rm MHz}_{74 \rm MHz} >$ 1.2. In all of these cases the 
signal-to-noise ratio is $\leq$ 7 at 151 MHz and $\leq$ 4 at 74 MHz meaning 
much lower true values of $\alpha^{151 \rm MHz}_{74 \rm MHz}$ are within the 
errors.

\addtocounter{figure}{0}

\begin{figure}
\begin{center}
\setlength{\unitlength}{1mm}
\begin{picture}(150,50)
\put(83,-20){\includegraphics{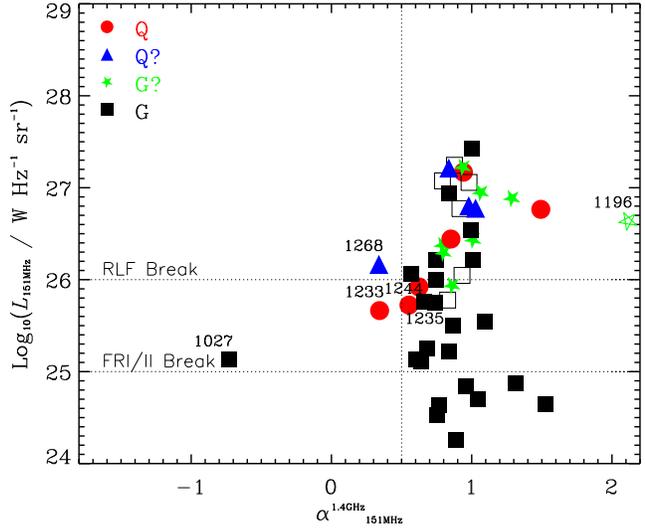}}
\end{picture}
\end{center}
\vspace{.5in}
{\caption[$L_{151 \rm MHz}$ vs $\alpha$]{\label{l151a}
Radio Luminosity at 151 MHz $L_{151 \rm MHz}$ versus the radio spectral 
index $\alpha$. Symbols are the same as in Figure~\ref{s74s151}; filled 
symbols denote objects with spectroscopic redshifts and open symbols represent 
objects with photometric redshifts. The horizontal lines show the RLF and 
FRI/FRII breaks calculated from the Willott et al. (2003) and the Fanaroff \& 
Riley (1974) values respectively using a typical steep-spectrum radio
spectral index of 0.8. The vertical line at $\alpha = 0.5$ shows the 
conventional border between steep and flat-spectrum sources.
}}
\end{figure}
\addtocounter{figure}{0}

\begin{figure*}
\begin{center}
\setlength{\unitlength}{1mm}
\begin{picture}(150,70)
\put(75,-15){\includegraphics{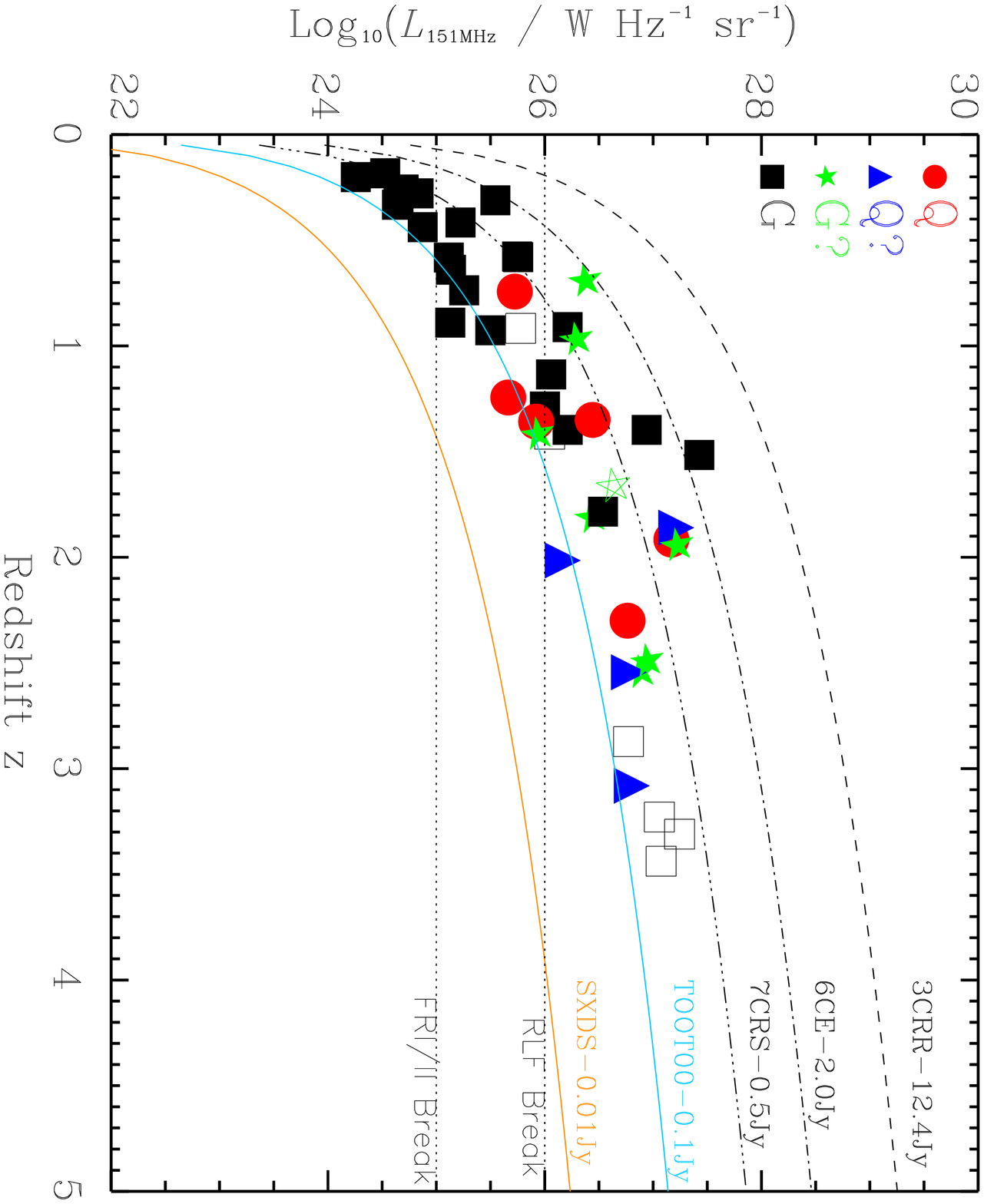}}
\put(165,-15){\includegraphics{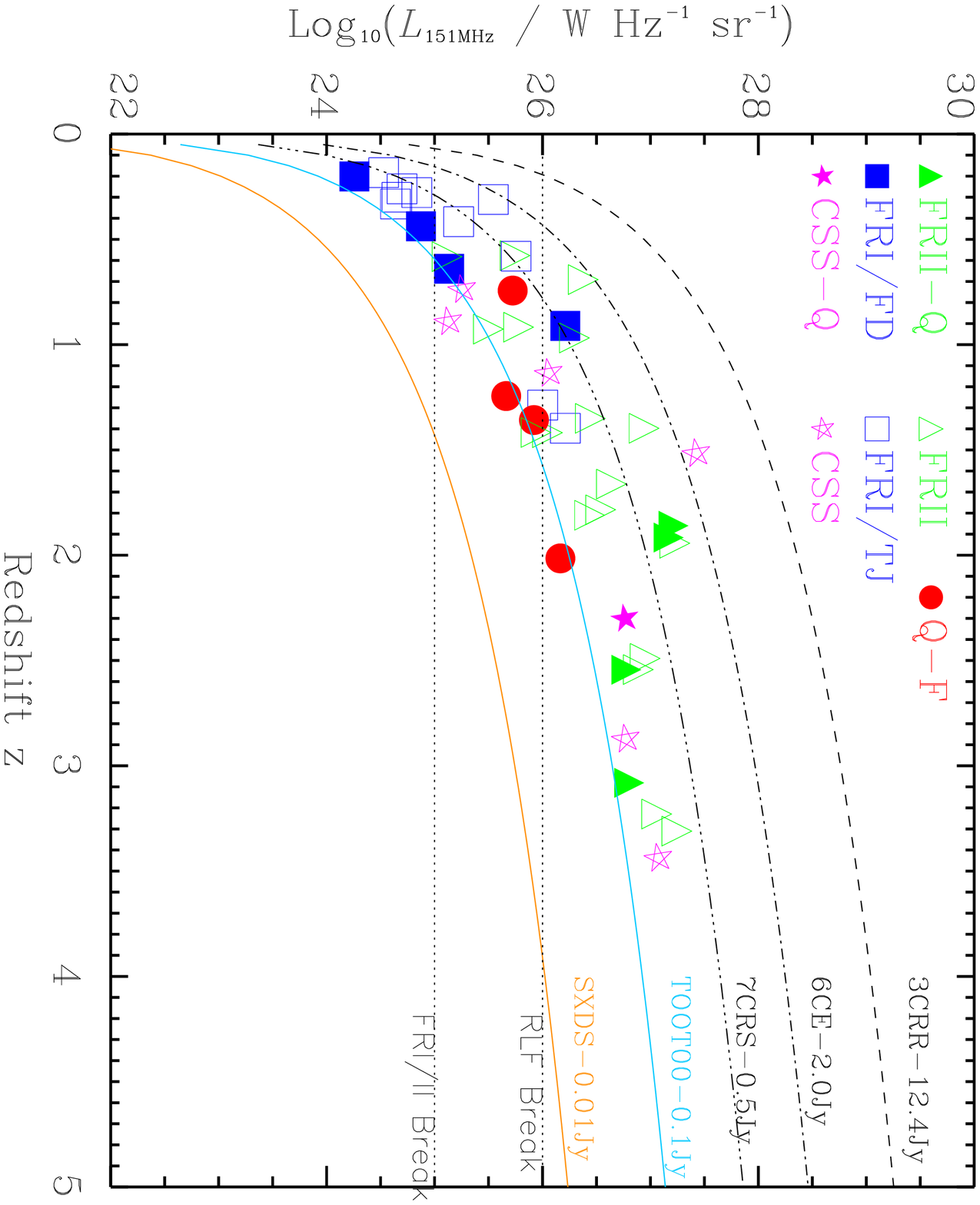}}
\end{picture}
\end{center}
\vspace{.5in}
{\caption[Radio Luminosity at 151 MHz $L_{151 \rm MHz}$ versus redshift 
$z$.]{\label{l151_z_ts}
Radio Luminosity at 151 MHz $L_{151 \rm MHz}$ versus redshift $z$ for the 
TOOT00 sample. 
{\bf a) Left}: Symbols are the same as in  Figure~\ref{s74s151}; filled symbols 
represent objects with spectroscopic redshift and open symbols denote objects 
with photometric redshift. {\bf b) Right}: Symbols denote radio-optical 
classification: green filled triangles denote FRII-Q objects; green 
open triangles denote FRIIs; blue filled boxes denote FD type FRIs; blue open 
boxes denote TJ type FRIs; red filled circles denote Q-F objects; purple 
filled stars denote CSS-Q objects; purple open stars denote CSS objects; a '?' 
denotes uncertainty in the radio-optical classification as stated in Table 3. 
In both plots, the horizontal lines show 
the RLF and FRI/FRII breaks calculated from the Willott et al. (2003) and the 
Fanaroff \& Riley (1974) values respectively using a typical steep-spectrum 
index of 0.8. The flux density limits of 3C, 6CE, 7CRS (Willott et al. 2001), 
TOOT00 and SXDS (Vardoulaki et al. 2008) samples 
are shown ($\alpha$ = 0.8 was used to convert to 151 MHz flux density).
}}
\end{figure*}

In Figure~\ref{l151a} we present the radio luminosity at 151 MHz versus the 
radio spectral index of our objects, calculated using the 1.4-GHz and 
151-MHz flux densities presented in Table 1. The median spectral index 
is $\alpha^{1.4 \rm GHz}_{151 \rm MHz}$ = 0.86. In Vardoulaki et al. 
(2008) we have shown that for a 1.4-GHz selected study of radio sources (SXDS) 
roughly 10 times fainter at 151 MHz than TOOT00 (see Figure~\ref{l151_z_ts}), 
the fraction of flat-spectrum objects is higher. Some important differences 
between the TOOT00 and SXDS radio-source samples can be attributed to the 
lower radio-frequency selection of TOOT00 radio sources. One such difference 
is that there seem to be quite a few objects in the 
SXDS\footnote{The radio spectral indices for the SXDS sources were calculated 
using 1.4 GHz and 610 MHz (or 325 MHz) data, whereas for the TOOT00 we use 1.4 
GHz and 151 MHz flux densities; note that is close enough for a rough 
comparison of the samples.} with flat radio 
spectral index (i.e. $\alpha <$ 0.5). The fraction of SXDS objects with 
$\alpha <$ 0.5 is 35 $\pm$ 6\% (13 out of 37), while $\sim$ 11 $\pm$ 3\% 
(4 out of 
37) have an inverted radio spectral index (i.e. $\alpha <$ 0). In the TOOT00 
sample, only 3 radio sources ($\sim$ 6 $\pm$ 3\%) have a flat radio spectral 
index, and only one (2 $\pm$ 1\%) of them (TOOT00\_1027), a possible 
gigahertz-peaked spectrum (GPS) object, has an inverted radio spectral index 
(see Table 1). The other two flat-spectrum objects, TOOT00\_1233 and 
TOOT00\_1268, are Q and Q? objects respectively. TOOT00\_1235 \& TOOT00\_1244 
have radio spectral indices $\alpha$ close to 0.5 and are probably core-jet 
objects (e.g. Augusto et 
al. 1998). These radio sources are usually moderately-flat-spectrum sources 
($\alpha \sim$ 0.5 corresponds to freshly injected electrons) often with 
double-sided kpc-scale jet-like structures. Both these objects with 
$\alpha \sim$ 0.5 have linear projected sizes $D \sim$ 100 kpc, as can be seen 
in Figure~\ref{l151d}a. The observed core-jet radio structure is probably 
caused by interactions with the intergalactic medium and projection 
effects (Paragi et al. 2000). The outlier, TOOT00\_1196, is either an unusual 
source or a spurious 7C detection (see Appendix A).

\addtocounter{figure}{0}

Figure~\ref{l151_z_ts} presents the radio luminosity at 151 MHz versus 
redshift for the TOOT00. The flux density limits of various radio surveys 
are shown: 3C, 6CE, 7CRS (12.4 Jy, 2.0 Jy and 0.5 Jy respectively; Willott et 
al. 2001), TOOT00 (0.1 Jy), and SXDS ($\simeq$ 0.01 Jy at 151 MHz\footnote{A 
typical 
steep spectrum radio spectral index $\alpha$ = 0.8 was used to convert from 
the 2 mJy flux density limit at 1.4 GHz to 151 MHz.}; Vardoulaki et al. 2008). 
We note that 3 out of 4 objects above redshift $z$ = 3 are galaxies with 
photometric redshifts derived from the $K_{\rm W}-z$ relation (see 
Section~\ref{sec:distribution}). The 3 galaxies with photometric redshift $z >$ 3 
are, TOOT00\_1224, TOTO00\_1252 \& TOOT00\_1152 and are very faint at $K$. 
According to Willott et al. (2003), the objects in the 7CRS are $\approx$ 0.3 
mag fainter than the ones in the 3CRS and 6CE. Since the $K_{\rm W}-z$ relation 
fits all three datasets, it will over-predict the redshift of radio-faint 
objects. Thus, if we re-calculate the redshift using that correction of 0.3 mag 
for TOOT00\_1224, TOOT00\_1252 \& TOOT00\_1152, then we get redshifts 
$z$ = 2.900, $z$ = 3.810 and $z$ = 2.747 respectively; only TOOT00\_1152 still 
has a photometric redshift $z_{\rm phot K-z} >$ 3.

The angular size $\theta$ of the TOOT00 objects is presented in Table 1. This 
value was measured from the radio maps, A- or B-array, depending on the object 
(see notes on the objects in Appendix A). In cases where the radio structure 
was compact in the B-array map but there was evidence of extended structure in 
the A-array data, the latter was used to measure $\theta$. In cases where the 
radio source had an one-sided structure, $\theta$ was measured from the core 
to the centre of that lobe. Lastly, in the case where the source presented a 
complex radio structure (e.g. TOOT00\_1107; see Figure~\ref{seds} in Appendix 
A), the distance from the core to each lobe was measured and then added 
together. The angular size $\theta$ was used in the calculation of the 
linear projected size $D$, which is plotted against the radio luminosity at 
151 MHz in Figures~\ref{l151d}a \&~\ref{l151d}b.

\subsection{The FRI/FRII structural divide}
\label{sec:radclass}

The structure of radio sources at low redshifts is directly connected to their 
radio luminosity, 
where objects with $\log_{10}(L_{151 \rm MHz} / \rm W Hz^{-1} sr^{1}) \lta$ 25 are 
FRI type radio sources and those above that value are FRIIs (Fanaroff \& Riley 
1974). Another 
classification used in Table 1 is based on the Owen \& Laing (1989) radio 
classification according to radio structure (see Section~\ref{sec:intro}); in 
that 
way we classify the TOOT00 radio sources as CDs, FDs, TJs and COM. Finally, 
we classify the TOOT00 objects based on both radio and optical data, as 
presented 
in Table 3: FRIIs correspond to CD objects; FRIs to FD and TJ; Q-F corresponds 
to quasars with flat radio spectral index ($\alpha \lta$ 0.5; a `?' denotes 
$\alpha$ close to 0.5); CSS-Q corresponds to quasars with compact radio 
structure; FRII-Q denotes a quasar that has FRII radio structure; CSS denotes 
a compact steep spectrum radio source. 

In Figure~\ref{l151d}a we present the radio luminosity at 151 MHz 
$L_{151 \rm MHz}$ versus largest projected linear size $D$, where we make use 
of the radio-optical classification (FRI, FRII, Q-F, CSS; see also 
Figure~\ref{l151_z_ts}b). Our aim is to 
investigate whether there is a significant number of FRI objects above the 
FRI/II and RLF break in radio luminosity. TOOT00 objects with radio 
luminosities at 
the FRI/II break are at redshift $z \sim$ 0.5, while the ones with 
$L_{151 \rm MHz} = L_{\rm RLF break}$ are at redshift $z \sim$ 1.5 as can be 
seen in Figure~\ref{l151_z_ts}. FRII type radio sources occupy the 
area above the FRI/II break, as is expected. It is interesting, however, that 
we find 7 FRI type (5 TJs and 2 FDs) radio sources above the FRI/II break with 
3 out of 7 of the FRIs also above the RLF break. Although there is an 
uncertainty in the TJ radio structure of several cases, e.g. TOOT00\_1099 and 
TOOT00\_1115 (see Table 1 and Figure~\ref{seds}), we seem able 
to conclude robustly that there is some cosmic evolution between $z \sim$ 0 
and $z \sim$ 1 in the FRI/II divide as has been commented on previously by 
Heywood, Blundell \& Rawlings (2007).

\subsection{The quasar fraction}
\label{sec:qfraction}

\begin{figure*}
\begin{center}
\setlength{\unitlength}{1mm}
\begin{picture}(150,70)
\put(75,-15){\includegraphics{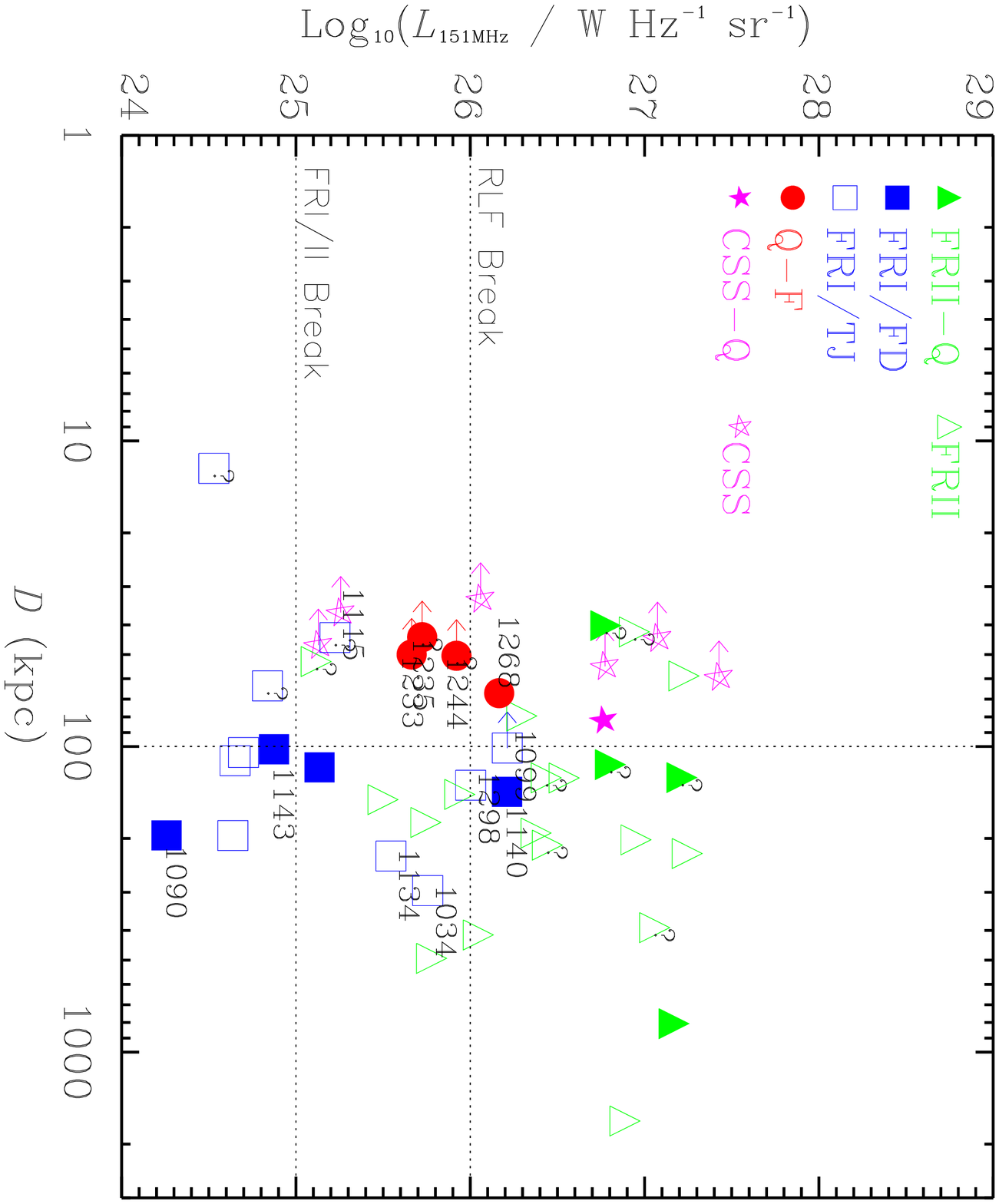}}
\put(165,-15){\includegraphics{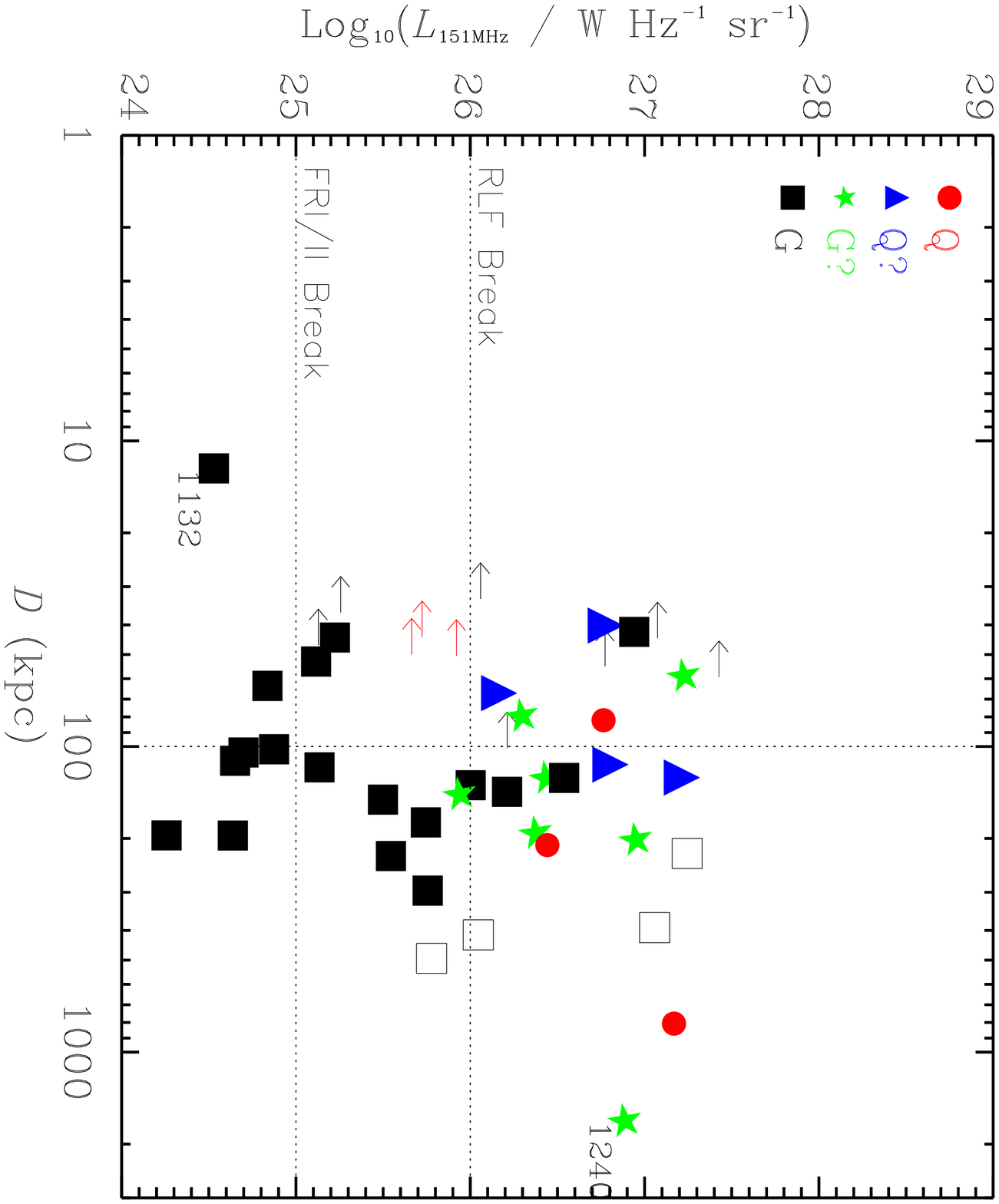}}
\end{picture}
\end{center}
\vspace{.4in}
{\caption[$L_{151 \rm MHz}$ vs $D$]{\label{l151d}
Radio Luminosity at 151 MHz $L_{151 \rm MHz}$ versus the largest projected 
linear size $D$. {\bf a) Left:} Symbols denote radio-optical classification 
as in Figure~\ref{l151_z_ts}; a `?' denotes uncertainty in the radio-optical 
classification as stated in Table 3. The horizontal lines show the RLF and 
FRI/FRII breaks calculated from the Willott et al. (2003) and the Fanaroff \& 
Riley (1974) values respectively using a typical steep-spectrum radio
spectral index 
of 0.8. Objects marked with their name are discussed in the text. 
{\bf b) Right:} Symbols denote optical classification and are the same as in 
Figure~\ref{s74s151}; filled symbols represent objects with spectroscopic 
redshifts and open symbols denote objects with photometric redshifts. The 
horizontal lines show the RLF and FRI/FRII breaks calculated from the Willott 
et al. (2003) and the Fanaroff \& Riley (1974) values respectively using a 
typical steep-spectrum radio spectral index of 0.8. Note that TOOT00\_1196 is 
not plotted since we cannot measure an angular size from the radio maps (see 
notes on the object in Appendix A).
}}
\end{figure*}

Figure~\ref{l151d}b shows the radio luminosity at 151 MHz versus 
the largest projected linear size $D$. The quasar fraction 
to the total number of objects, is $f_{\rm q}$ = 0.25 (10 out of 40 objects) 
above the FRI/FRII break (we include Q and Q? type objects as quasars in this 
calculation; $f_{\rm q}$ = 0.13 if we include only Q objects). Above the RLF 
break the quasar fraction is 0.27 (7 out of 26 objects); the quasar fraction 
drops to 0.12 if we exclude the Q? objects. Such values agree with previous 
studies (e.g. Willott et al.\ 2000) where the quasar fraction is 
$\sim 0.1 \rightarrow 0.4$ over the relevant range of radio luminosity 
$10^{25} \rm W Hz^{-1} sr^{-1} \ltsim $ $L_{151 \rm MHz}$ 
$\ltsim 10^{27} \rm W Hz^{-1} sr^{-1}$. 

Below the FRI/FRII break, the quasar fraction is $f_{\rm q}$ = 0, which also 
agrees with previous studies. These values increase if one studies 
objects using 24 $\mu \rm m$ emission as an indicator of quasar activity, 
as has been done by Vardoulaki et al.\ (2008) on the SXDS radio-source sample. 
This is because hot dust emission at 24 $\mu \rm m$ is much harder to hide 
than optical broad lines which can be extinguished by small amounts of dust. 
The `quasar-mode fraction' $f_{\rm QM}$, i.e.\ 
the fraction of objects with `significant' accretion rates, as determined 
from the observed 24 $\mu \rm m$ flux density, to the total number of objects, 
was found to be $\sim 0.5 \rightarrow 0.9$ above the FRI/FRII break and 
$f_{\rm QM} \ltsim$ 0.1; some high-accretion-rate objects do exist at low 
radio luminosities. There are no 24 $\mu \rm m$ data available in the TOOT00 
region on which to make a study of $f_{\rm QM}$ in the TOOT00 sample.

\addtocounter{figure}{0}

\section{The redshift distribution of TOOT00}
\label{sec:distribution}

\begin{figure*}
\begin{center}
\setlength{\unitlength}{1mm}
\begin{picture}(150,60)
\put(78,-20){\includegraphics{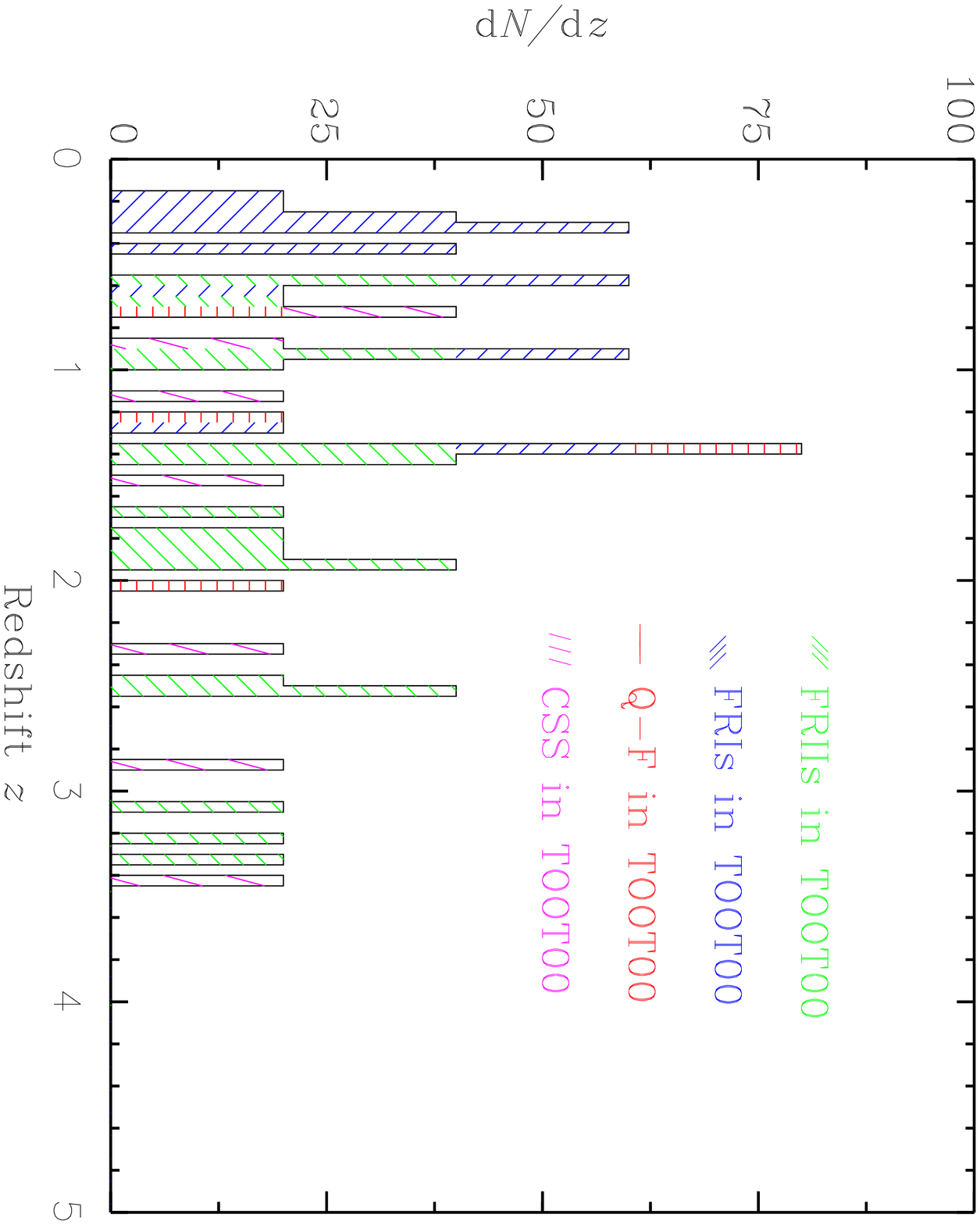}}
\put(165,-20){\includegraphics{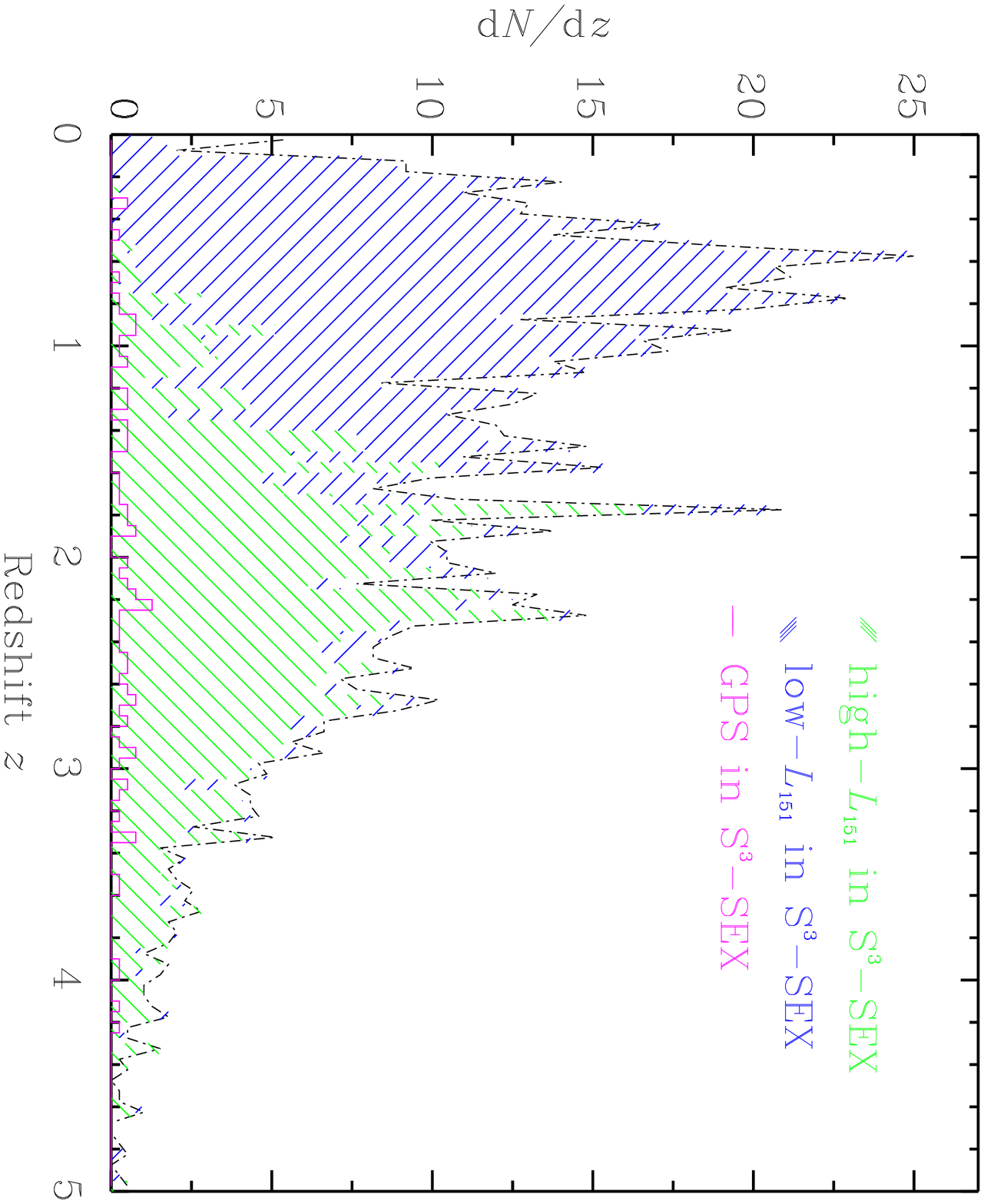}}
\end{picture}
\end{center}
\vspace{0.5in}
{\caption[Histogram of redshift distribution of the TOOT00 objects and the 
S$^{3}$-SEX distribution]{\label{n_z_frclass}
{\bf a) Left:} Histogram of the redshift distribution for the 47 TOOT00 radio 
sources according to radio/optical class (see Table 3). The binning on the 
redshift axis is d$z$ = 0.05. The green colour represents FRIIs, blue 
represents FRIs, red represents flat-spectrum quasars (Q-F) and purple 
represents CSS objects; the black histogram shows the TOOT00 distribution. 
FRIIs correspond to CD radio sources, FRIs to FD and TJ, Q-F to quasars with 
flat radio spectral index, Q-F? to quasars with $\alpha$ close to the critical 
value 0.5 (0.5 $< \alpha^{1.4 \rm GHz}_{151 \rm MHz} <$ 0.7), Compact 
Steep Spectrum (CSS) to objects with steep radio spectral index ($\alpha >$ 
0.7) and compact radio structure, and CSS-Q to quasars associated with a 
compact steep spectrum ($\alpha >$ 0.7) object. {\bf b) Right:} Model 
predictions 
from the S$^{3}$-SEX simulation (Wilman et al. 2008) for an area of 0.1212 sr, 
which is scaled down to the TOOT00 area of 0.0015 sr. The binning on the 
redshift axis is d$z$ = 0.05. Objects are marked 
according to their radio source classification as stated in Wilman et al. 
(2008); blue denotes high-$L_{151}$ objects, green denotes low-$L_{151}$ 
objects and purple stands for GPS sources in the S$^{3}$-SEX simulation; RQQ 
and normal galaxies are not presented for clarity.
}}
\end{figure*}
\addtocounter{figure}{0}

Figure~\ref{n_z_frclass}a presents the redshift distribution of the 47 
TOOT00 radio sources, with a binning of d$z$ = 0.05. The median spectroscopic 
redshift is $z_{\rm spec}$ = 1.135, whereas the median redshift including the 
photometric redshifts is $z_{\rm med}$ = 1.287. This value is larger than the 
median redshift of the 6CE and 7CRS radio source surveys (Rawlings, Eales \& 
Lacy 2001; 
Willott et al. 2003) of $z \sim 1.1$, but we have to take into account that the 
high-redshift objects of our sample tend to have uncertain photometric 
redshifts (see Section~\ref{sec:optclass}). If the $K_{\rm W}-z$ photometric 
redshifts are re-calculated for an increase of $\Delta$m = 0.3 mag in all 
magnitudes (i.e. assuming a similar offset from the mean $K_{\rm W}-z$ 
relation for TOOT00 objects and 7CRS objects), then the median redshift drops 
to $z_{\rm med}$ = 1.244; we adopt $z_{\rm med} \sim$ 1.25. We note that only 
TOOT00\_1252 could plausibly be at $z \gta$ 3.5.

A comparison of the TOOT00 radio-source distribution to any other distribution 
involves correcting for the 9 objects that were excluded from the original 
sample and for objects that are missing since the 7C data have significant 
incompleteness approaching the $S_{151 \rm MHz}$ = 100 mJy flux density limit 
(see Figure 8 of McGilchrist et al. 1990). The weighting factors needed to 
correct the sensitivity variations across the TOOT00 7C data area not 
available, so we can only approximately compare our source counts to those 
in 7C.

We compare the distribution of the TOOT00 radio sources to a simulation 
obtained from the SKADS Simulated Skies Semi-Empirical 
Extragalactic Database S$^{3}$-SEX (Wilman et al. 2008). This is a set of 
simulations of the radio sky suitable for planning science with the proposed 
Square Kilometre Array (SKA) radio telescope (Wilman et al. 2008), that are 
constructed to match results from the 3CRR, 6CE and 7CRS redshift surveys, as 
well as radio source counts. 
This simulation of the extragalactic radio continuum sky allows for 
large-scale structure in the cosmological distribution of radio 
sources, and the types included in the simulations are radio-quiet AGN, 
radio-loud AGN of `FRI' and `FRII' classes and star-forming galaxies. Core-jet 
radio sources are not explicitly included in the simulation. Furthermore, 
these `FR classes' are determined directly by the radio 
luminosity\footnote{In S$^{3}$-SEX (Wilman et al. 2008), the integration 
limits for the radio luminosity function of the FRI radio sources are 
$20.0 \lta \log_{10}(L_{151 \rm MHz / W Hz^{-1}sr^{-1}}) \lta 28.0$ and those of the 
FRII radio sources 
are $25.5 \lta \log_{10}(L_{151 \rm MHz / W Hz^{-1}sr^{-1}}) \lta 30.5$.} and not the 
radio structure of the objects (see Wilman et al. 2008). Due to these 
imperfections, and the fact that our radio-optical classification of the 
TOOT00 radio sources is structural (see Section~\ref{sec:alpha}), we will name 
the `FRI class' in S$^{3}$-SEX as the low-$L_{151}$ sub-population and the `FRII 
class' as the high-$L_{151}$ sub-population of radio sources. The 
simulation covers a sky area of 20$\times$20 deg$^{2}$ or 0.1212 sr, 
out to a cosmological 
redshift of $z$ = 20, and down to flux density limit of 10 nJy at 151 MHz, 
610 MHz, 1.4 GHz, 4.86 GHz and 18 GHz.

We searched for objects in the S$^{3}$-SEX database with simulated 
151 MHz total flux density greater than 100 mJy. We found 3139 radio sources 
above the flux density limit after rejecting those whose core components 
lay outside the 20$\times$20 deg$^{2}$ area of the simulation. The total flux 
densities of multiple component radio sources (i.e. FRIs and FRIIs) 
were determined by summing together all of the individual components of 
each source, including the few components which were outside the 
20$\times$20 deg$^{2}$ area. Using this full list of radio sources we then 
constructed 100 circular sub-fields with the same area as TOOT00, namely 
0.0015 sr (radius 1.27 deg.) centred at random positions within the 
simulation. The mean number of objects $y$ from these 100-simulations is 
$<$y$>$ = 40.1 with a standard deviation, dominated by Poisson errors and the 
effects of large-scale structure, $\sigma_{\rm y}$ = 8.82.

The flux density scale of TOOT00 was bootstrapped from 6C, so should be 
accurate at the $\pm$ 10\% level. This is supported by Figure~\ref{s74s151}, 
which shows that the sources detected at both 74 MHz and 151 MHz have spectral 
indices $\sim$ 0.5, as expected at low frequencies where the spectral index 
should reflect the injected, i.e. un-aged, power-law index of the underlying 
distribution in electron energies (e.g. Alexander \& Leahy 1987).

There has been no quantitative estimate of the incompletenesses and 
Eddington bias (e.g. Teerikorpi 2004) in the lower flux density bins of the 
TOOT00 7C data, which are extremely important effects given the low 
signal-to-noise thresholds adopted in the catalogue (see Table 1). The raw 
TOOT00 7C source counts are significantly higher than those inferred in the 7C 
survey of McGilchrist et al. (1990), largely because of lower significance 
sources. At the $S_{151 \rm MHz} \simeq$ 100 mJy level, corrections for 
incompleteness and Eddington bias are highly uncertain. Most of the TOOT00 
radio sources, apart from 9 (see Table 1) and the 9 `off-edge' objects, have a 
signal-to-noise ratio $\sim$ 5.5, a criterion similar to that adopted by 
McGilchrist et al. (1990); note that there is only one possible 
spurious source in our final list (TOOT00\_1196; see Appendix A) as we have 
securely detected all bar this one source with the VLA. We find 56 sources in 
5.0 deg$^{2}$ in the TOOT00 region. The source counts in the TOOT00 7C region 
using only $>$ 5.5 $\sigma$ detections are 24.7 $\pm$ 1.6, and thus within the 
errors of the corrected McGilchrist et al. source counts.

\begin{figure}
\begin{center}
\setlength{\unitlength}{1mm}
\begin{picture}(150,60)
\put(81,-20){\includegraphics{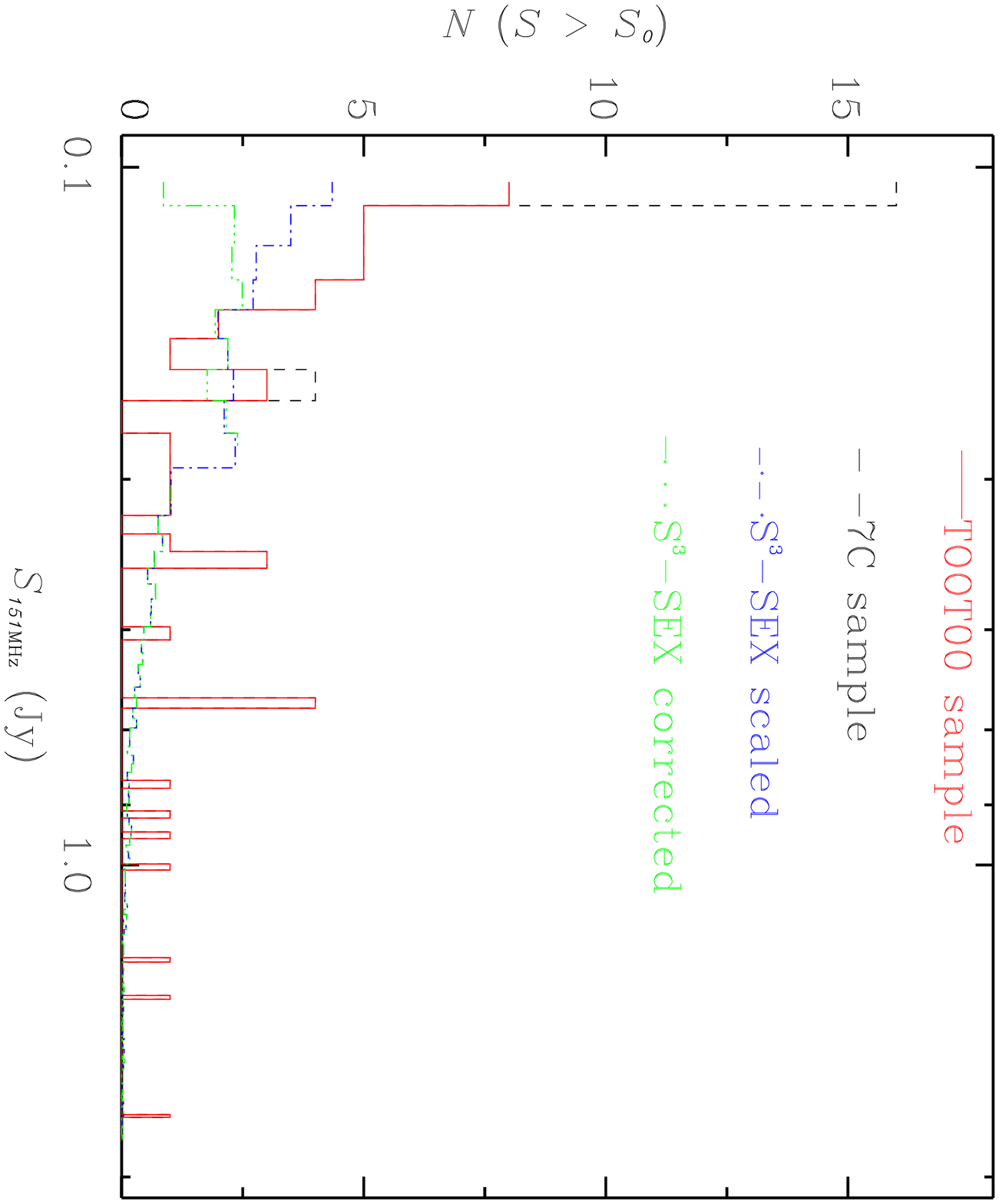}}
\end{picture}
\end{center}
\vspace{0.5in}
{\caption[Histogram of flux density distribution for the TOOT00 
objects]{\label{n_s}
Histogram of the flux density distribution for objects in the TOOT00 sky area 
with flux density $S \gta $ 100 mJy at 151 MHz. The black dashed line shows the 
56 sources in the 7C sample that lie inside the area marked by the red circles 
in Figure~\ref{fig:radec}. The red solid line indicates the 47 TOOT00 
radio sources. The blue dotted-dashed histogram is the S$^{3}$-SEX simulation 
scaled down to 0.0015 sr. This includes $\sim$ 40 objects. The green 
3-dotted-dashed line presents the S$^{3}$-SEX simulation scaled down to 0.0015 
sr in which the lower bins ($S_{151 \rm MHz} <$ 0.25 Jy) are weighted according 
to Table 5 of the McGilchrist et al. (1990), plus an additional correction 
takes into account the missing 7C objects in TOOT00. The final number of 
objects in the scaled and corrected S$^{3}$-SEX simulation is $\sim$ 33.
}}
\end{figure}
\addtocounter{figure}{0}

\begin{figure}
\begin{center}
\setlength{\unitlength}{1mm}
\begin{picture}(150,60)
\put(83,-20){\includegraphics{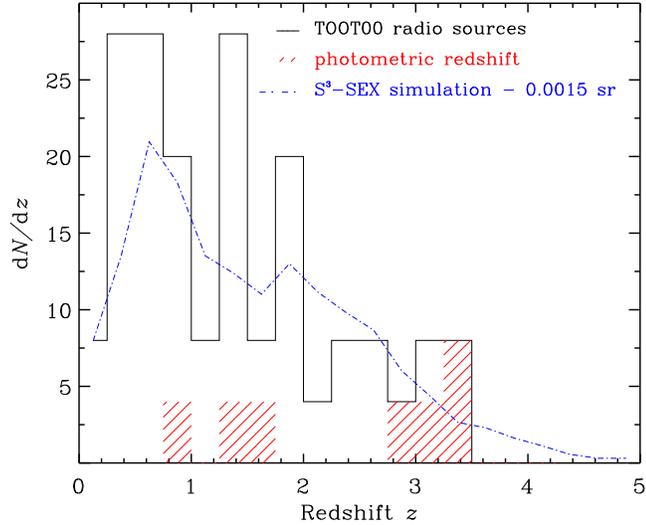}}
\end{picture}
\end{center}
\vspace{0.5in}
{\caption[Histogram of the redshift distribution of the TOOT00 
objects]{\label{nz_scaled}
Histogram of the redshift distribution for the 47 TOOT00 radio sources with 
median $z_{\rm med}$ = 1.287 (85\% spectroscopic and 17\% photometric 
redshifts, black \& red lines respectively). The binning on the redshift axis 
is d$z$ = 0.25. We overplot model predictions from the S$^{3}$-SEX simulation 
for an area of 0.1212 sr, which is scaled down to an area of 0.0015 sr 
(blue dotted-dashed line) to match the TOOT00 sky area, and includes $\sim$ 
40 objects. 
}}
\end{figure}
\addtocounter{figure}{0}

Figure~\ref{n_s} presents the number of objects in the TOOT00 sample compared 
to the 7C objects in the same sky area that have flux density limits above 
100 mJy at 151 MHz. This histogram shows that 47 out of the 56 objects 
from the 7C that have $S_{151 \rm MHz} \gta$ 100 mJy are included in TOOT00 
sample. Figure~\ref{n_s} shows that almost all the difference is confined to the 
lowest (95 mJy $< S_{151 \rm MHz} <$ 115 mJy) bin for reasons explained in 
Section~\ref{sec:sample} and caption to Figure~\ref{fig:radec}.

In Figure~\ref{n_s} we also plot the flux density distribution from the 
S$^{3}$-SEX simulation scaled down to the 0.0015-sr sky area of the TOOT00 
sample. We note that the S$^{3}$-SEX simulation, with or without corrections 
similar to the McGilchrist et al. (1990) work, significantly under-predicts 
the 7C objects in the TOOT00 area in the lowest flux density bins 
(95 mJy $< S_{151 \rm MHz} <$ 115 mJy). 

We suspect that the Eddington bias in 
the 7C sample (which includes sources extracted from the 7C maps with 
signal-to-noise ratios as low as $\sim$ 3; see Table 1) causes the spike in the 
lowest-$S_{151 \rm MHz}$ bin, meaning that the exclusion of faint `off-edge' 
sources (Section~\ref{sec:sample}) in the construction of the TOOT00 survey, 
brings the flux density distribution closer in line between data and 
simulation. Modelling of the Eddington bias would be possible, given a known 
source count significantly deeper than the 7C survey, but this will only 
become available with the advent of LOFAR (e.g. R\"{o}ttgering 2007).

We hereafter assume that the TOOT00 sample is close to the true distribution 
(i.e. the one in the TOOT00 area in the limit of infinite signal-to-noise 
ratio in the low-frequency survey data). We compare this to the S$^{3}$-SEX 
distribution in flux density and other properties without any, in any case 
highly uncertain, corrections for incompleteness and Eddington bias. With 
these  assumptions, there is a reasonable agreement between the 47 TOOT00 
radio sources and the 40 $\pm$ 8.8 objects expected from the simulation
(the standard deviation is a combination of Poisson errors and genuine 
field-to-field variations due to the large scale structure enclosed in the 
S$^{3}$-SEX simulation). 

In Figure~\ref{nz_scaled} we compare the measured TOOT00 redshift distribution 
and that predicted by the S$^{3}$-SEX simulation. We calculated 
the two-sided Kolmogorov-Smirnov (K-S test) statistic and associated 
probability to investigate whether data and simulation are consistent
(Press et al. 1993). The maximum deviation between the cumulative distribution 
of the data and simulation is $\sim$ 0.12 and the significance level of the 
K-S test is $\sim$ 44.46\%. This result suggests that the real TOOT00 and 
simulated S$^{3}$-SEX samples are consistent.

As a further note on the radio-source population, Figure~\ref{n_z_frclass} 
demonstrates that there is a broad agreement between the TOOT00 radio/optical 
classes of the TOOT00 data and the S$^{3}$-SEX simulation, allowing for known 
limitations in the simulation. In the TOOT00 sample (Figure~\ref{n_z_frclass}a) 
we have 47 objects: 22 FRIIs, 14 FRIs, 4 Q-Fs and 7 CSS radio sources; one of 
the CSS objects is in fact a GPS radio source. The distribution on the 
right of Figure~\ref{n_z_frclass} shows the S$^{3}$-SEX simulation, scaled 
down to 0.0015 sr, which includes $\sim$ 40 objects: $\sim$ 
18 high-$L_{151}$ objects, $\sim$ 21 low-$L_{151}$ objects, $\sim$ 0.1 RQQs and 
$\sim$ 0.2 normal galaxies and $\sim$ 1 GPS; the normal galaxies and RQQs are 
not plotted in this figure for clarity.

\section{Conclusions}
\label{sec:conclusions}

In this paper we presented optical spectroscopy and optical and near-IR 
imaging on the 47 TOOT00 radio-source sample from a survey covering 0.0015 sr. 
We have 85\% spectroscopic completeness and we estimate photometric redshifts 
for the rest of the sample using the photometric redshift code HyperZ or 
the $K_{\rm W}-z$ relation from Willott et al. (2003). The key points of our 
analysis are:\\
\\
{\bf $\bullet$}  The median redshift for the whole sample is 
$z_{\rm med} \sim$ 1.25. This value is slightly higher than the 
the $z \sim 1.1$ of the 6CE and 7CRS surveys.\\
{\bf $\bullet$} The fraction of `naked' quasars is $0.13 < f_{\rm q} < 0.25$ 
above the FRI/FRII break in the 151-MHz radio luminosity and below that break, 
the quasar fraction drops towards zero. These are similar results to the 3CRR, 
6CE, 7CRS redshift surveys.\\
{\bf $\bullet$} The total number of TOOT00 objects and their distribution are 
consistent with a simulated distribution of radio sources from the SKADS 
Simulated Skies Semi-Empirical Extragalactic Database (S$^{3}$-SEX), 
simulations extrapolated from the 3CRR/6CE/7CRS datasets.\\
{\bf $\bullet$} The radio structures of the TOOT00 sample are broadly in line 
with those in the S$^{3}$-SEX sample with small exceptions, like the lack of 
core-jet sources in the simulation.\\
{\bf $\bullet$} There is observational evidence from TOOT00 that the `FRI/II' 
structural divide depends on cosmic epoch. We conclude that there must be some 
cosmic evolution between redshifts $z$ = 0 and $z$ = 1. 
It is interesting that the S$^{3}$-SEX `FRIs' (or low-$L_{151}$ sub-population) 
include objects in-between the classic FRI/II break and the RLF, which in real 
samples include some CD sources as well as the FD sources counted as FRI in 
this study. \\
{\bf $\bullet$} The low fraction of flat-spectrum objects in TOOT00 of 6\%
can be attributed to the low-frequency selection of the radio sample. The 
TOOT00 radio-source sample also includes 2 core-jet objects with radio 
spectral indices $\alpha \sim$ 0.5.

\section*{Acknowledgements}

EV would like to thank her parents for private funding. We would like 
to thank the anonymous referee for useful comments. We would like to thank 
Cristina Fernandes for help with WHT observing in 2009, and 
Matt Jarvis, Chris Simpson and Dan Smith for allowing this to happen.
We also thank Gavin Dalton for help accessing ODT data.

\appendix

\section{Notes on the TOOT00 radio sources}
\label{sec:notes}

\oddsidemargin 0.0in
\evensidemargin 0.0in

{\bf TOOT00\_1140} is a galaxy `G' and is resolved in the $K$-band. This FD 
radio source (the two components are not compact in the A-array map) is at a 
spectroscopic redshift $z$ = 0.911. The NVSS data suggest the presence of 
diffuse radio emission that is absent in the A/B-array maps. Due to lack of 
ODT optical photometry, we measure the optical magnitudes directly from the 
spectrum, as explained in Section~\ref{sec:analysis}. The photometric 
redshift from HyperZ seems to be a good fit to the data (see 
Figure~\ref{seds}), 
although it does not agree perfectly with the spectroscopic one.

{\bf TOOT00\_1099} is a possible TJ radio source, and a galaxy resolved in 
the near-infrared $K$-band. The optical spectrum reveals only one clear 
emission line at 8937$\rm \AA$, which is extended over 
$\sim$ 5 arcsec $\approx$ 43 kpc. We believe that this is 
$\rm [OII]_{\lambda 3727}$ and not $\rm [OIII]_{\lambda 5007}$, since the weaker 
of that doublet, $\rm [OIII]_{\lambda 4959}$, does not appear in the spectrum. 
The [OII] emission line suggests a redshift $z$ = 1.397. We are not concerned 
that other strong lines, like [CII], are not visible, since they fall near 
strong sky lines in the blue part of the spectrum. No other expected weak 
lines can be identified in the spectrum. The photometric redshift values 
estimated from HyperZ and the $K_{\rm W}-z$ differ from the spectroscopic 
value implying an under-luminous galaxy in which the 4000 $\rm \AA$ break has 
not been securely identified in the SED.

{\bf TOOT00\_1125} is a spectroscopically confirmed quasar at redshift 
$z$ = 1.916, which is unresolved in the $K$-band. The B-array radio 
map reveals a CD radio source. The photometric redshift from HyperZ does not 
agree with the spectroscopic, and the $K_{\rm W}-z$ relation underestimates the 
photometric redshift, presumably, in both cases, because of a non-stellar 
contribution to the $K$-band.

{\bf TOOT00\_1143} is a FD radio galaxy (the hot spots at the A-array radio 
map are not compact) at redshift $z$ = 0.438, and is 
resolved in the $K$-band. We use the peak flux density at 151 MHz and 1.4 GHz 
in our analysis, since we believe the integrated flux density is affected 
by confusion as indicated by a NVSS map of the area. There is a fairly good 
agreement between spectroscopic and photometric redshifts.

{\bf TOOT00\_1233} is a quasar at redshift $z$ = 0.438 with a flat radio 
spectral index, and probably a 
COM radio source. A comparison of the VLA B-array and NVSS flux 
densities suggests time variability. This is possibly a broad-absorption-line 
QSO (BALQSO), although the fact that the blue and red spectra were joined 
where the MgII line is, causes some concern regarding this classification. The 
offset between the tentative 6000$~ \rm \AA$ absorption and the MgII resonance 
line indicates a speed of $\approx$ 0.1c of the outflow, which is not 
unreasonable. The BC galaxy templates obviously provide a very 
poor fit to the photometric data, since the SED is flat, as expected in an 
unobscured quasar.

{\bf TOOT00\_1022} is a COM radio source. The optical spectrum taken on 
August 2000 (PA = 9$^{\rm o}$) is completely blank and not presented here. The 
optical spectrum taken on August 2006 (PA = 140$^{\rm o}$) shows a featureless 
faint continuum down to 4000 $\rm \AA$, putting a limit on the redshift of 
$\lta$ 3.4. HyperZ provides a poor fit to the photometric 
data; more photometric points are needed in the near-infrared to estimate a 
more accurate photometric redshift. We use the $K_{\rm W}-z$ redshift for 
this object.

{\bf TOOT00\_1204} is a FD radio galaxy at $z$ = 0.6395. The high value of 
the NVSS flux density is not due to confusion, since there are no 
nearby radio sources in a 2-arcmin radius; significant amount of flux is 
invisible in the A- and B-array data. Due to the absence of ODT optical 
photometry we measure the optical flux directly 
from the spectrum (see Section~\ref{sec:alpha}). There is a fairly good agreement 
between spectroscopic and photometric redshifts.

{\bf TOOT00\_1235} is a spectroscopically confirmed quasar and probably a COM 
radio source. In Figure~\ref{seds} we present the A-array VLA map, where the 
radio structure looks stretched along one axis; this is an artifact caused by 
bandwidth smearing due to the position of the radio source at the edge of 
the VLA primary beam (see Figure~\ref{fig:radec}). We classify this quasar 
as a Q-F?, since the value of $\alpha$ is close to 0.5, which also suggests 
that this is a core-jet radio source. There is a fairly good agreement 
between spectroscopic and $z_{K_{\rm W}-z}$ redshifts, contrary to the 
$z_{\rm BC}$ one; this result is expected since the BC templates do not fit a 
quasar SED well.

{\bf TOOT00\_1224} is a compact radio source; the A-array map is bandwidth 
smeared since the object is at the edge of the VLA primary beam (see 
Figure~\ref{fig:radec}). Two optical spectra taken at different PAs and with 
different slit widths (Table 2) are completely blank. As the 
photometric redshift indicates, this is probably due to the faintness of the 
object. We use the $z_{K_{\rm W}-z}$ redshift for our calculations since the 
photometric $z_{\rm BC}$ redshift is highly uncertain. As can be seen from 
Figure~\ref{seds}, the optical magnitudes from ODT are `bright', but yet the 
spectrum is invisible. This can be explained by the fact that in 
spectroscopy the optical slit only captures a part of the light from the 
object, while in photometry more of the light gets observed due to the bigger 
apertures. The optical photometry goes down to $\sim$ 4000 $\rm \AA$, which 
puts a limit on the redshift of the object of $\lta$ 3.4. 

{\bf TOOT00\_1291} is a CD? radio source, since the eastern hotspot is 
compact in the A-array radio map. The optical spectrum is blank, so 
we use HyperZ to calculate a photometric redshift. The BC templates 
do not fit the data well; more near-IR photometric points are needed for a 
more accurate fit. We use the photometric redshift 
calculated from the $K_{\rm W}-z$ relation for the analysis.

{\bf TOOT00\_1107} is a TJ radio galaxy at $z$ = 0.300. The radio map shows 
a very diffuse source. From the radio overlay in Figure~\ref{seds} we notice 
that the extended radio structure is bent backwards as expected if the 
source of the jets is moving with respect to an intra-cluster medium. 
We present the secondary best-fit photometric redshift from HyperZ, since 
that value is closer to the spectroscopic redshift; no reddening law was used 
to calculate the photometric redshift. The $z_{K_{\rm W}-z}$ does not agree with 
either spectroscopic or photometric redshifts, implying an intrinsically faint 
host galaxy.

{\bf TOOT00\_1180} is probably a rather one-sided CD radio source and a 
reddened quasar at $z = 1.810$, since the expected broad emission lines (like 
Ly$\alpha$ and CIV) appear to be narrow in the spectrum of Figure~\ref{seds}. 
We present the A-array radio map since it reveals the extended structure of 
the radio source; the radio structure is not bandwidth smeared since the 
source is close to the VLA pointing position.

{\bf TOOT00\_1027} is a compact radio source at $z$ = 0.890, a point 
source at $K$ and a possible GPS object, since it has an inverted radio 
spectral index. This spectral turnover might be caused by synchrotron 
self-absorption (e.g. Snellen et al. 2000) or free-free absorption (e.g. 
Sawada-Satoh et al. 2000). Apparent extended radio components in the maps of 
this source are almost certainly artifacts due to imperfect calibration. The 
photometric redshift calculated from the $K_{\rm W}-z$ is very high compared to 
the other two redshifts, implying a faint host galaxy. This agrees with 
the findings of Snellen (2008) that GPS radio sources correspond to fainter 
hosts than other radio sources.

{\bf TOOT00\_1195} is a CD radio source with prominent hot spots at the 
end of the radio structure. The optical spectrum shows only a faint 
continuum without clear emission lines or any break. HyperZ provides a 
very poor fit to the data. We use $z_{K_{\rm W}-z}$ in the analysis.

{\bf TOOT00\_1069} is a possible COM 
(or margi- nally extended) 
radio source and 
a spectroscopically confirmed quasar at $z$ = 2.300 with prominent broad 
emission lines. Obtaining this redshift was a triumph of `blind spectroscopy' 
(Rawlings, Eales \& Warren 1990) as a wide (4-arcsec) slit was placed 
North-South in an attempt to cover all plausible radio components. The 
A-array radio map is bandwidth smeared, which is not helpful in making a 
reliable radio classification. We classify it as a compact-steep-spectrum 
quasar (CSS-Q), as indicated by the steep radio spectral index, 
$\alpha^{1.4 \rm GHz}_{151 \rm MHz}$ = 1.49, and the optical spectrum of a quasar 
(see Figure~\ref{seds}). Both photometric redshifts are unrealistic as expected 
if the SED is dominated by non-stellar emission. We assign an upper limit to 
the $K$ magnitude, since the signal-to-noise ratio is not high enough to 
distinguish any associated $K$ object from the background noise.

{\bf TOOT00\_1094} is a compact radio source with faint continuum and a 
probable emission line 
at 7040 $\rm \AA$ in the red part of the optical spectrum. The photometric 
redshifts suggest that this emission line could be MgII, giving a redshift 
$z$ = 1.516. The photometric redshift fit from HyperZ could be improved 
if more photometric points between $I$ and $K$ were available.

{\bf TOOT00\_1149} has a TJ radio structure. This radio galaxy is at 
$z$ = 0.260, where spectroscopic and photometric redshifts seem to be in 
excellent agreement. We don't have a $K$-band image from UKIRT for this 
object, so we use the $K$ image from the ODT survey. 

{\bf TOOT00\_1298} is a TJ radio source probably at $z$ = 1.287 from likely 
[OII] and associated red continua in the spectrum. We do not 
trust the photometric redshift estimated from HyperZ, since only three 
photometric points were used in the fit. The $z_{K_{\rm W}-z}$ 
value is slightly different from the spectroscopic and photometric redshift.

{\bf TOOT00\_1200} is a CD radio source with prominent hot spots at both ends 
of the radio structure, that are visible in both B- and A-array VLA maps. 
The optical spectrum has a plethora of narrow emission lines, placing the 
object at redshift $z$ = 0.691. There is a blue excess in the SED.

{\bf TOOT00\_1215} is a galaxy at $z$ = 0.278 with a compact radio component 
coincident with the galaxy and radio emission slightly elongated to the west; 
hence we classify it as TJ?. The value of the photometric redshift from the 
$K_{\rm W}-z$ is very close the spectroscopic one. The redshift calculated 
from HyperZ has a much higher value, probably because more photometric 
points are needed for an accurate fit; no reddening 
law was used to estimate $z_{\rm BC}$. We don't have a $K$-band image from 
UKIRT for this object, so we use the $K$ image from the ODT survey.

{\bf TOOT00\_1240} is a large CD radio source $\sim$ 2.5 arcmin long, 
possibly at spectroscopic redshift $z$ = 2.543?. The 
exposure times used in the first attempt at spectroscopy (August 2000) of this 
object were accidentally far 
too short to detect it. The fact that the $K$ position of the object is 
close to a spectroscopically confirmed M-star (RA: 00 13 05.2 \& Dec: +36 01 
48.1; marked with `M' in Figure~\ref{seds}) also makes it hard to get a 
photometric redshift. We managed to get a spectrum of the M-star and the 
bright galaxy south-west of the object ID, marked with `G' in 
Figure~\ref{seds}, at $z$ = 0.191. In August 2004 we targeted the 
object north of the west radio lobe with an associated 
compact radio component (see Figure~\ref{seds}) which is a galaxy at 
$z$ = 0.556. The blue spectrum from the 
August 2009 observing run has continuum emission blueward of 4290 $\rm \AA$, 
but, as this becomes increasingly contaminated by leakage from the M-star at 
redder wavelengths, it is hard to assess the colour of the object. 
We show in Figure~\ref{seds} only the blue-arm spectrum 
in which there are possible Ly$\alpha$ and NV$_{\lambda\lambda 1240}$ lines. 
Along the slit we measured a redshift of $z$ = 0.193 for a galaxy close to the 
western hotspot at RA: 00 13 01.36 \& Dec: +36 01 57.04. 
The $K-$band image of this area excludes the area where the east 
lobe is located. In Figure~\ref{seds} we present the $K-$band/radio B-array 
overlay. Below it, we give a DSS image of the area, where the double radio 
structure is clearly visible. The photometric redshift from HyperZ is not 
reliable: more photometric points are needed to provide 
a better fit to the SED. The value of photometric redshift from the 
$K_{\rm W}-z$ relation ($z_{K_{\rm W}-z}$ = 2.185) is close to the tentative 
spectroscopic redshift. 

{\bf TOOT00\_1289} is CD radio source, since the radio hot spots do not 
disappear in the A-array VLA map. The optical spectrum indicates a possible 
redshift of $z$ = 1.784 although, if true, there are no hints of Ly$\alpha$ 
at the expected location. The photometric redshift from HyperZ is highly 
uncertain; more photometric points are needed for an accurate fit.

{\bf TOOT00\_1072} is a giant elliptical galaxy with a CD radio structure, 
since the compact hot spots are also visible in the A-array map, and 
a clear absorption spectrum, suggesting a spectroscopic redshift $z$ = 0.577. 
We have fairly good agreement between photometric and spectroscopic redshifts, 
and the best-fit BC template fits the photometric data points well.

{\bf TOOT00\_1134} is a radio galaxy at $z$ = 0.311. The photometric redshift 
from HyperZ is a good fit to the data and agrees with the spectroscopic value; 
no reddening law was used in the calculations. The integrated radio flux 
densities include a contribution from a separate radio source to the 
south-east (as indicated by a NVSS map of the area); this second object is 
associated with a galaxy that, from its SED, may well be in the same cluster. 
We use the peak flux densities at 74 MHz, 151 MHz for our analysis. For the 
1.4-GHz flux density we use the value measured from the B-array VLA map using 
the package {\textsc AIPS}: we use a box around the radio source, excluding 
the radio source to the south-east of the object.

{\bf TOOT00\_1244} is a spectroscopically confirmed quasar at $z$ = 1.358 and 
a compact radio source. We classify it as a Q-F?, since the radio spectral 
index has a value close to 0.5, which also suggests this may be a core-jet 
radio source. The photometric redshift is not reliable, as is expected 
if its blue SED is dominated by non-stellar emission. We don't have a $K$-band 
image from UKIRT for this object, so we use the $K$ image from the ODT survey.

{\bf TOOT00\_1066} is a FD radio source, since the double radio structure 
seen in Figure~\ref{seds} disappears in the A-array radio map. The 
spectroscopic 
redshift is probably $z$ = 0.926, although different from the photometric 
redshifts. We use the peak flux densities at 151 MHz and 1.4 GHz, due to 
confusion as indicated by a NVSS map of the area.

{\bf TOOT00\_1261} is a spectroscopically confirmed quasar that appears 
to be reddened, since the expected broad emission lines, like Ly$\alpha$ and 
CIV, are broad only in their wings. The A-array radio map reveals a CD radio 
structure. The apparent offset between the $K$ and radio positions 
(Figure~\ref{seds}) is within 
the uncertainty of the astrometry. The SED secondary best-fit is surprisingly, 
and presumably fortuitously, good considering the BC templates are not 
expected to fit quasar SEDs very well. We use the peak flux densities at 151 
MHz and 1.4 GHz, due to confusion as indicated by the NVSS map of the area. 

{\bf TOOT00\_1252} is a possibly one-sided radio source; we classify 
it as CD?. The west `lobe' is $\sim$ 45 arcsec from the `core' which may be a 
true core plus a short eastern lobe   
($S_{1.4 \rm GHz  int}$ = 18.9mJy and $S_{1.4 \rm GHz  peak}$ = 14.4 mJy). 
The $K$ detection is 2.7$''$ South-West of the peak radio position. We 
assign an upper limit to the $K$ 
magnitude measured from the $K$-band image, since the signal-to-noise ratio is 
not high enough in order to distinguish any $K$ object from the background. 
The optical spectrum taken on August 2000 (at PA = 109$^{\rm o}$) is blank in 
the blue and shows signs of continuum in the red. The spectrum taken on July 
2004 (at PA = 265$^{\rm o}$) is completely blank, with confidence in correct 
pointing as the slit was pointed in a way to observe as many objects as 
possible along the slit. Deeper observations could provide 
a reliable optical spectrum and an accurate optical position for the object. 
The HyperZ fit gives only a lower limit 
to the photometric redshift since only upper limits are used in the 
calculations; more photometric points are needed to improve the fit. We use 
the photometric redshift from the $K_{\rm W}-z$ relation, since the photometric 
redshift from HyperZ is unreliable.

{\bf TOOT00\_1214} is a one-sided radio source classified as CD?. It is a 
point source at $K$ and a seemingly reddened quasar at $z$ = 3.081. In the 
optical spectrum we can see evidence of a Ly$\alpha$ absorption system below 
5000 $\rm \AA$. There is a very poor agreement between photometric and 
spectroscopic redshifts, as expected for a quasar SED. We use the peak flux 
densities at 151 MHz and 1.4 GHz, due to confusion as indicated by a NVSS map 
of the area.

{\bf TOOT00\_1152} is a CD radio source with prominent hot spots at both 
ends of the radio structure, that do not disappear in the A-array map. The 
object most likely associated with the radio structure gives a weak 
featureless faint optical continuum in the spectroscopy, which puts a limit 
on the redshift of the object of $\lta$ 3.  We use the 
$K_{\rm W}-z$ redshift for our calculations 
since the formal output of the HyperZ code is clearly a very poor fit.

{\bf TOOT00\_1129} is a CD radio source, since the hot spots do not disappear 
in the A-array map, and a possible reddened quasar, as indicated by the narrow 
Ly$\alpha$ and CIV lines. Neither of the two photometric redshifts agree with 
the spectroscopic 
value, as expected if non-stellar (e.g.\ dust scattered light from the obscured 
quasar) light is dominating the SED. The formal output of the HyperZ code 
is clearly a very poor fit.

{\bf TOOT00\_1090} is a FD radio source, as can be seen by the diffuse radio 
structure in Figure~\ref{seds}. This galaxy is at redshift $z$ = 0.201. There 
is a very good agreement between spectroscopic and photometric redshifts.

{\bf TOOT00\_1267} is a CD radio source, since the radio hot spots are 
present in both the A- and B-array radio maps. The optical spectrum places 
this object at redshift $z$ = 0.968. There is a fairly good agreement between 
photometric and spectroscopic redshifts.

{\bf TOOT00\_1188} is a CD radio source (the radio hot spots do not disappear 
in the A-array map) with a spectroscopic redshift $z$ = 1.417. The HyperZ fit 
is fairly good; more near-IR photometric points are needed for a more accurate 
fit. The redshift calculated from the $K_{\rm W}-z$ relation is slightly higher 
than the spectroscopic value.

{\bf TOOT00\_1196} is the only object in the sample that may be a spurious 7C 
radio source. Its unusual properties means that if it is real it is a 
very diffuse `FD?' radio source since 
there is no bright radio core nor a bright hot spot near the 7C radio 
position, as can be 
seen by the B-array map presented in Figure~\ref{seds}. Due to this 
fact, we cannot measure the angular size of the radio source. Any radio 
emission is probably associated with the galaxy shown in Figure~\ref{seds}, 
close to the 7C radio position. Since any radio source in the B-array map has a 
flux density below the detection limit, we use a 2$\sigma$ detection from the 
NVSS, recalling that the rms noise of the NVSS is 0.45 mJy beam$^{-1}$. This 
gives a very steep radio spectral index of $>$ 2.1. The object is not detected 
at 74 MHz, marginally surprising given a predicted detection at $\sim$ 
4.5$\sigma$. We cannot rule out the possibility that the 4.4$\sigma$ 7C 
detection is spurious. Assuming the source is real and the galaxy ID shown in 
Figure~\ref{seds}, optical photometry provides only $R$- and $I$-band 
images, 
while it was not observed in the $B$-band, so the HyperZ fit is uncertain. The 
optical spectrum is blank in the 
blue, but shows a featureless continuum in the red with no clear lines. This 
object is classified as a possible galaxy G?, and we use the redshift 
calculated from the $K_{\rm W}-z$ for our analysis.

{\bf TOOT00\_1173} is a TJ radio source at $z$ = 0.332, with diffuse radio 
structure. The HyperZ fit is good and there is a good agreement between 
photometric and spectroscopic redshifts.

{\bf TOOT00\_1228} is a COM radio source probably at redshift $z$ = 1.135, 
and a point source at $K$. There is poor agreement between spectroscopic and 
photometric redshifts.

{\bf TOOT00\_1034} is TJ radio source at $z$ = 0.580, where the east radio 
emission 
is very diffuse. The photometric redshift agrees with the spectroscopic one, 
although more photometric points could improve the HyperZ estimate.

{\bf TOOT00\_1255} is probably a CD radio source since the hot spots do not 
disappear in the A-array radio map. Here we present the B-array map 
(see Figure~\ref{seds}). This galaxy is at redshift $z$ = 0.582, which does not 
agree well with the photometric redshift. The one estimated from the 
$K_{\rm W}-z$ relation has a higher value, presumably because it is an 
unusually underluminous galaxy.

{\bf TOOT00\_1048} is a CD radio source since the radio hot spots do not 
disappear in the A-array map, with a secure redshift $z$ = 1.943. The optical 
spectrum indicates a reddened quasar and it is unresolved in the $K$-band. The 
ID in the $K$ is 0.6 arcsec away from the centre of the classical double radio 
structure, which is within the uncertainty of the astrometry and the radio 
position. We classify it as a G?. We do not trust the photometric redshift from 
HyperZ since we have only four photometric points for the calculation.

{\bf TOOT00\_1029} is probably a COM radio source; the A-array map is 
bandwidth smeared and isn't presented here. The redshift estimated from HyperZ 
is close to the spectroscopic redshift; no reddening law was used in that 
calculation. The $z_{\rm K-z}$ estimate does not agree with the spectroscopic 
redshift.

{\bf TOOT00\_1115} is probably a TJ radio source, since both A- and  B-array 
radio maps are extended; here we present the A-array map. McLure et al.\ 
(2004) classify this object as a low-excitation-galaxy LEG (see Jackson \& 
Rawlings 1997). The optical spectrum gives a secure redshift $z$ = 0.416, and 
there is very good agreement with the photometric redshift.

{\bf TOOT00\_1132} is possibly a TJ? radio source and a galaxy at $z$ = 0.183. 
There is a fairly good agreement between the spectroscopic and photometric 
$z_{\rm BC}$ redshifts, although the $z_{\rm K-z}$ value is slightly higher.

{\bf TOOT00\_1250} is a spectroscopically confirmed quasar at $z$ = 1.350, 
with broad emission lines in the optical spectrum. The A- \& B-array radio maps 
reveal a CD? radio structure; the A-array map shows clearly that both hot spots 
are compact (see Figure~\ref{seds}). The object is unresolved in the $K$-band. 
The value estimated from HyperZ disagrees with the spectroscopic and 
the $z_{\rm K-z}$ redshifts due to a poor fit to too few photometric data points.

{\bf TOOT00\_1268} is a COM radio source at redshift $z$ = 2.015, and it 
might be a reddened quasar. We classify it as a G? optically and a Q-F in the 
radio for a quasars with a flat radio spectral index. The A-array radio map is 
bandwidth smeared at the source position; we do not present here. 
The fact that there is no visible sign of a 4000 $\rm \AA$ break 
spectroscopically and that the SED is relatively flat, makes 
it hard to estimate an accurate photometric redshift. 

{\bf TOOT00\_1251} is a CD radio source at $z$ = 2.490, since the radio 
hotspots do not disappear in the A-array map. The Ly$\alpha$ line is obvious 
in the 2D optical spectrum. We do not trust the 
photometric redshift from HyperZ due to large errors on the 
photometry. The $z_{K_{\rm W}-z}$ value agrees with the tentative spectroscopic 
value.

{\bf TOOT00\_1203} is a small CD? radio source at redshift $z$ = 1.397. The 
photometric redshifts are slightly lower than the spectroscopic value. More 
photometric points could improve the HyperZ fit.

\clearpage

\addtocounter{table}{0}
\scriptsize
\begin{table*}
\begin{center}
 {\caption[Table~\ref{tab:imaginglog}]{
Optical and near-IR photometry for the 47 TOOT00 radio sources that was 
used to calculate photometric redshifts with HyperZ: 
$U$, $B$, $V$, $R$ and $I$, 2$''$ Vega magnitudes are from the ODT 
survey (MacDonald et al. 2004) or they are measured from the spectra of each 
object (Figure~\ref{seds}) due to lack of imaging; $K-$ photometry was obtained 
by UKIRT; $J^{*}$, $H^{*}$ and $K^{*}$ magnitudes are provided by 
K. Inskip (complete photometry in five different apertures is shown 
in Table A2); errors on the magnitudes are in brackets below the values. 
{\bf Column 1} shows the name of the TOOT00 object. In {\bf Columns 2-7} we 
give $U-$, $B-$, $V-$, $R-$, $I-$band magnitude respectively; the character 
`s' shows measurement from the spectrum of each object; the character 
`o' denotes ODT data. The errors on the magnitudes are shown below each 
magnitude value. {\bf Columns 7 \& 8} show photometry from the `Inskip' 
catalogue on the $J^{*}-$ and $H^{*}-$band, respectively, where they were 
measured with a 4$''$ diameter aperture. {\bf Column 9} presents $K-$band 
photometry from UKIRT measured with a 4$''$ diameter aperture or the one from 
the `Inskip' catalogue (see Section ~\ref{sec:sample}) marked with an `*'; the 
character `o' denotes the value is provided by the ODT. Note: In the HyperZ 
photometric redshift calculation, all errors less than 10\% on the near-IR 
magnitudes were taken as 10\%.
}}
\begin{picture}(50,180)
\put(-260,-530){\includegraphics{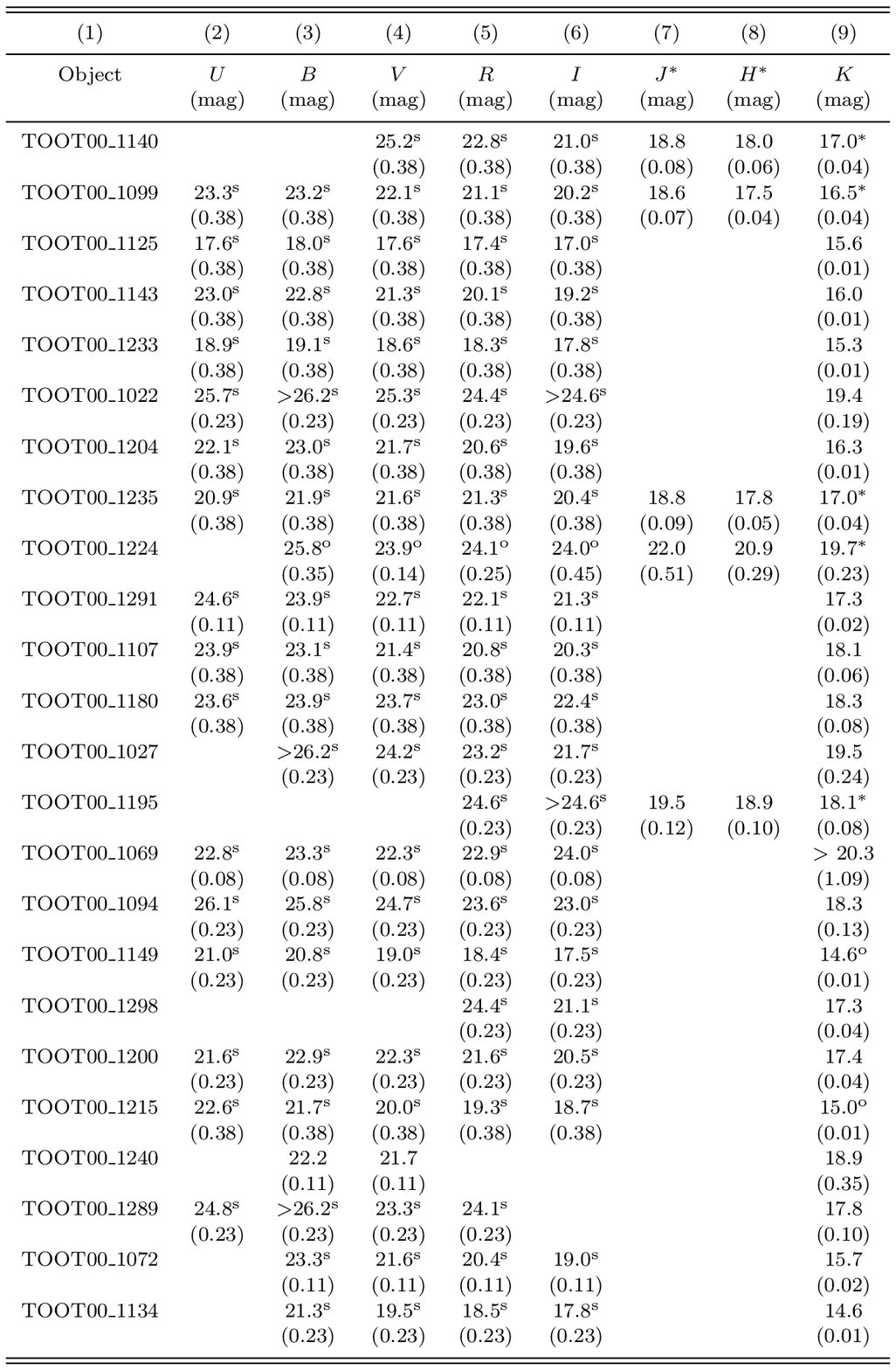}}
\end{picture}
\end{center}
 \end{table*}
\normalsize

\clearpage

\addtocounter{table}{-1}
\scriptsize
\begin{table*}
\begin{center}
 {\caption[junk]{(continued)
}}
\begin{picture}(50,180)
\put(-260,-535){\includegraphics{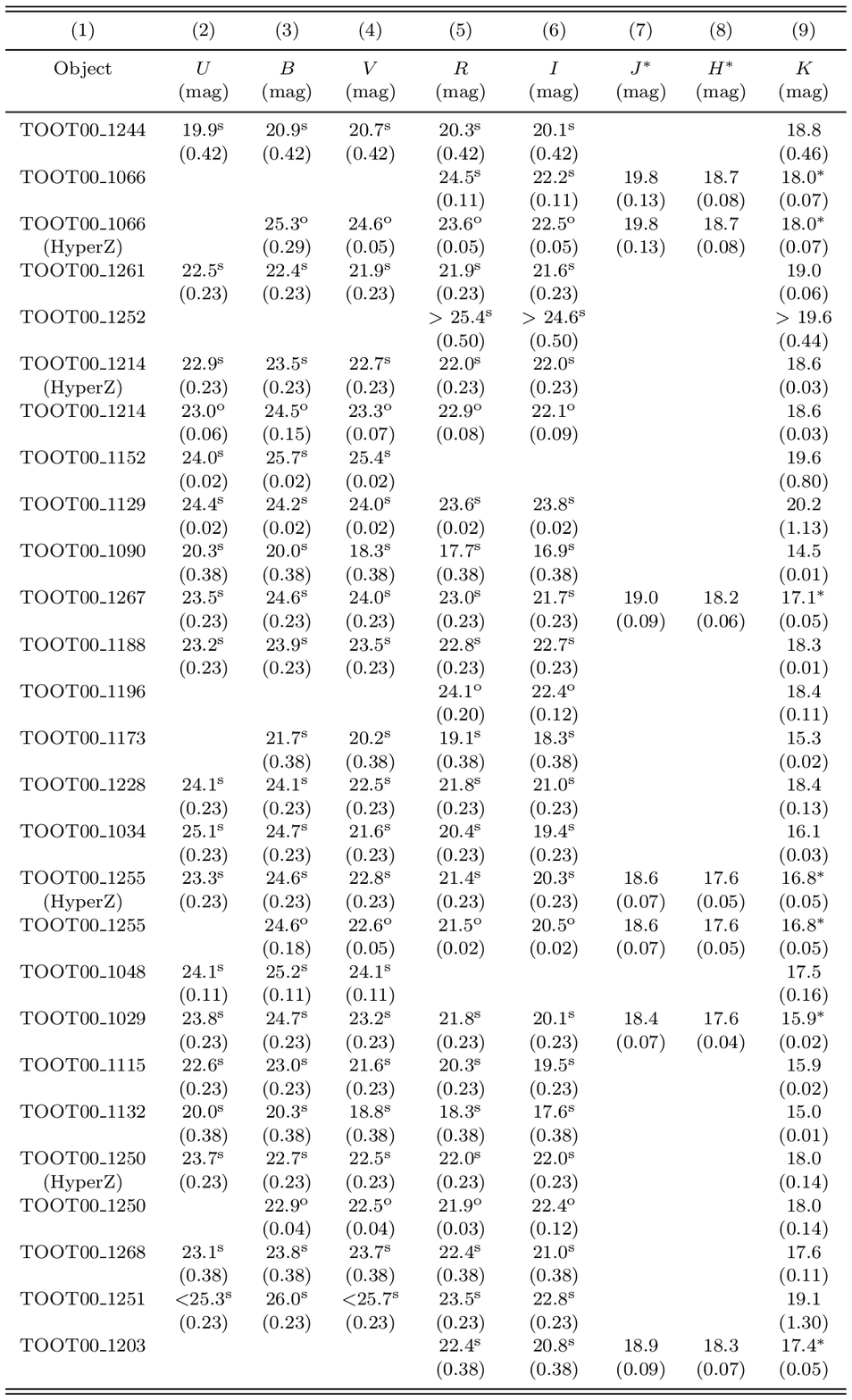}}
\end{picture}
\end{center}
 \end{table*}
\normalsize

\clearpage

\addtocounter{table}{0}
\scriptsize
\begin{table*}
\begin{center}
 {\caption[junk]{\label{tab:imaginglog2}
Optical and near-IR photometry for the 47 TOOT00 radio sources; errors 
are in brackets below the values. {\bf Column 1} shows the name of the TOOT00 
object. {\bf Column 2} gives the photometric band, where for the $K-$band 
from UKIRT we also provide the observing date; the $J^{*}-$, $H^{*}-$ and 
$K^{*}-$band data are from the Inksip catalogue; the character `o' denotes 
data from the ODT. {\bf Column 3} gives the near-IR position of the 
object (RA in $\rm h$ $\rm m$ $\rm s$ \& Dec in $\rm ^{o}$ $'$ $''$; J2000.0). 
{\bf Columns 4-8} present the photometry in different diameter aperture 
measurements, 3$''$, 4$''$, 5$''$, 8$''$ and 9$''$ respectively, for near-IR 
data only. Note that for the optical photometry (see MacDonald et al. 2004) the 
magnitudes given are not measured in a 3-arcsec diameter aperture, but are 
given in the same column due to lack of space.
}}
\begin{picture}(50,180)
\put(-260,-535){\includegraphics{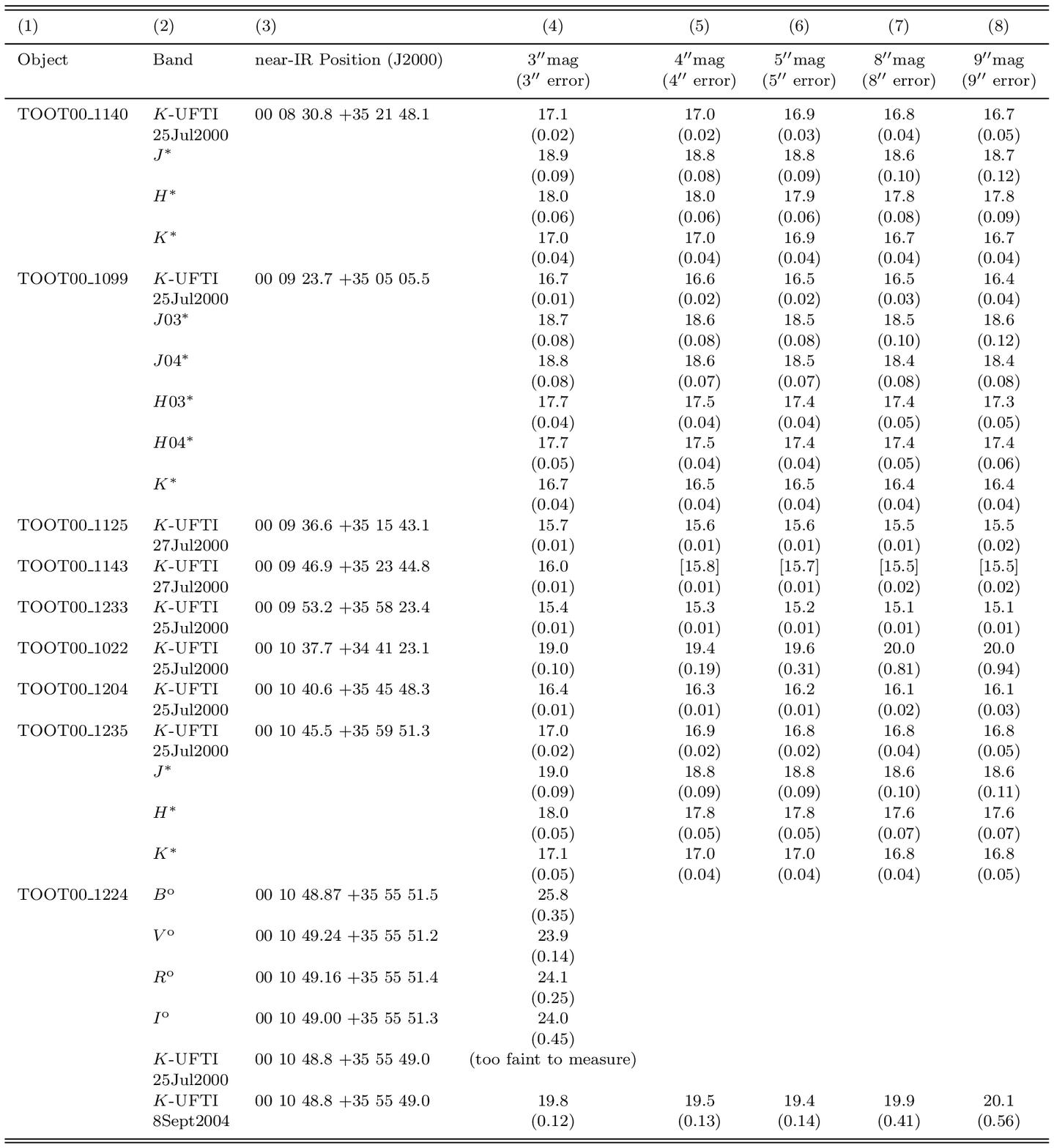}}
\end{picture}
\end{center}
 \end{table*}

\normalsize

\clearpage

\addtocounter{table}{-1}
\scriptsize
\begin{table*}
\begin{center}
 {\caption[junk]{(continued)
}}
\begin{picture}(50,180)
\put(-260,-535){\includegraphics{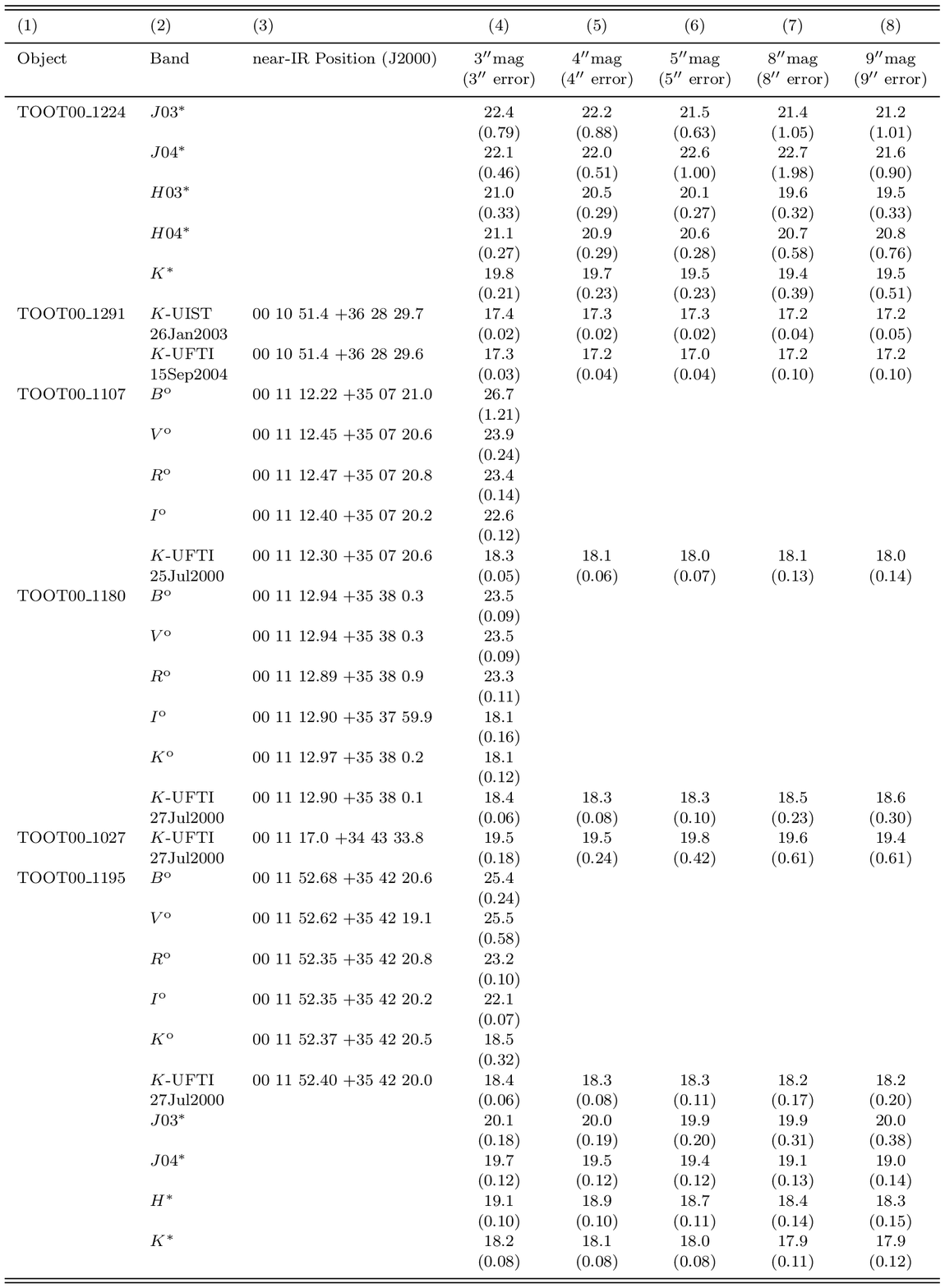}}
\end{picture}
\end{center}
 \end{table*}

\normalsize
\clearpage

\addtocounter{table}{-1}
\scriptsize
\begin{table*}
\begin{center}
 {\caption[junk]{(continued)
}}
\begin{picture}(50,180)
\put(-260,-530){\includegraphics{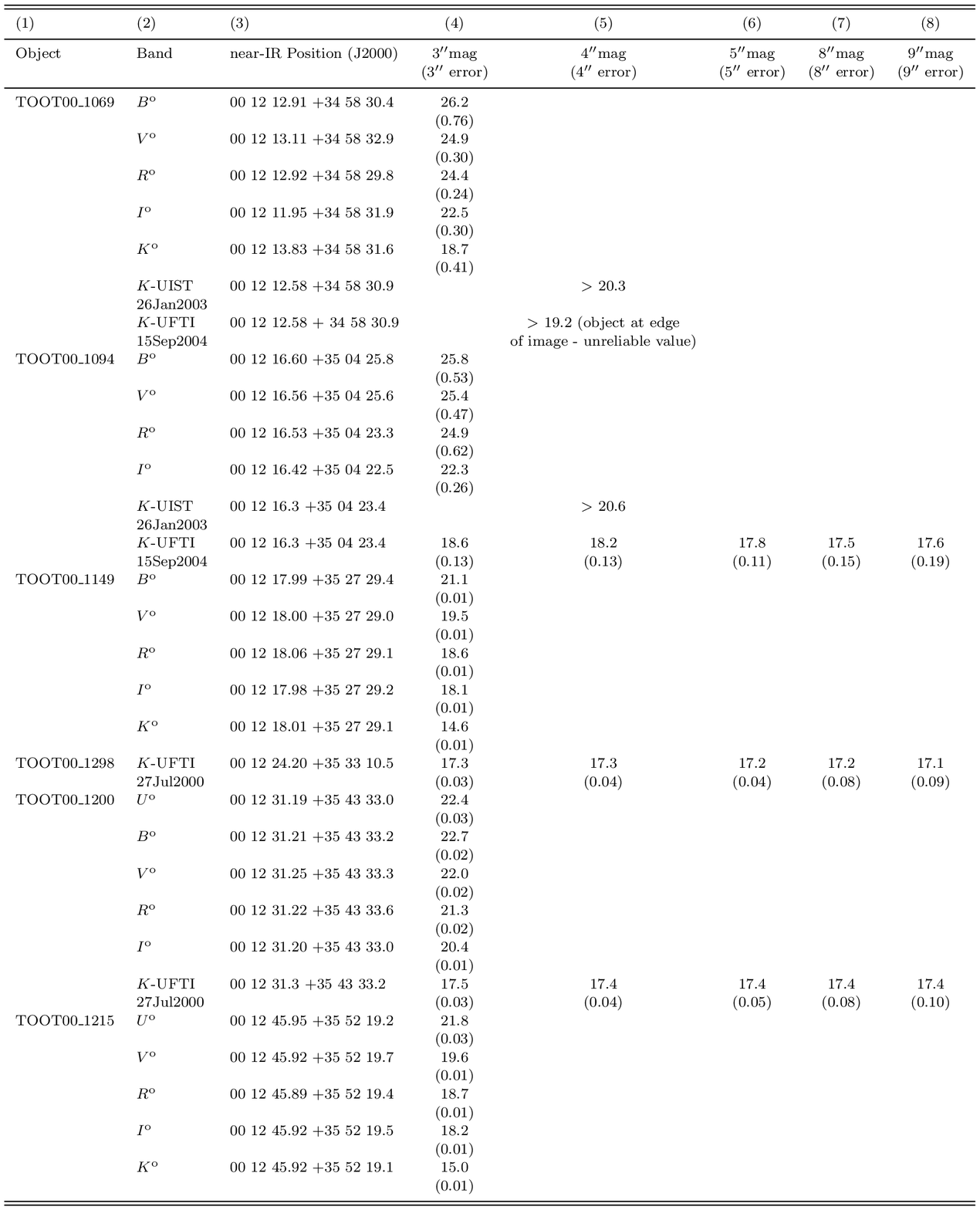}}
\end{picture}
\end{center}
 \end{table*}

\normalsize
\clearpage

\addtocounter{table}{-1}
\scriptsize
\begin{table*}
\begin{center}
 {\caption[junk]{(continued)
}}
\begin{picture}(50,180)
\put(-260,-535){\includegraphics{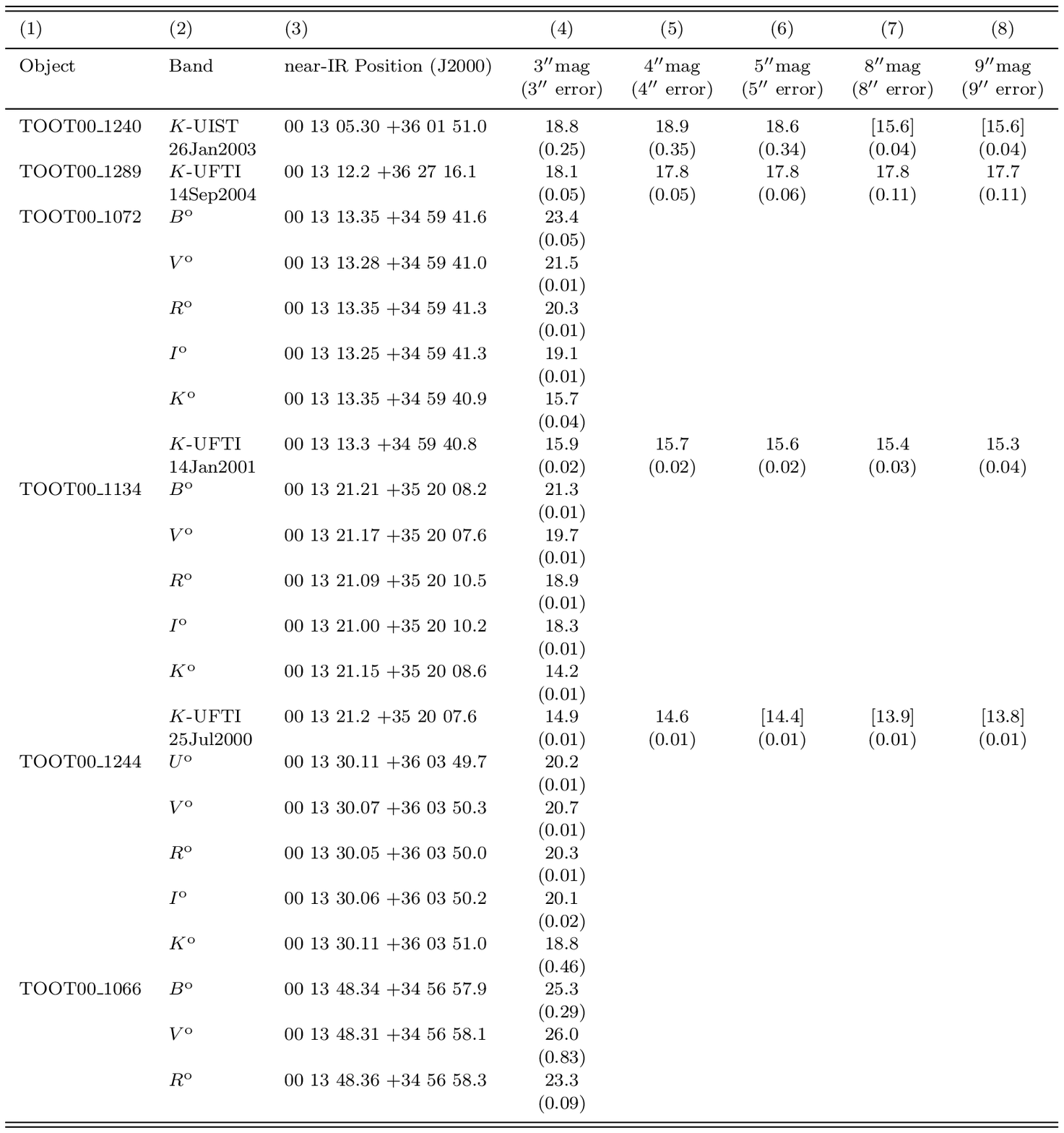}}
\end{picture}
\end{center}
 \end{table*}

\normalsize
\clearpage

\addtocounter{table}{-1}
\scriptsize
\begin{table*}
\begin{center}
 {\caption[junk]{(continued)
}}
\begin{picture}(50,180)
\put(-260,-535){\includegraphics{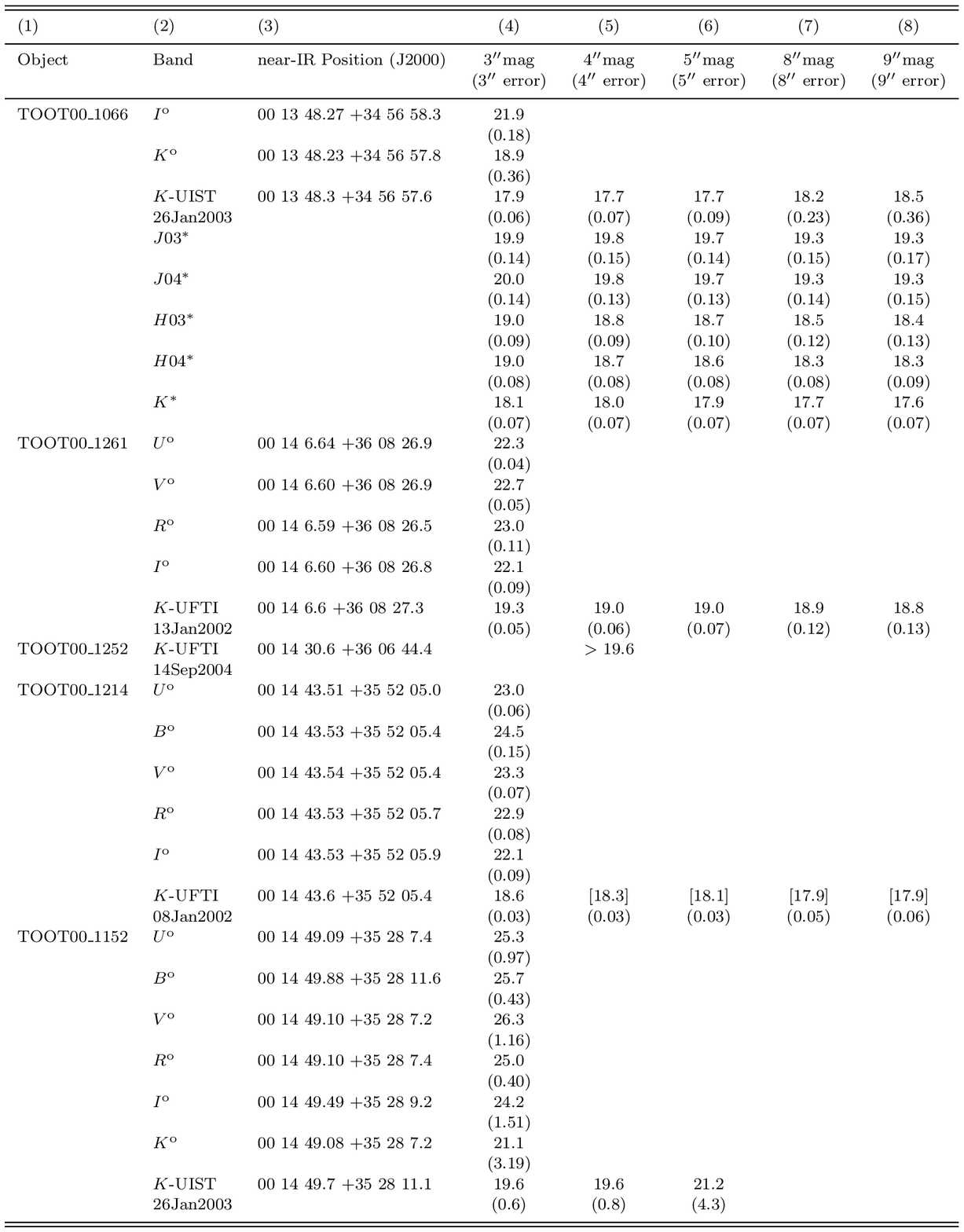}}
\end{picture}
\end{center}
 \end{table*}

\normalsize
\clearpage

\addtocounter{table}{-1}
\scriptsize
\begin{table*}
\begin{center}
 {\caption[junk]{(continued)
}}
\begin{picture}(50,180)
\put(-260,-530){\includegraphics{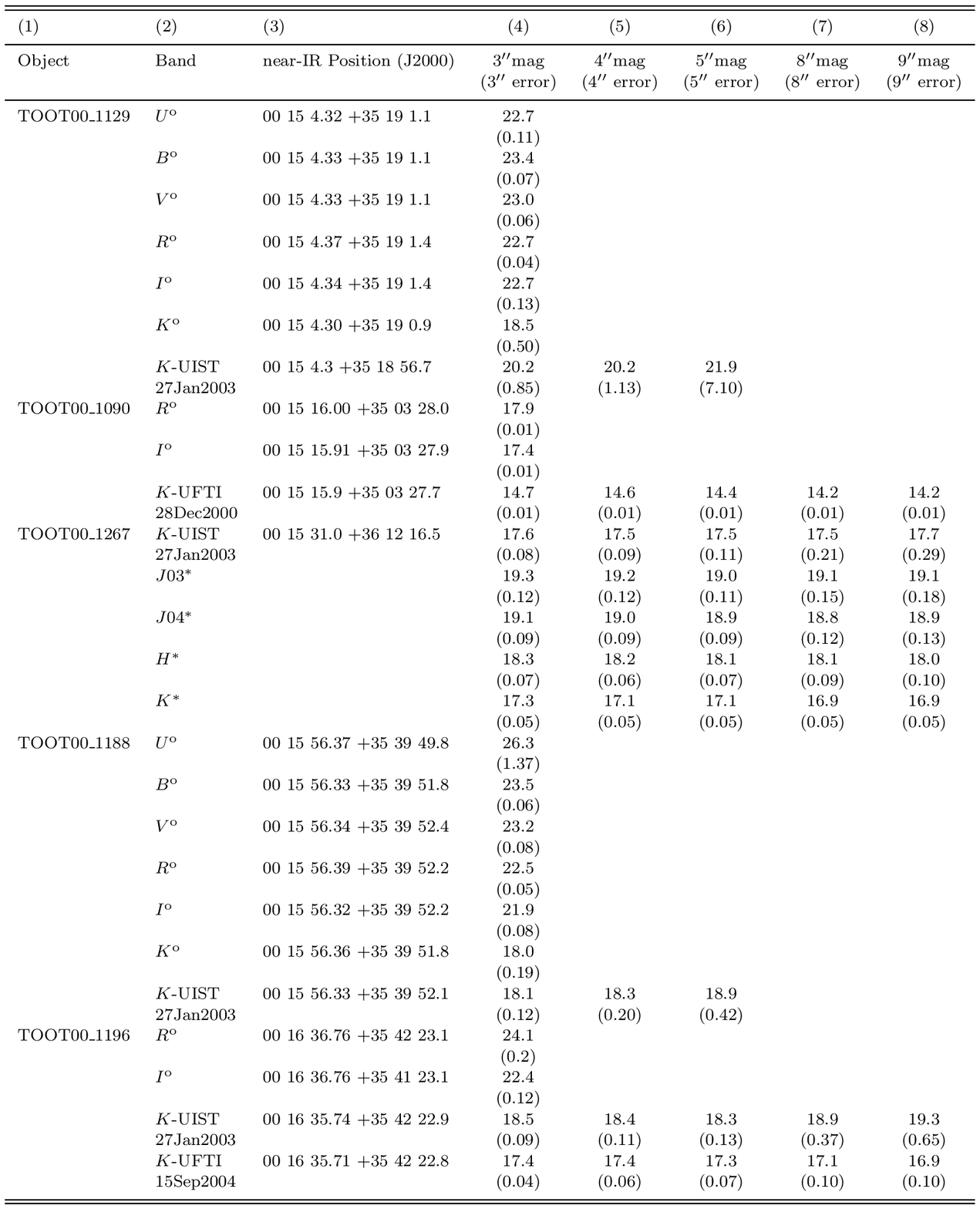}}
\end{picture}
\end{center}
 \end{table*}

\normalsize
\clearpage

\addtocounter{table}{-1}
\scriptsize
\begin{table*}
\begin{center}
 {\caption[junk]{(continued)
}}
\begin{picture}(50,180)
\put(-260,-535){\includegraphics{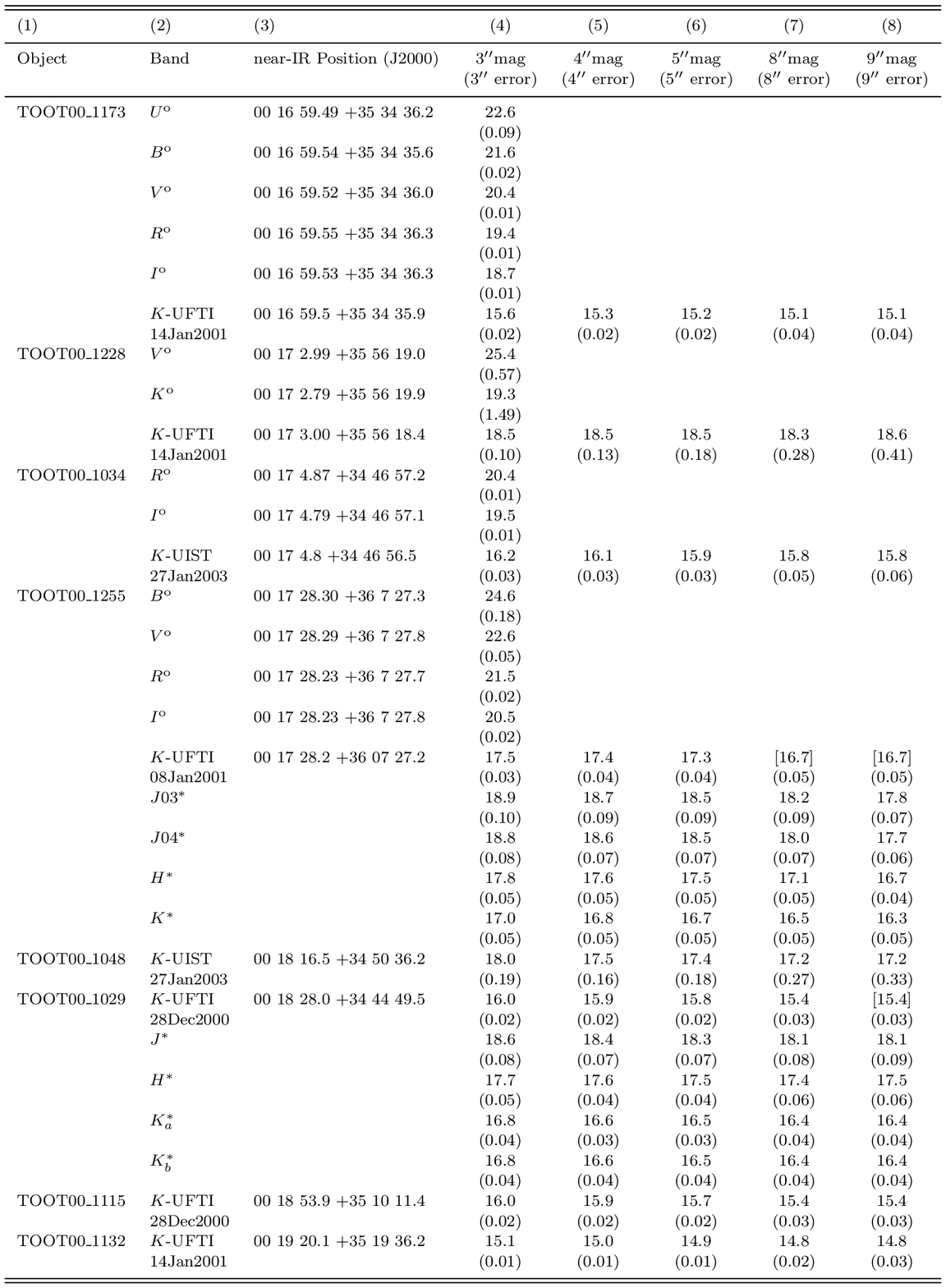}}
\end{picture}
\end{center}
 \end{table*}

\normalsize
\clearpage

\addtocounter{table}{-1}
\scriptsize
\begin{table*}
\begin{center}
 {\caption[junk]{(continued)
}}
\begin{picture}(50,180)
\put(-260,-535){\includegraphics{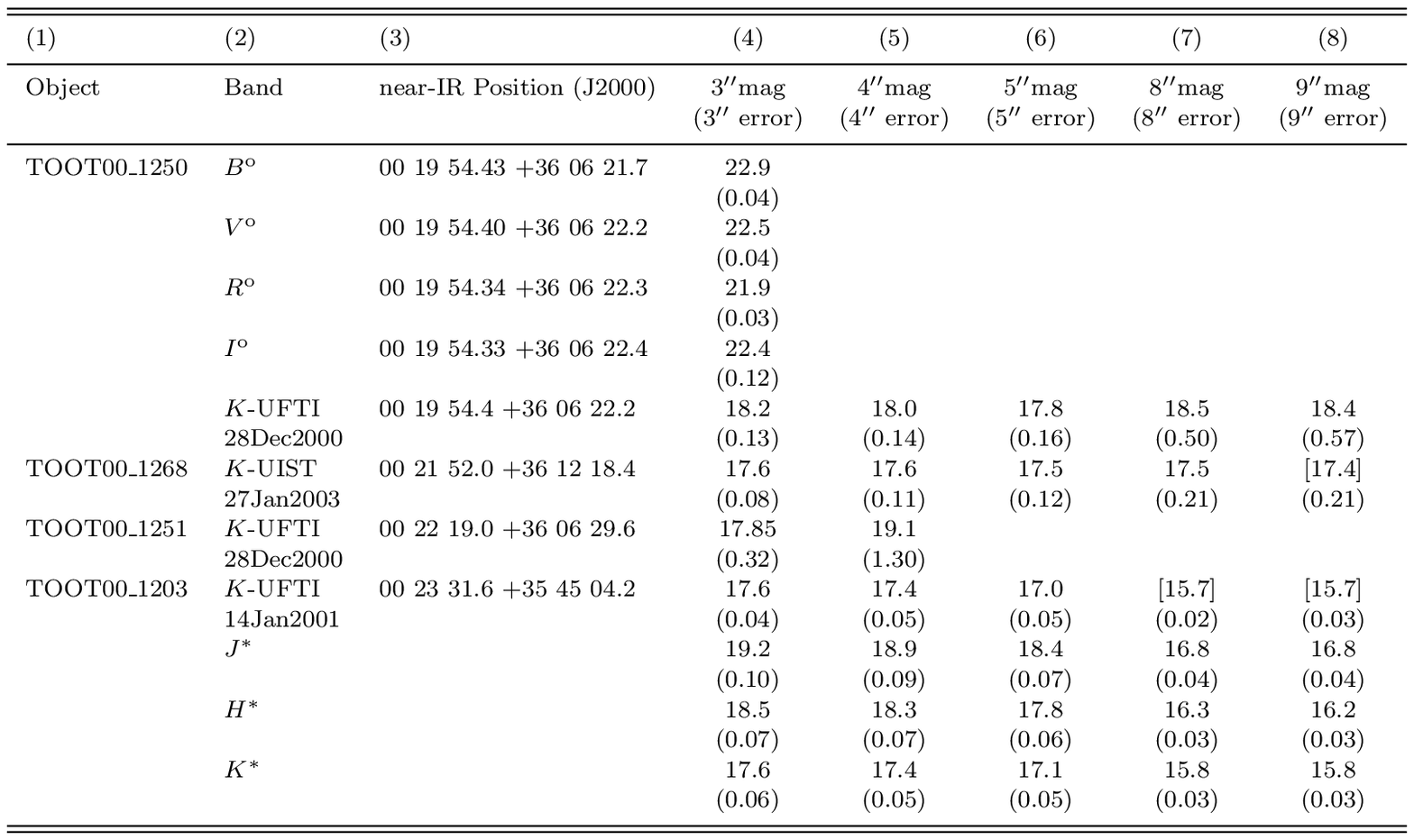}}
\end{picture}
\end{center}
 \end{table*}

\normalsize

\clearpage

\addtocounter{figure}{0}

\begin{figure*}
\begin{center}
\setlength{\unitlength}{1mm}
\begin{picture}(150,165)
\put(-47,-95){\includegraphics{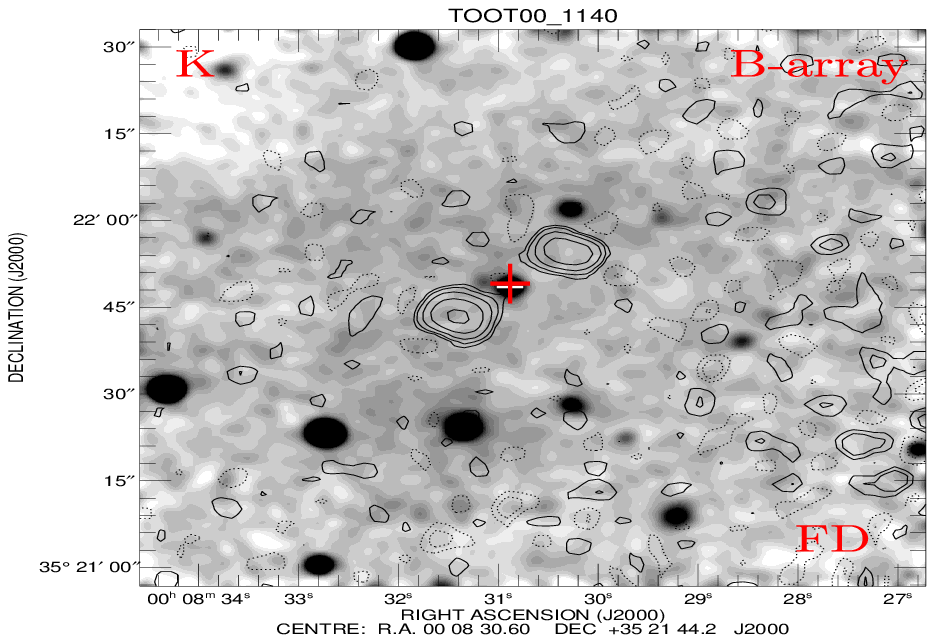}}
\put(160,30){\includegraphics{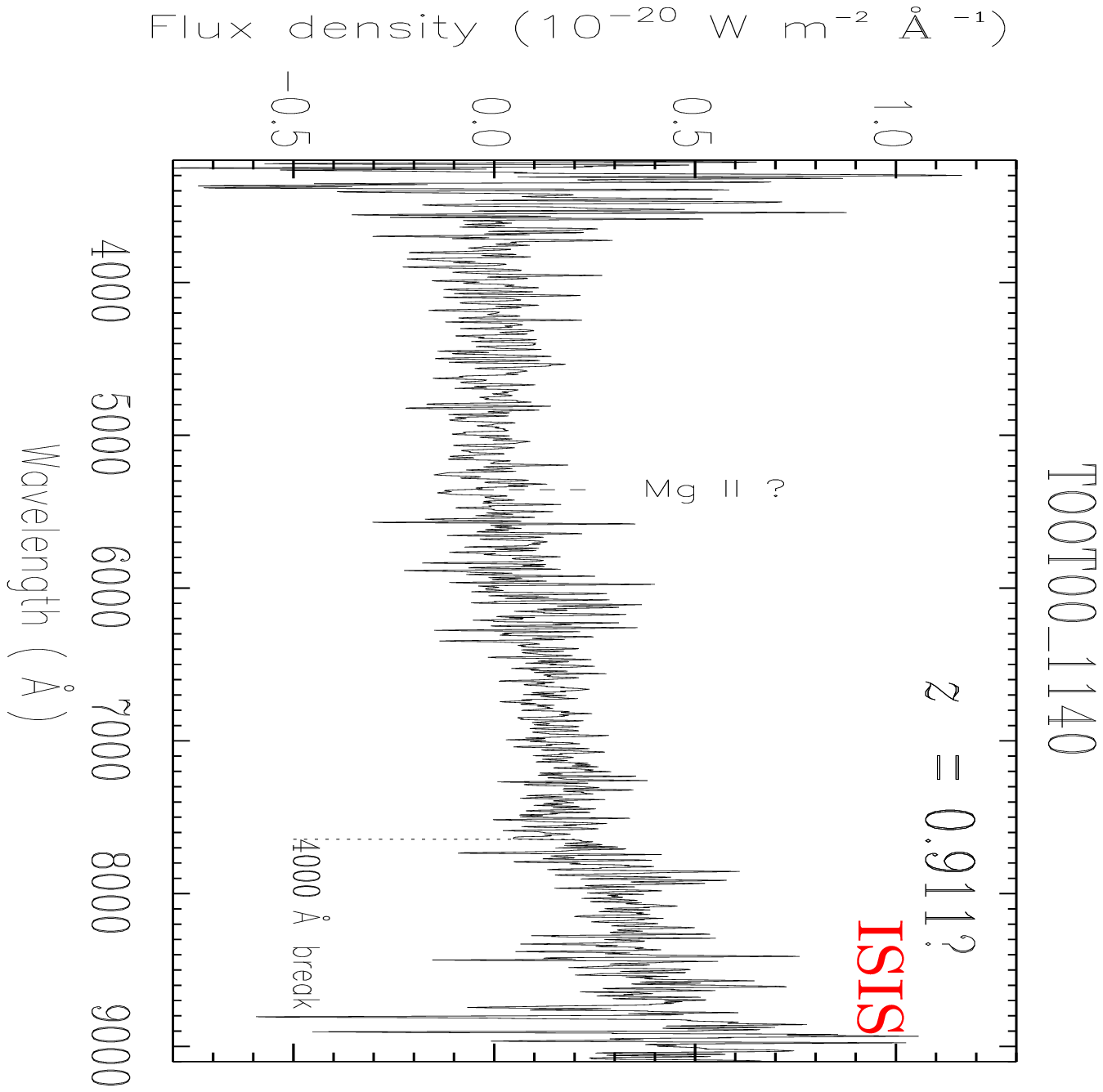}}
\put(80,-20){\includegraphics{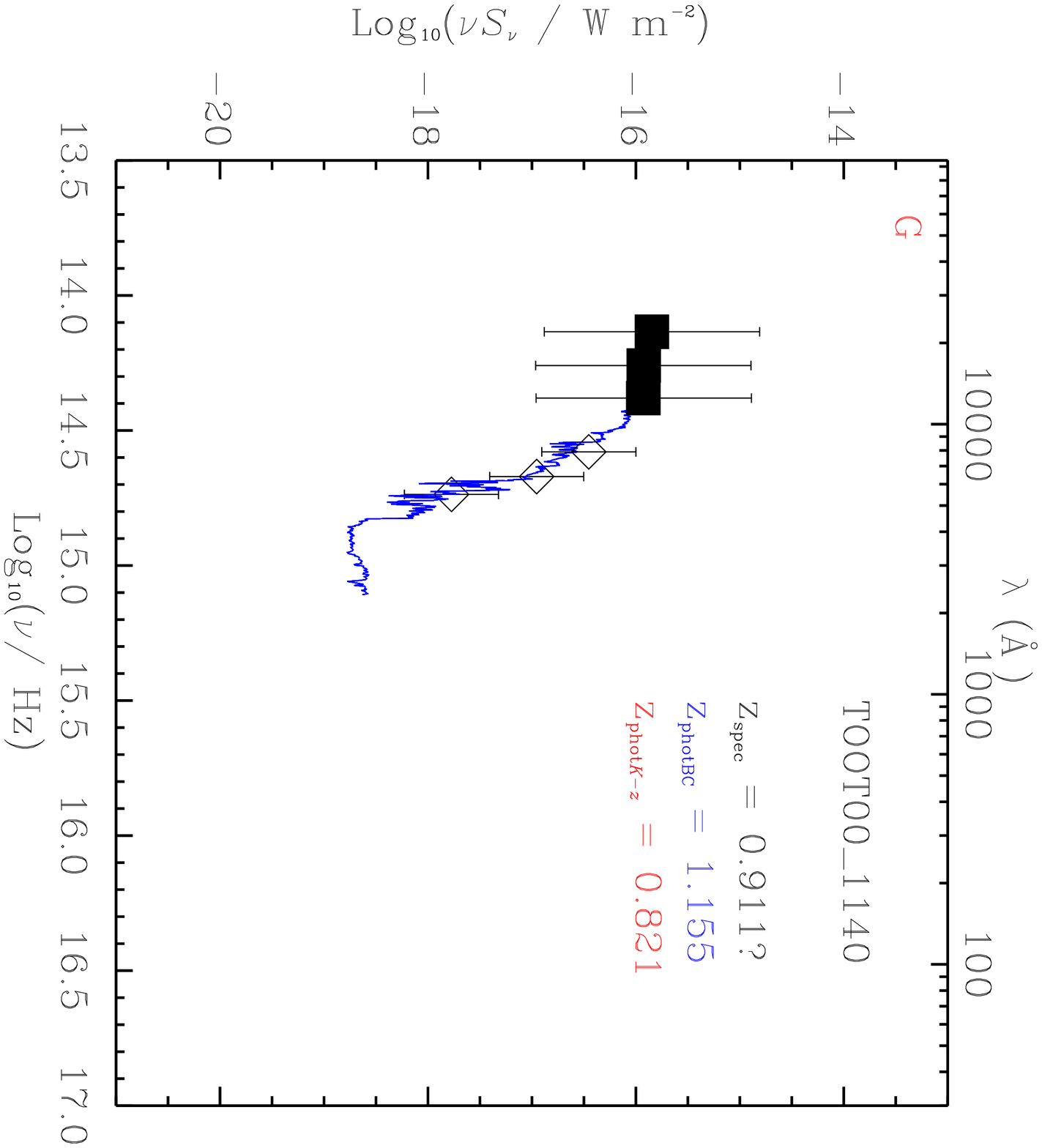}}
\put(47,-95){\includegraphics{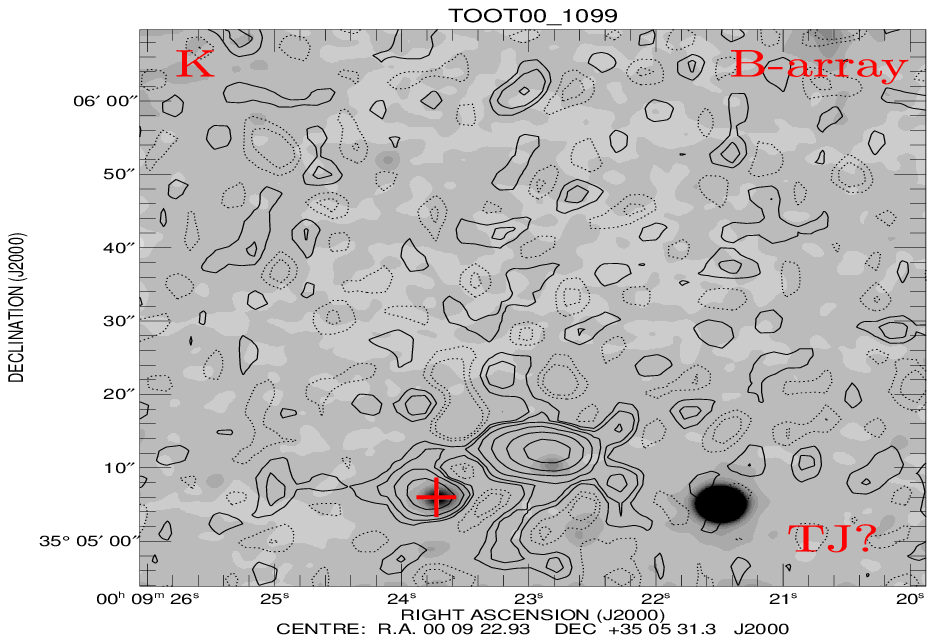}}
\put(260,30){\includegraphics{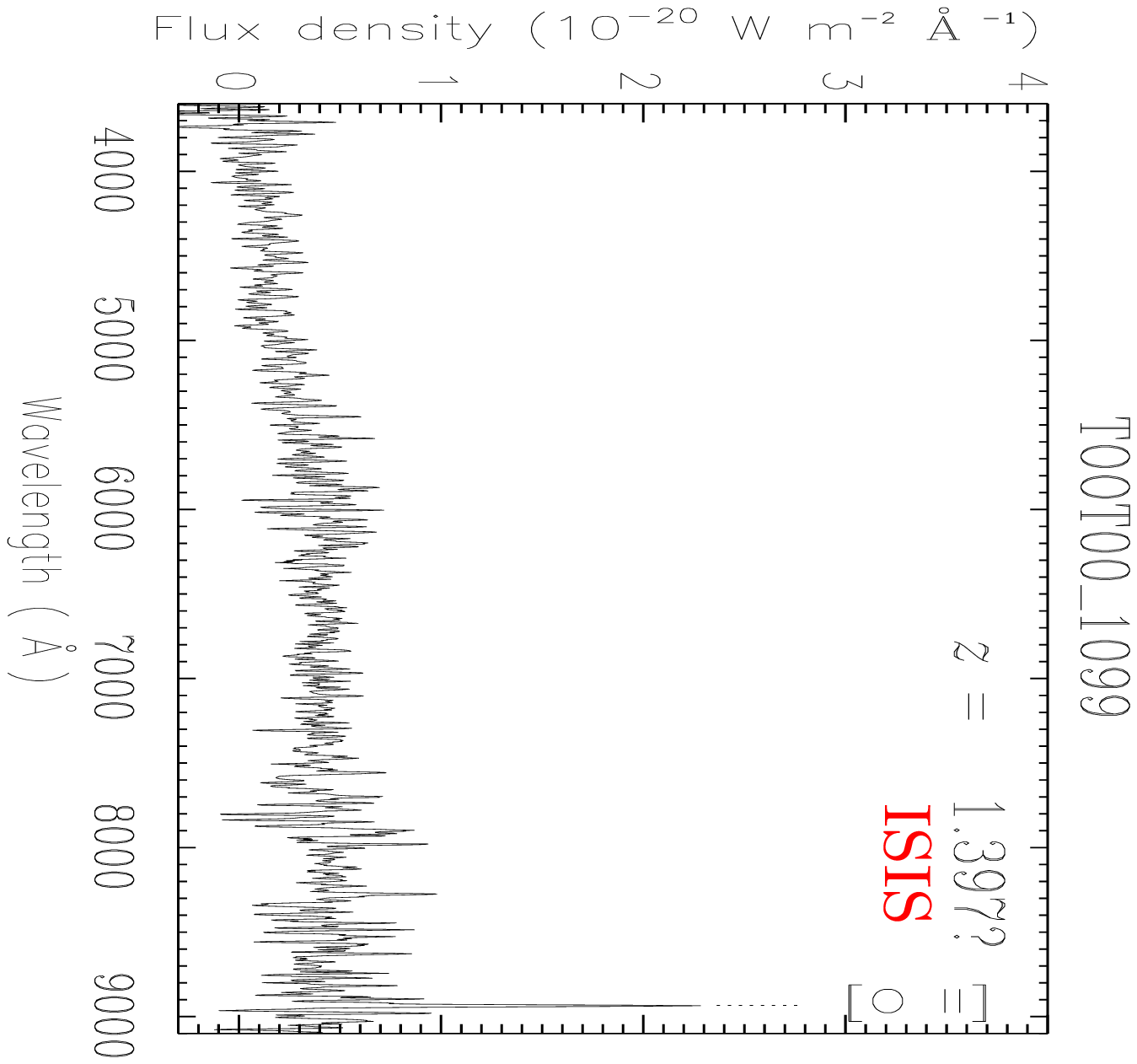}}
\put(180,-20){\includegraphics{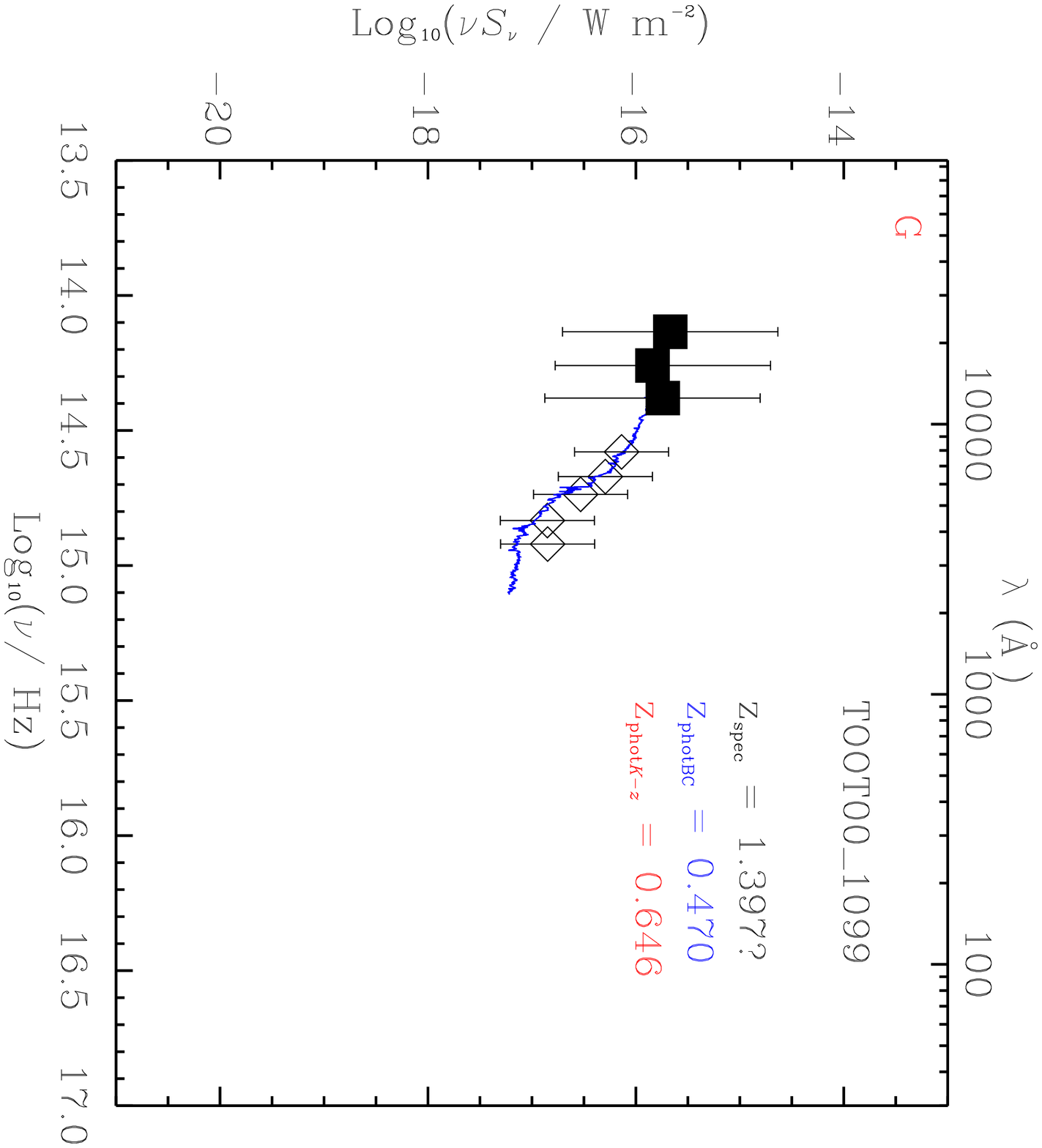}}
\end{picture}
\end{center}
\vspace{.6in}
{\caption[junk]{\label{seds}
At the {\bf top} of each panel we present the $K$-band images and overlaid 
radio contours from the VLA (B- or A-array as stated). The contour levels 
follow: $2^{n_{\rm contour}-1}\times \sigma$, where $\rm n_{\rm contour}$ is the 
number of contours and $\sigma$ the noise level. A white/red cross indicates 
the position of the assumed near-ID. 
In the {\bf middle} we give the optical spectrum of the object, as described 
in Section~\ref{sec:sample}. The blue and red part of each spectrum have been 
combined at $\sim$ 6000 $\rm \AA$. We haven't corrected for atmospheric 
absorption, which is labelled as `A' in the spectrum; `S' denotes a sky line; 
`F' denotes fringing; and `C' a cosmic ray effect. 
At the {\bf bottom} we give the SED and the results from the HyperZ fit 
on the photometric data. Symbols: filled squares denote photometric points, 
whereas open diamonds denote that the fluxes were estimated from the optical 
spectra. Note: for object TOOT00\_1240 we also provide a zoomed-in 
$K-$band/radio overlay and a DSS/radio overlay. 
The positions of an M-star and a galaxy G ($z$ = 0.191) close to the position 
of TOOT00\_1240 are marked with an `M' and a `G' respectively. For 
TOOT00\_1252 both A- and B-array radio maps are presented; the $K$ position is 
not marked since it is a limit at $K$. The red arrow in the $K$-band/B-array 
overlay in TOOT00\_1196 shows the 7C radio position.
}}
\end{figure*}

\addtocounter{figure}{-1}

\clearpage

\begin{figure*}
\begin{center}
\setlength{\unitlength}{1mm}
\begin{picture}(150,220)
\put(-25,-40){\includegraphics{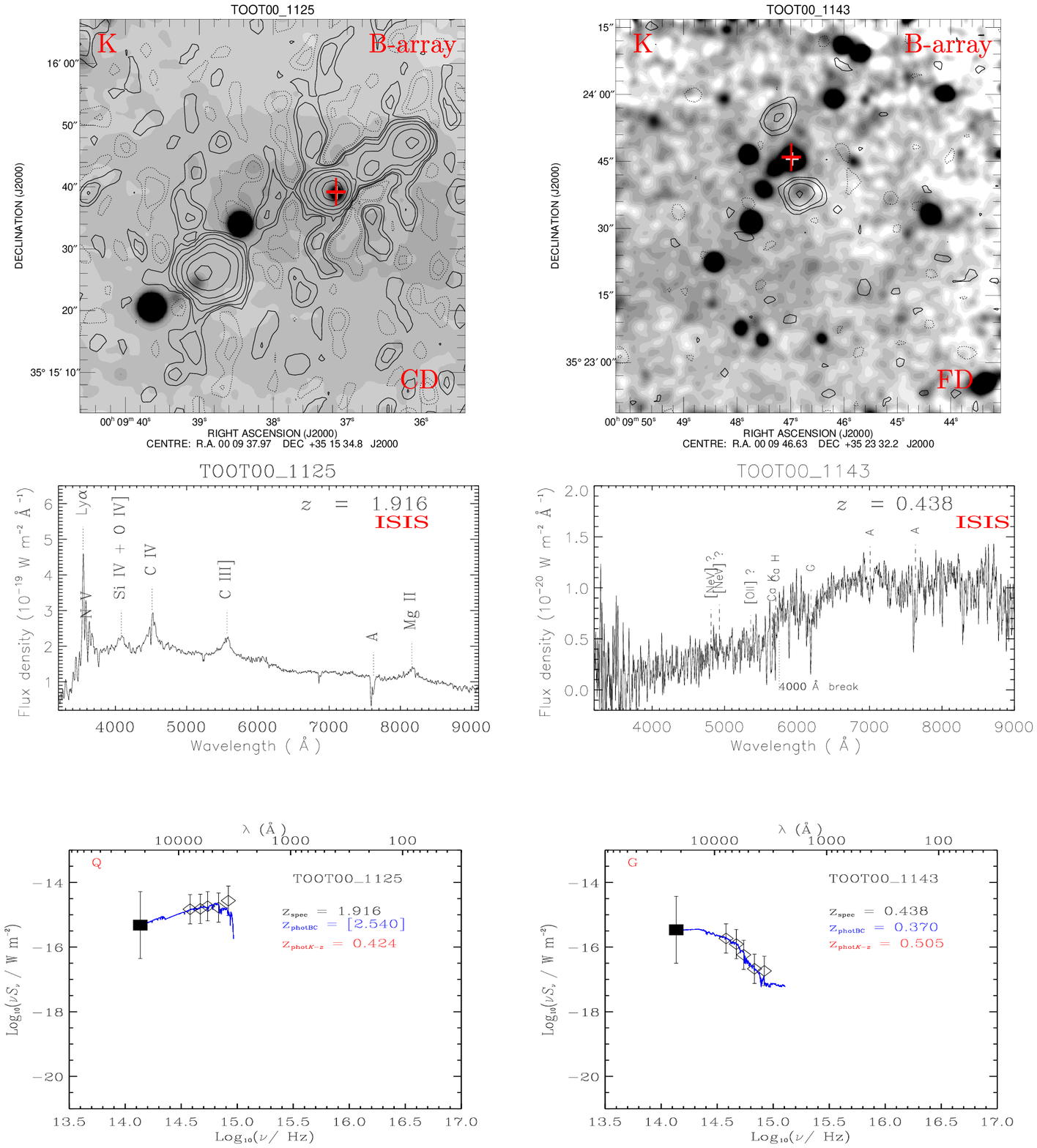}}
\end{picture}
\end{center}
\vspace{0.1in}
{\caption[junk]{(continued)
}}
\end{figure*}

\addtocounter{figure}{-1}

\clearpage

\begin{figure*}
\begin{center}
\setlength{\unitlength}{1mm}
\begin{picture}(150,220)
\put(-25,-40){\includegraphics{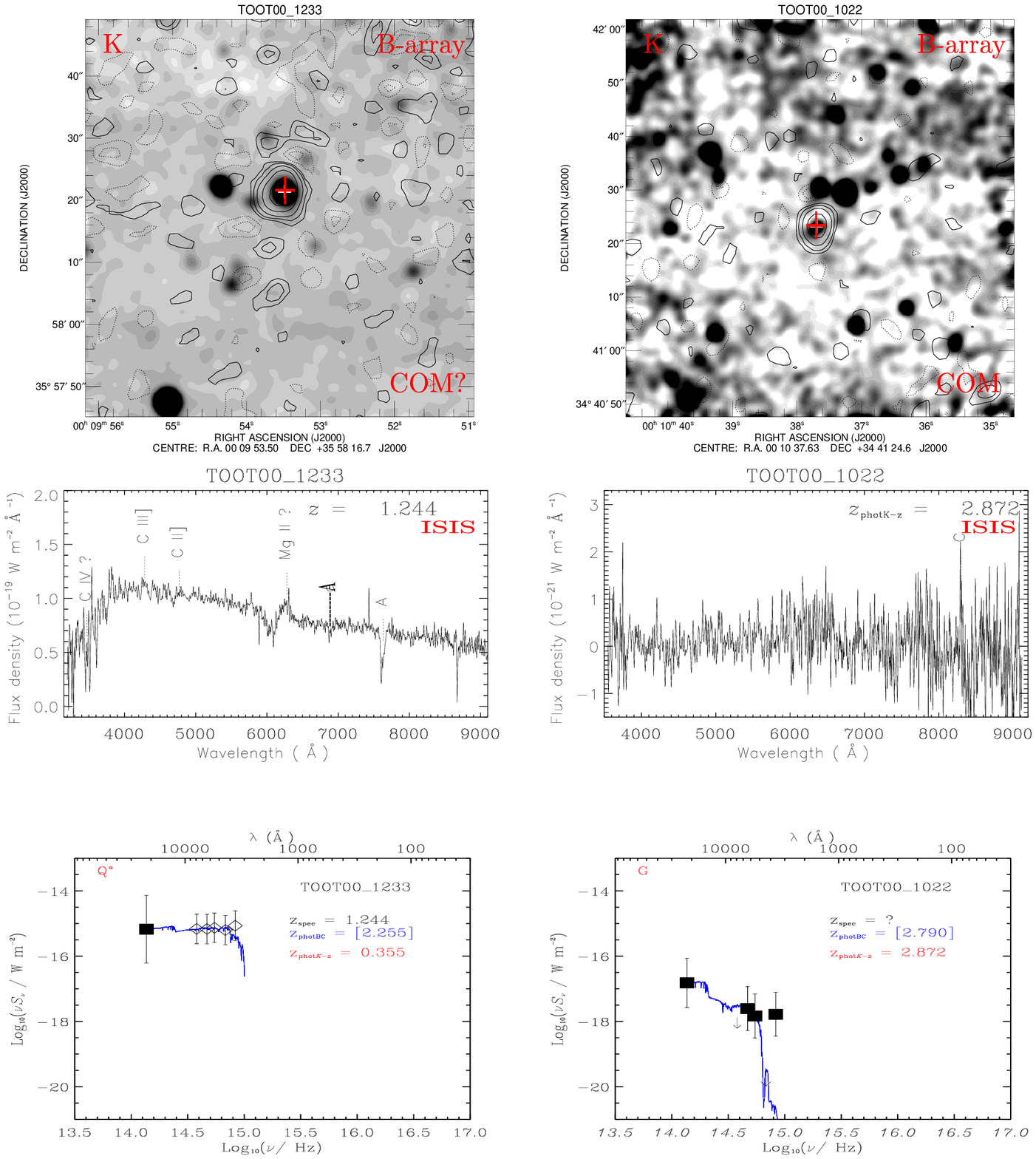}}
\end{picture}
\end{center}
\vspace{0.1in}
{\caption[junk]{(continued)
}}
\end{figure*}

\addtocounter{figure}{-1}

\clearpage

\begin{figure*}
\begin{center}
\setlength{\unitlength}{1mm}
\begin{picture}(150,220)
\put(-25,-40){\includegraphics{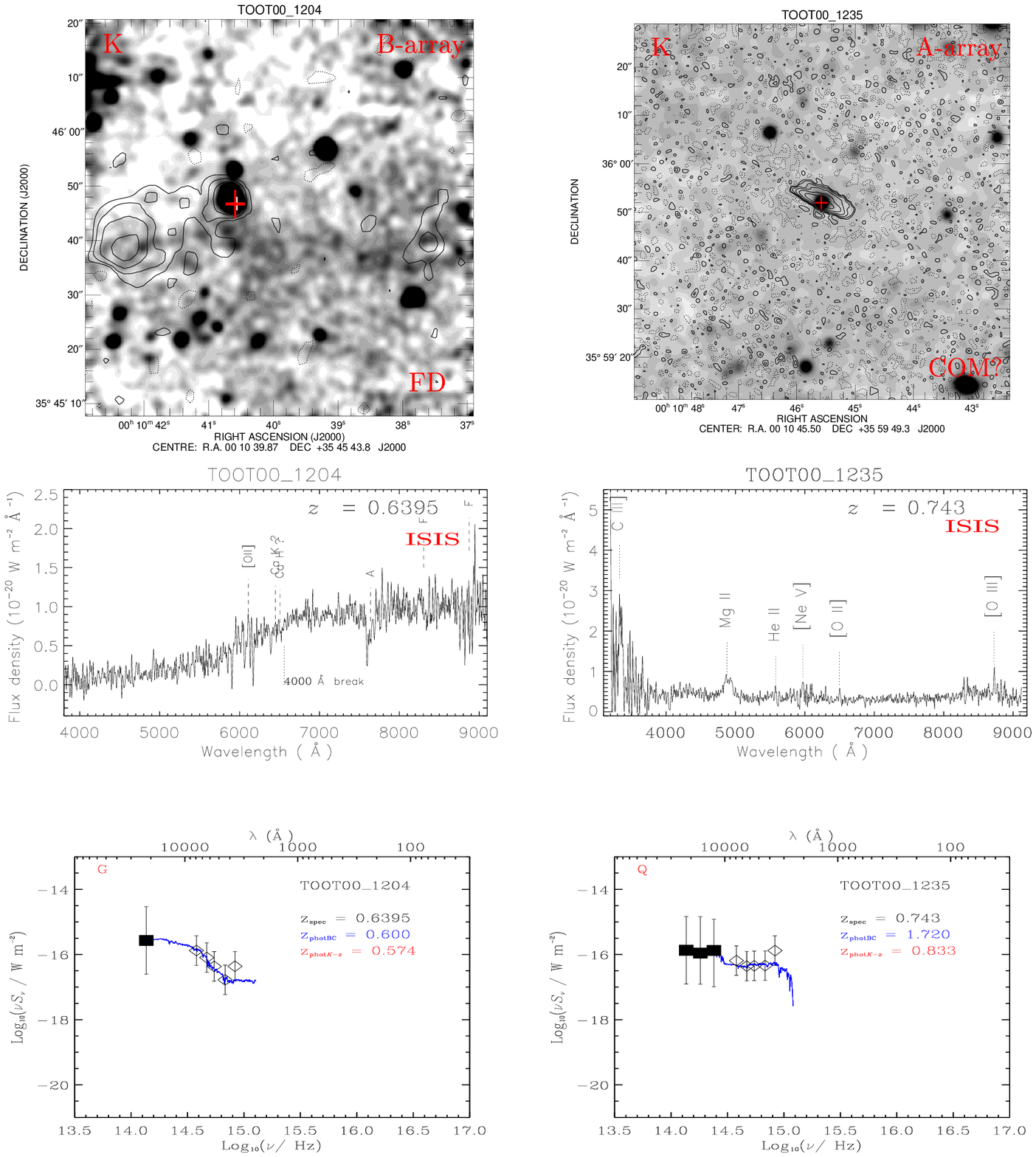}}
\end{picture}
\end{center}
\vspace{0.1in}
{\caption[junk]{(continued)
}}
\end{figure*}

\addtocounter{figure}{-1}

\clearpage

\begin{figure*}
\begin{center}
\setlength{\unitlength}{1mm}
\begin{picture}(150,220)
\put(-25,-40){\includegraphics{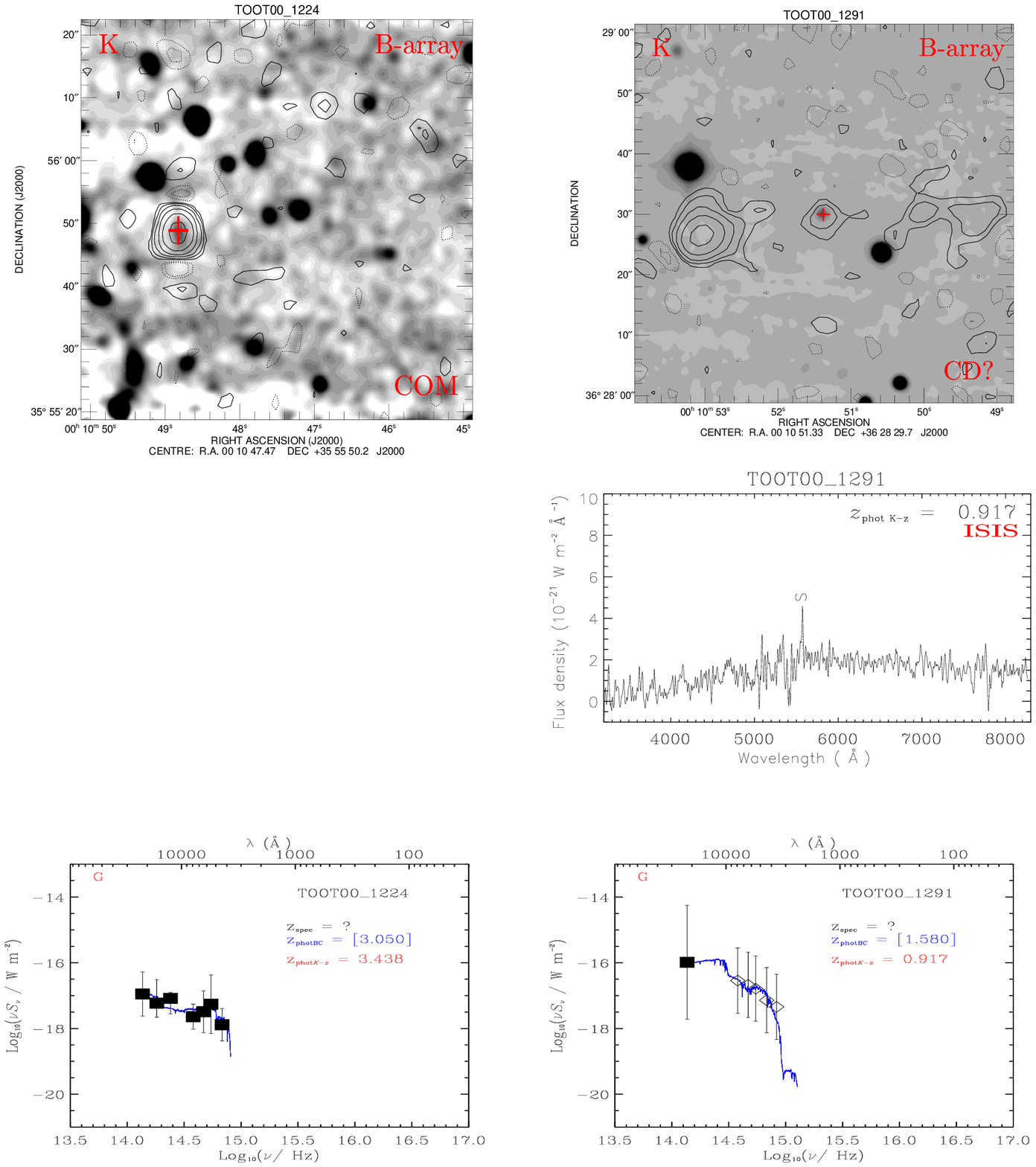}}
\end{picture}
\end{center}
\vspace{0.1in}
{\caption[junk]{(continued)
}}
\end{figure*}

\addtocounter{figure}{-1}

\clearpage

\begin{figure*}
\begin{center}
\setlength{\unitlength}{1mm}
\begin{picture}(150,220)
\put(-25,-40){\includegraphics{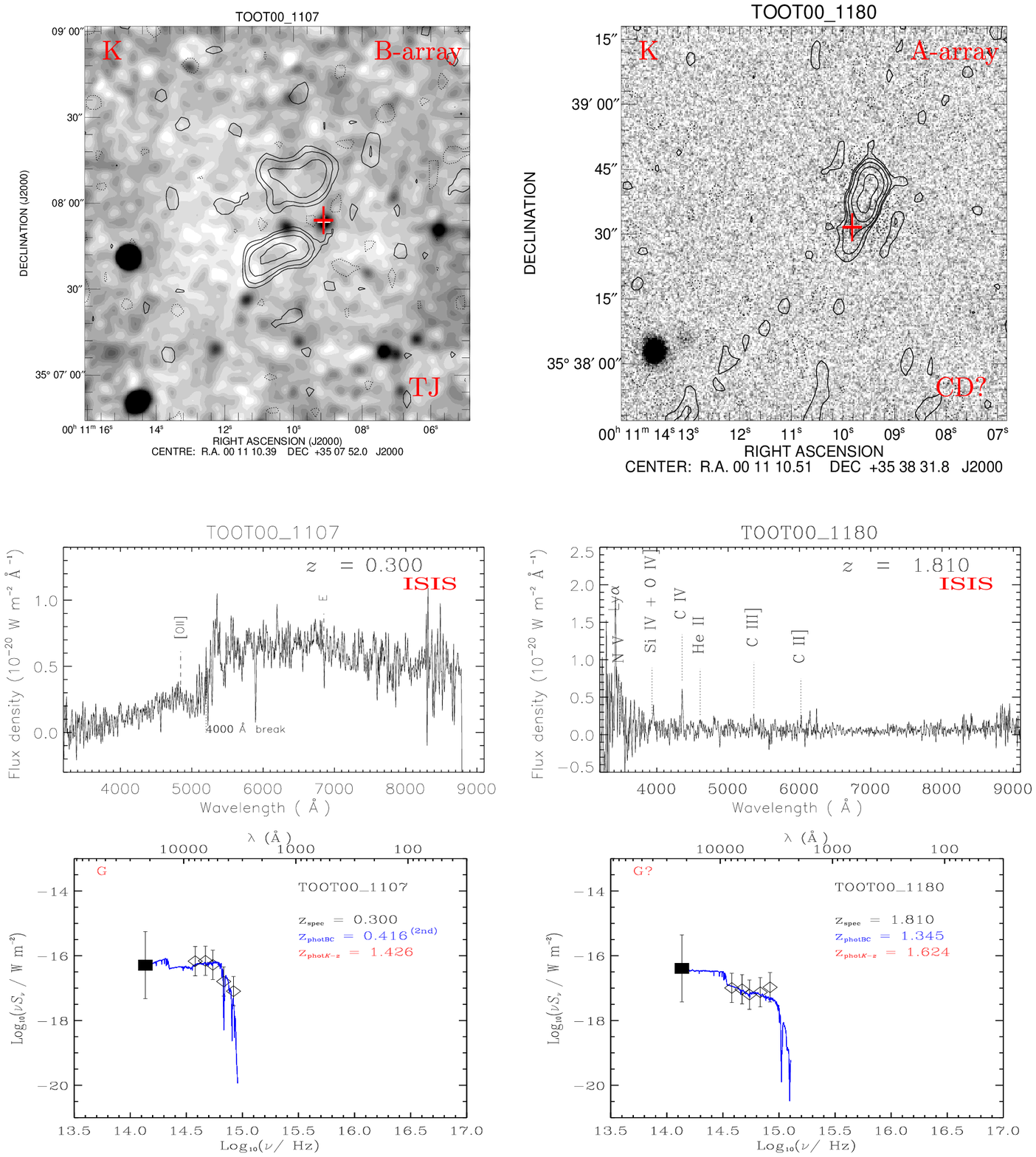}}
\end{picture}
\end{center}
\vspace{0.1in}
{\caption[junk]{(continued)
}}
\end{figure*}

\addtocounter{figure}{-1}

\clearpage

\begin{figure*}
\begin{center}
\setlength{\unitlength}{1mm}
\begin{picture}(150,220)
\put(-25,-40){\includegraphics{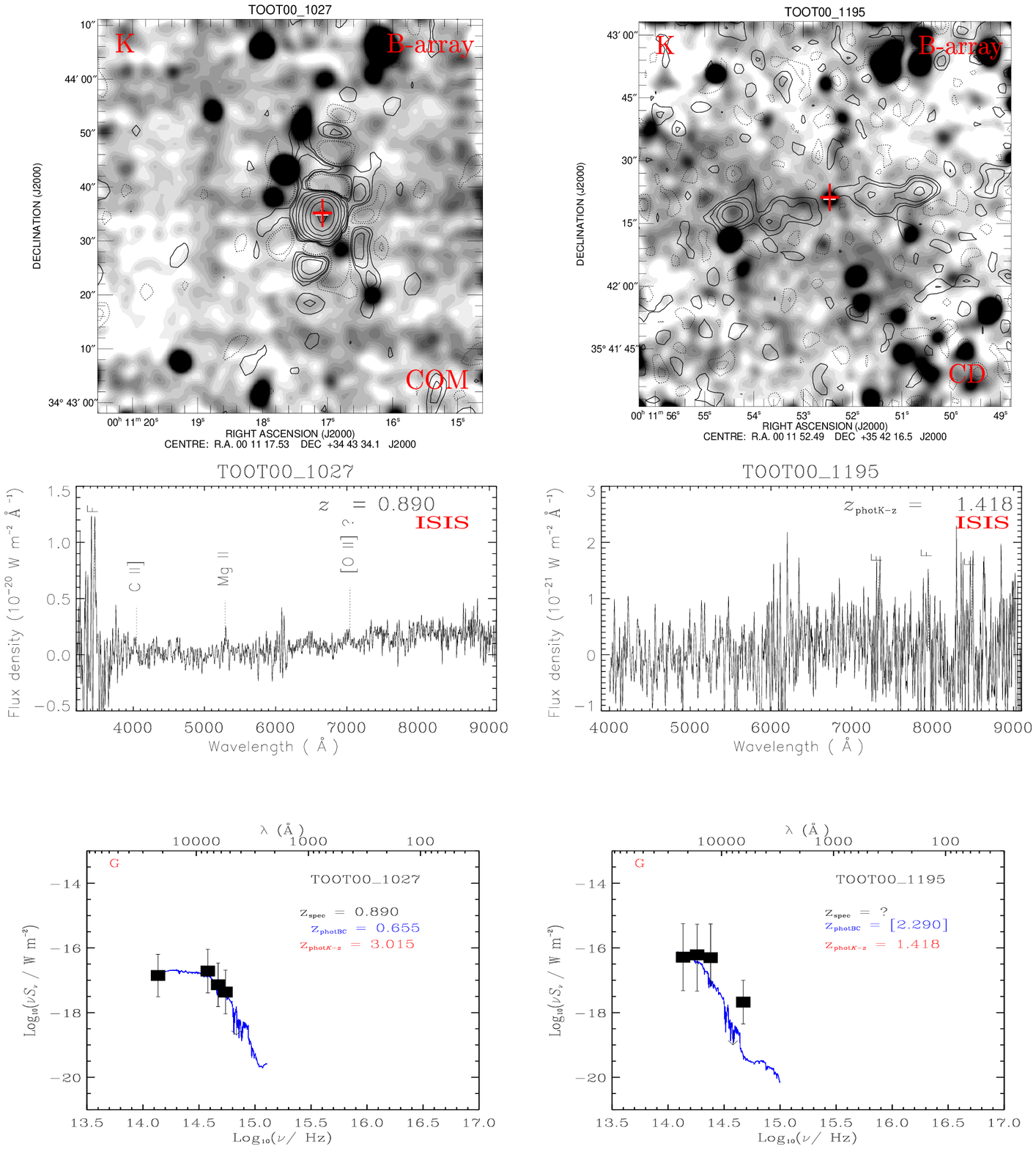}}
\end{picture}
\end{center}
\vspace{0.1in}
{\caption[junk]{(continued)
}}
\end{figure*}

\addtocounter{figure}{-1}

\clearpage

\begin{figure*}
\begin{center}
\setlength{\unitlength}{1mm}
\begin{picture}(150,220)
\put(-25,-40){\includegraphics{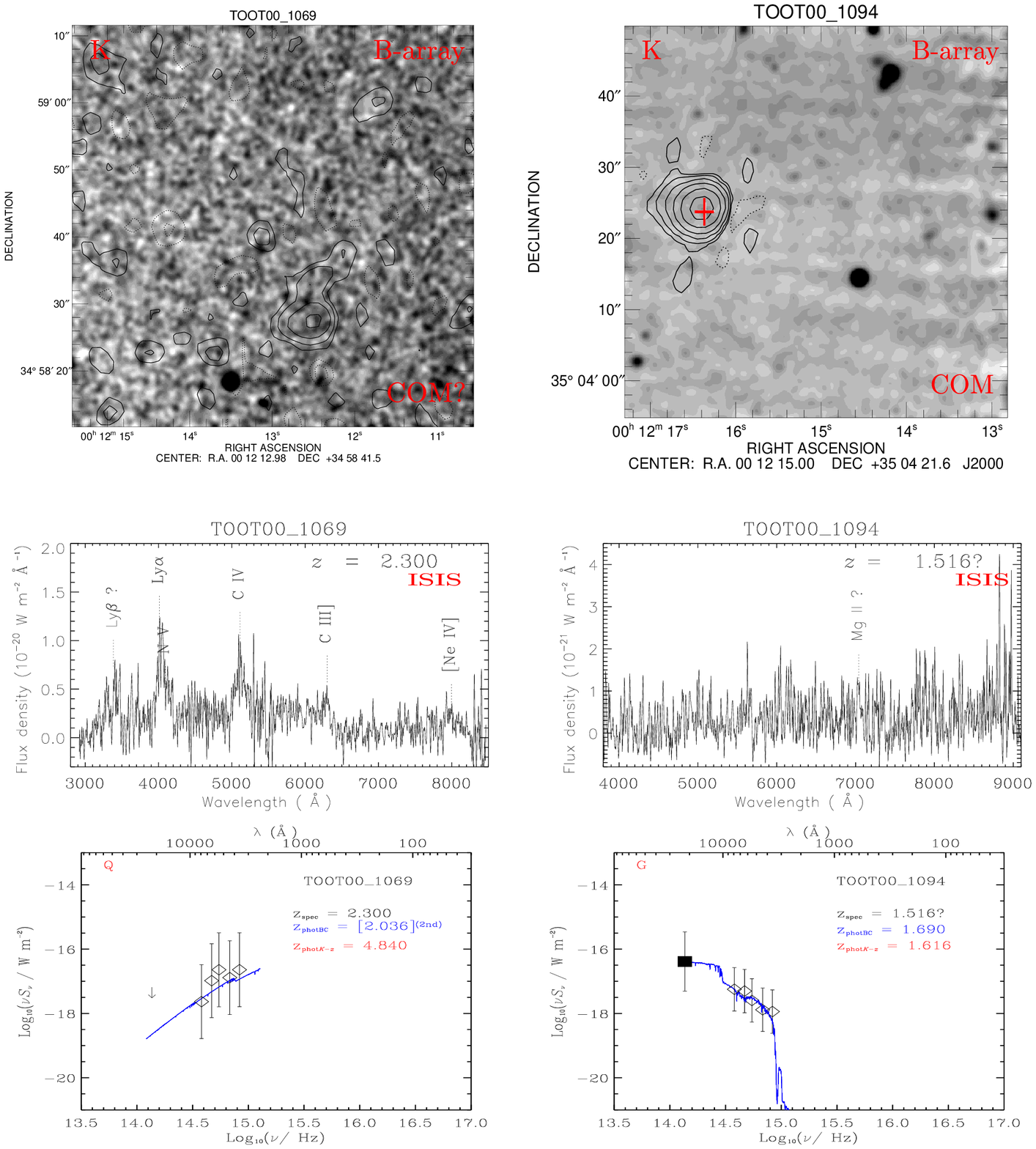}}
\end{picture}
\end{center}
\vspace{0.1in}
{\caption[junk]{(continued)
}}
\end{figure*}

\addtocounter{figure}{-1}

\clearpage

\begin{figure*}
\begin{center}
\setlength{\unitlength}{1mm}
\begin{picture}(150,220)
\put(-25,-40){\includegraphics{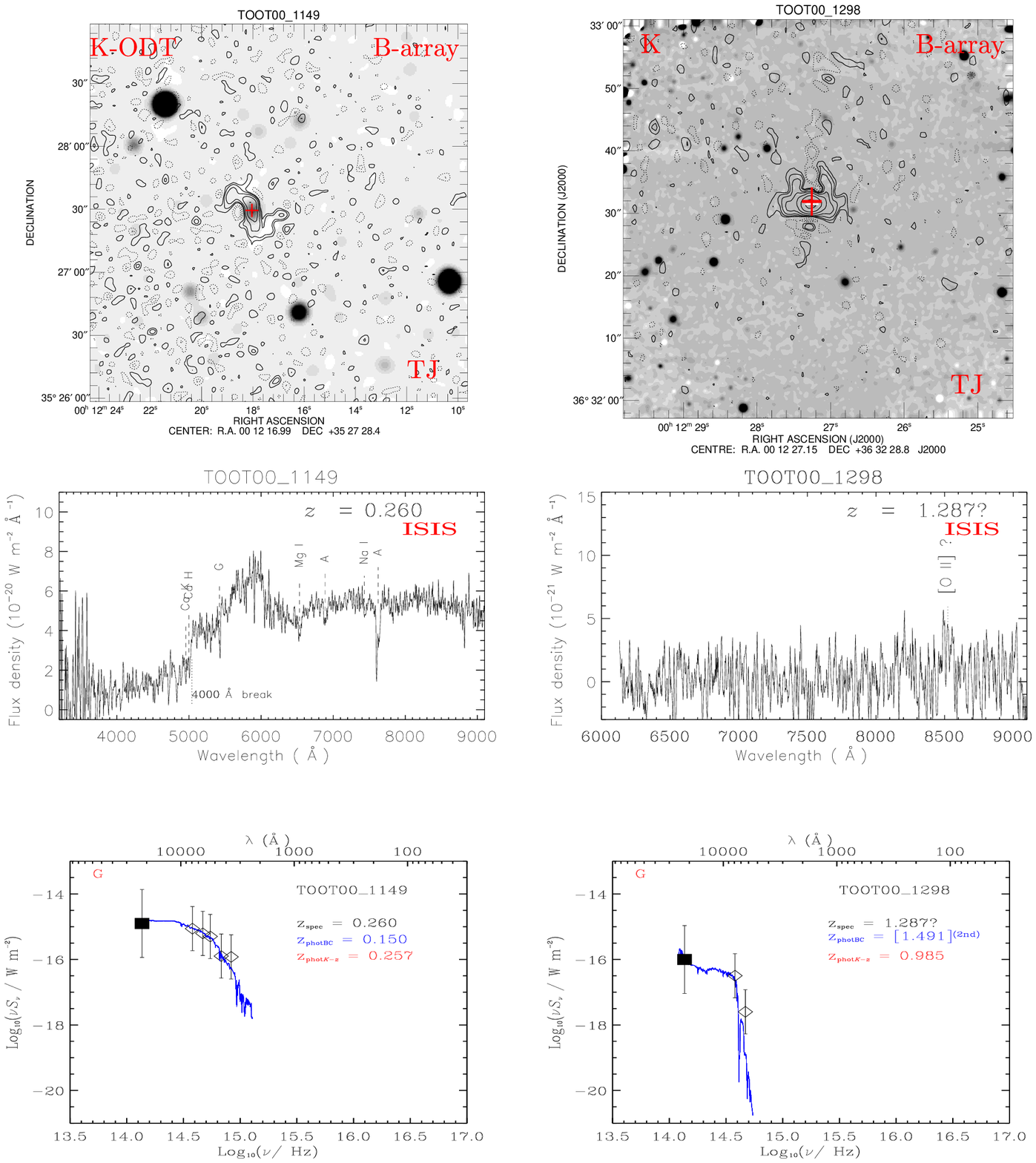}}
\end{picture}
\end{center}
\vspace{0.1in}
{\caption[junk]{(continued)
}}
\end{figure*}

\addtocounter{figure}{-1}

\clearpage

\begin{figure*}
\begin{center}
\setlength{\unitlength}{1mm}
\begin{picture}(150,220)
\put(-25,-40){\includegraphics{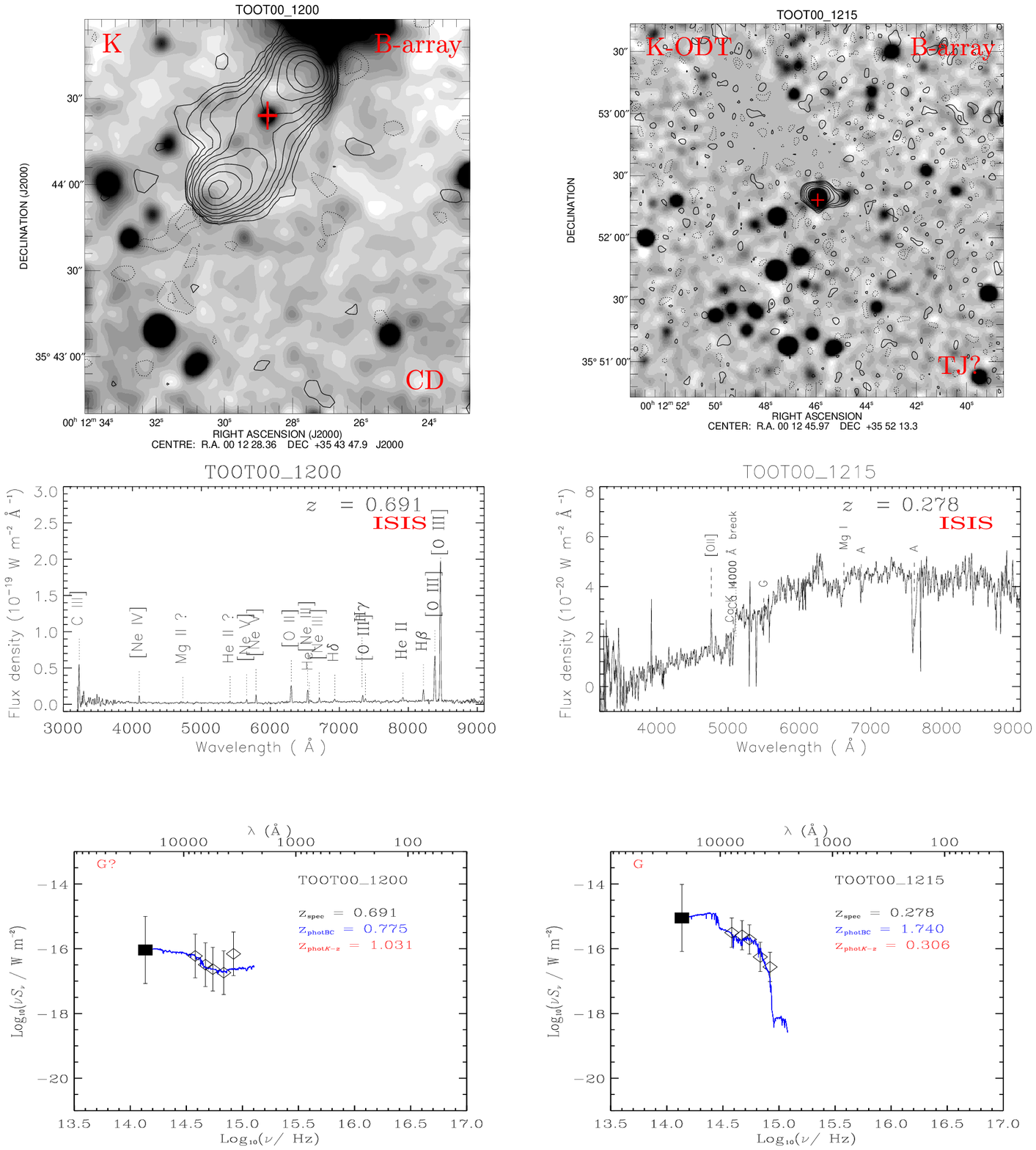}}
\end{picture}
\end{center}
\vspace{0.1in}
{\caption[junk]{(continued)
}}
\end{figure*}

\addtocounter{figure}{-1}

\clearpage

\begin{figure*}
\begin{center}
\setlength{\unitlength}{1mm}
\begin{picture}(150,220)
\put(-25,-40){\includegraphics{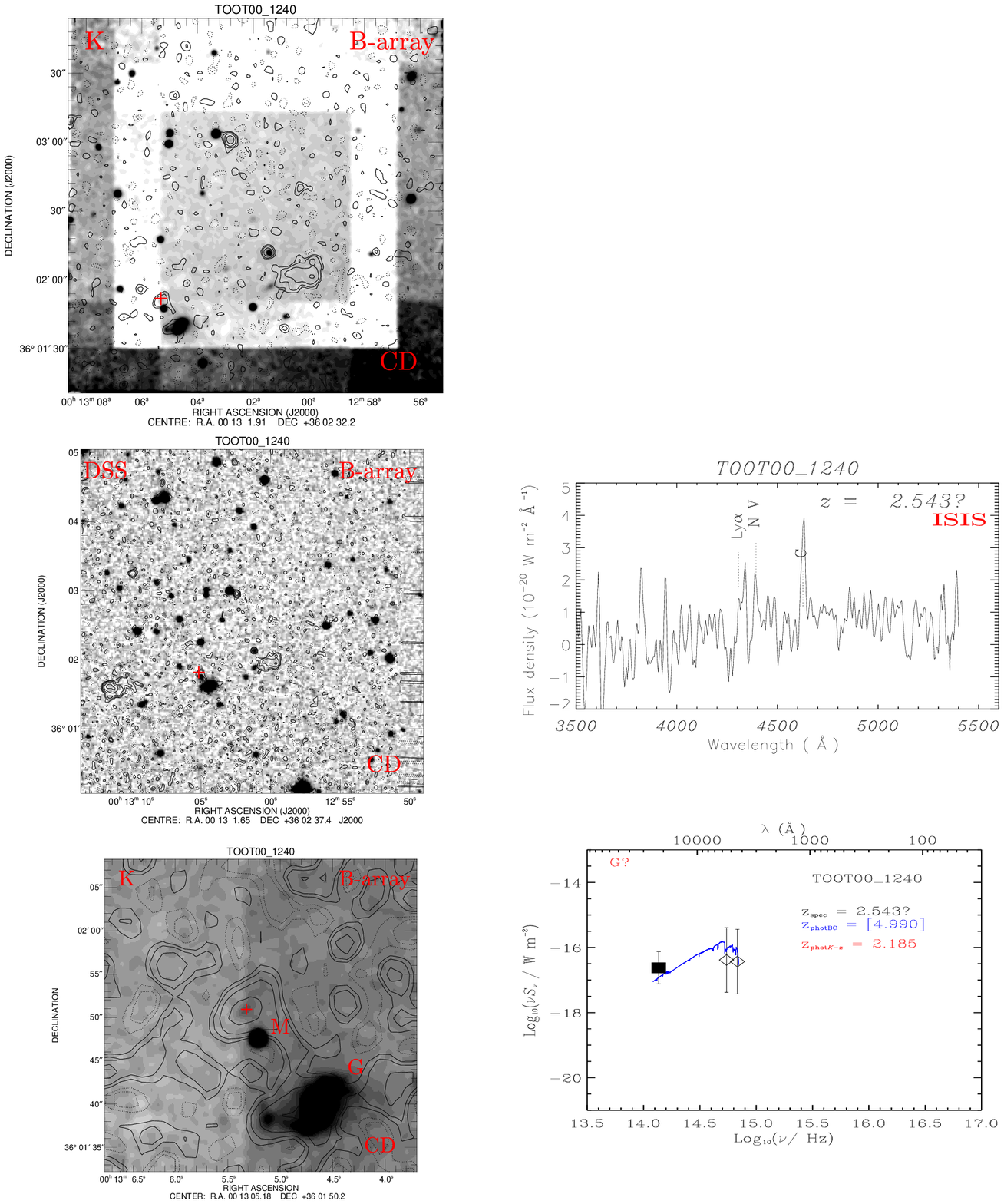}}
\end{picture}
\end{center}
\vspace{0.1in}
{\caption[junk]{(continued)
}}
\end{figure*}

\addtocounter{figure}{-1}

\clearpage
\begin{figure*}
\begin{center}
\setlength{\unitlength}{1mm}
\begin{picture}(150,220)
\put(-25,-40){\includegraphics{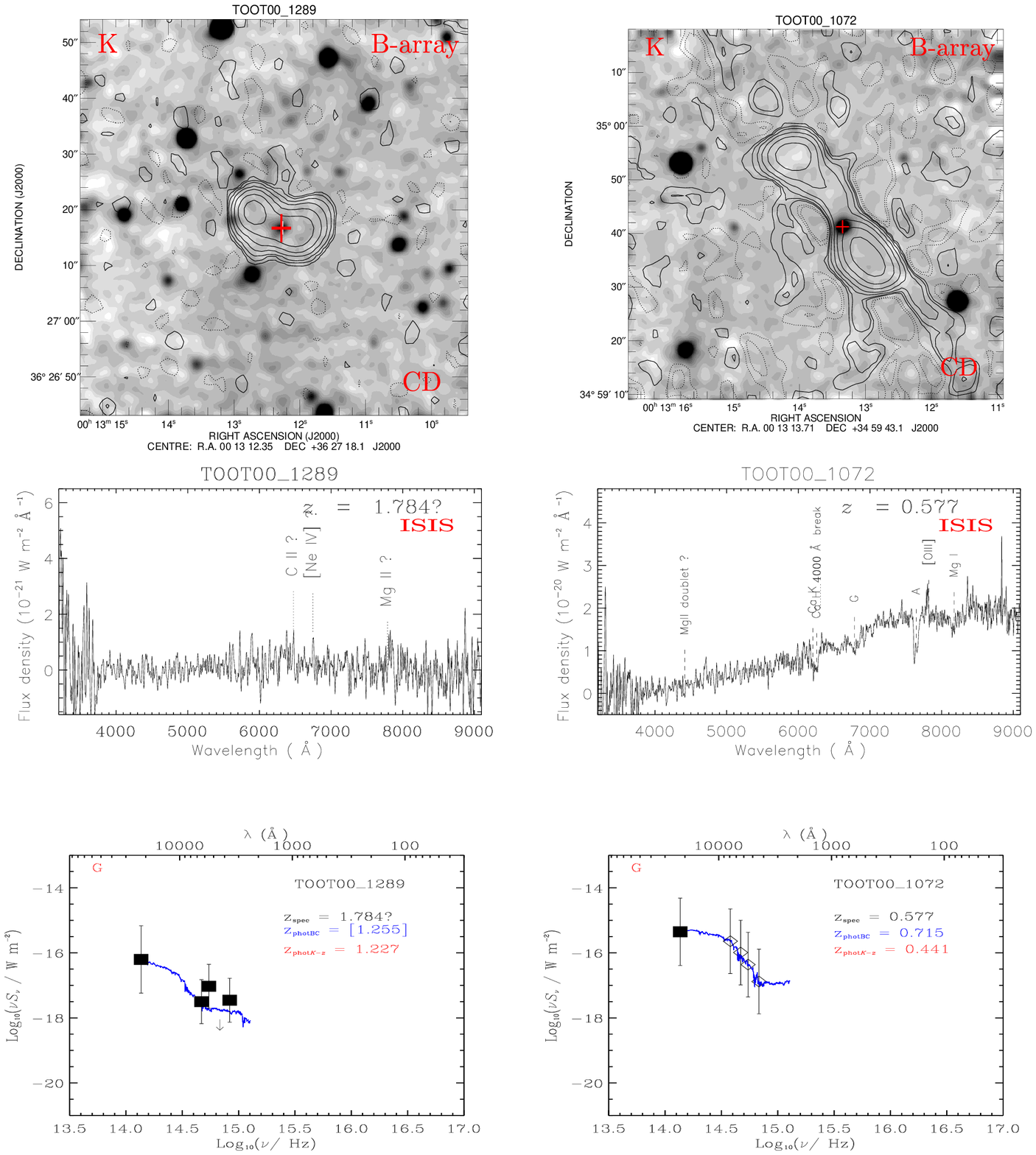}}
\end{picture}
\end{center}
\vspace{0.1in}
{\caption[junk]{(continued)
}}
\end{figure*}

\addtocounter{figure}{-1}

\clearpage

\begin{figure*}
\begin{center}
\setlength{\unitlength}{1mm}
\begin{picture}(150,220)
\put(-25,-40){\includegraphics{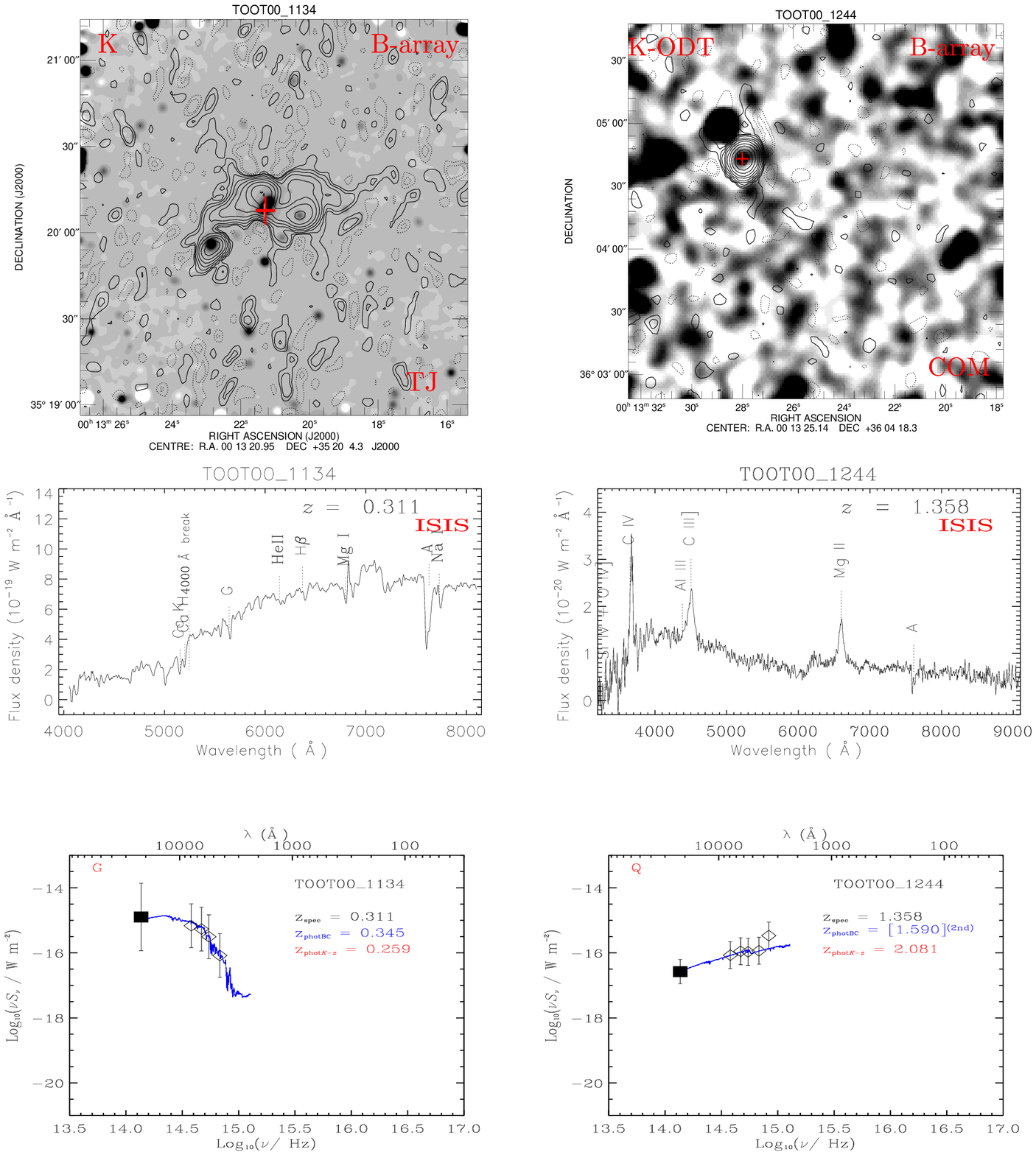}}
\end{picture}
\end{center}
\vspace{0.1in}
{\caption[junk]{(continued)
}}
\end{figure*}

\addtocounter{figure}{-1}

\clearpage

\begin{figure*}
\begin{center}
\setlength{\unitlength}{1mm}
\begin{picture}(150,220)
\put(-25,-40){\includegraphics{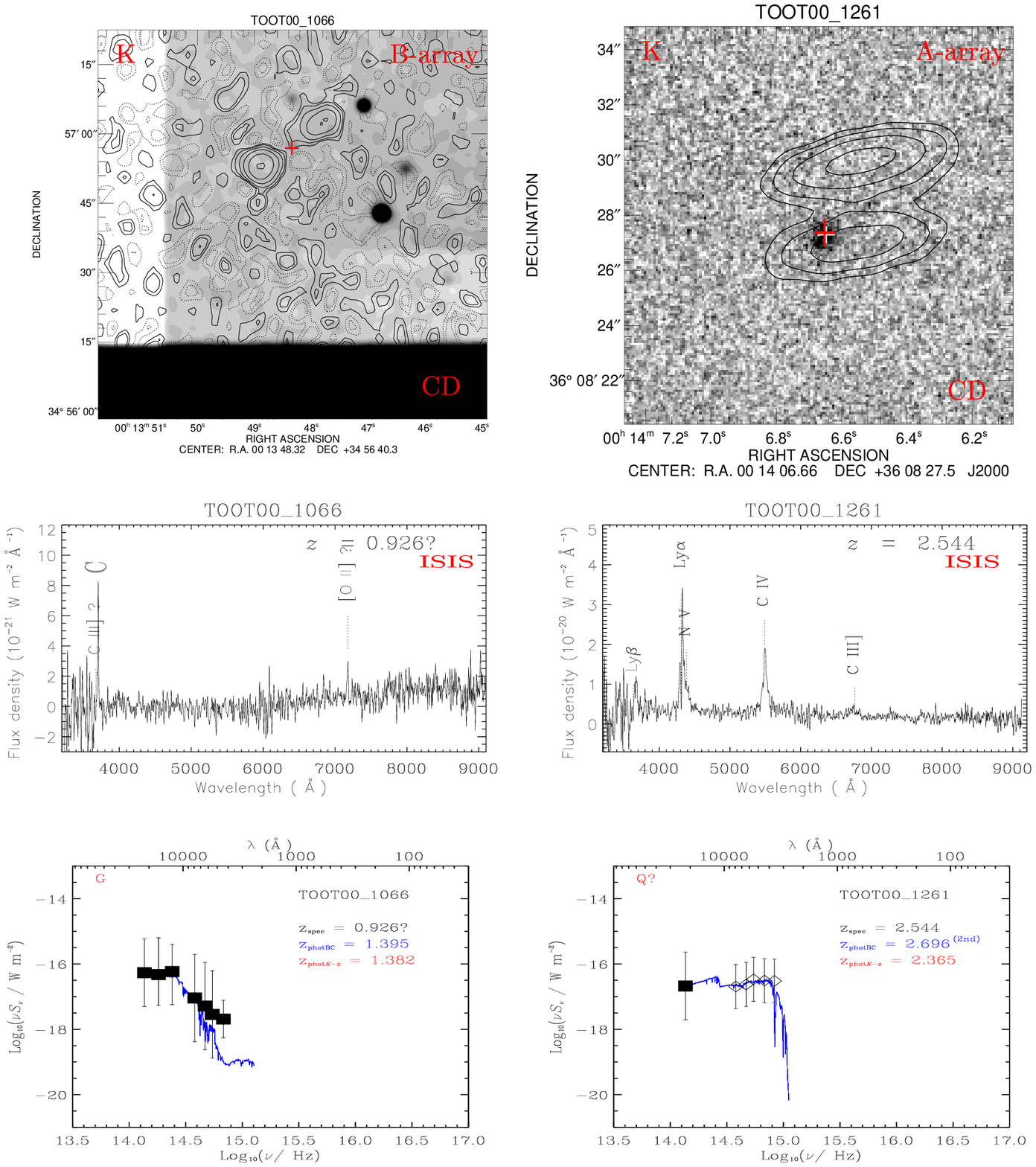}}
\end{picture}
\end{center}
\vspace{0.1in}
{\caption[junk]{(continued)
}}
\end{figure*}

\addtocounter{figure}{-1}

\clearpage

\begin{figure*}
\begin{center}
\setlength{\unitlength}{1mm}
\begin{picture}(150,220)
\put(-25,-40){\includegraphics{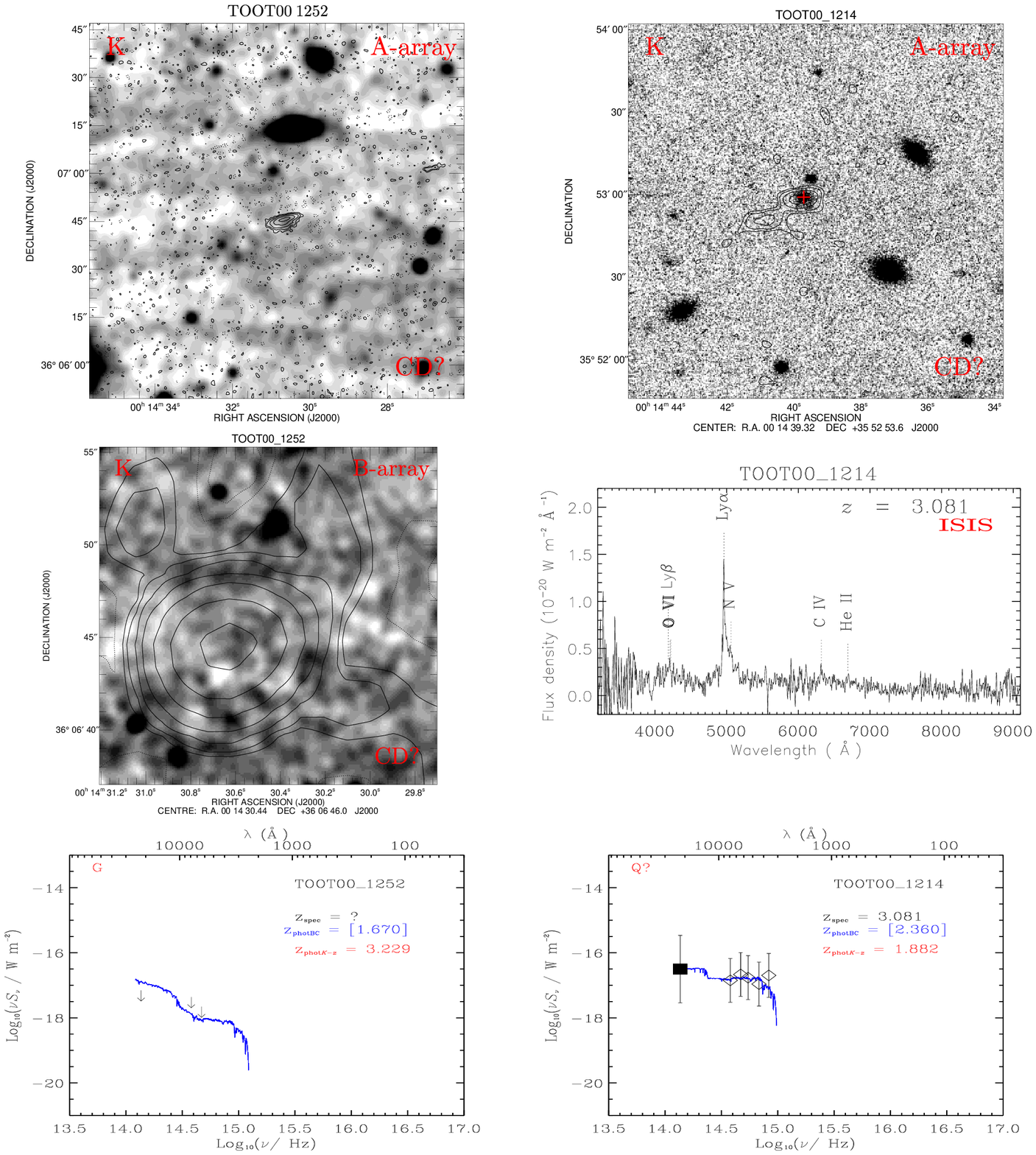}}
\end{picture}
\end{center}
\vspace{0.1in}
{\caption[junk]{(continued)
}}
\end{figure*}

\addtocounter{figure}{-1}

\clearpage

\begin{figure*}
\begin{center}
\setlength{\unitlength}{1mm}
\begin{picture}(150,220)
\put(-25,-40){\includegraphics{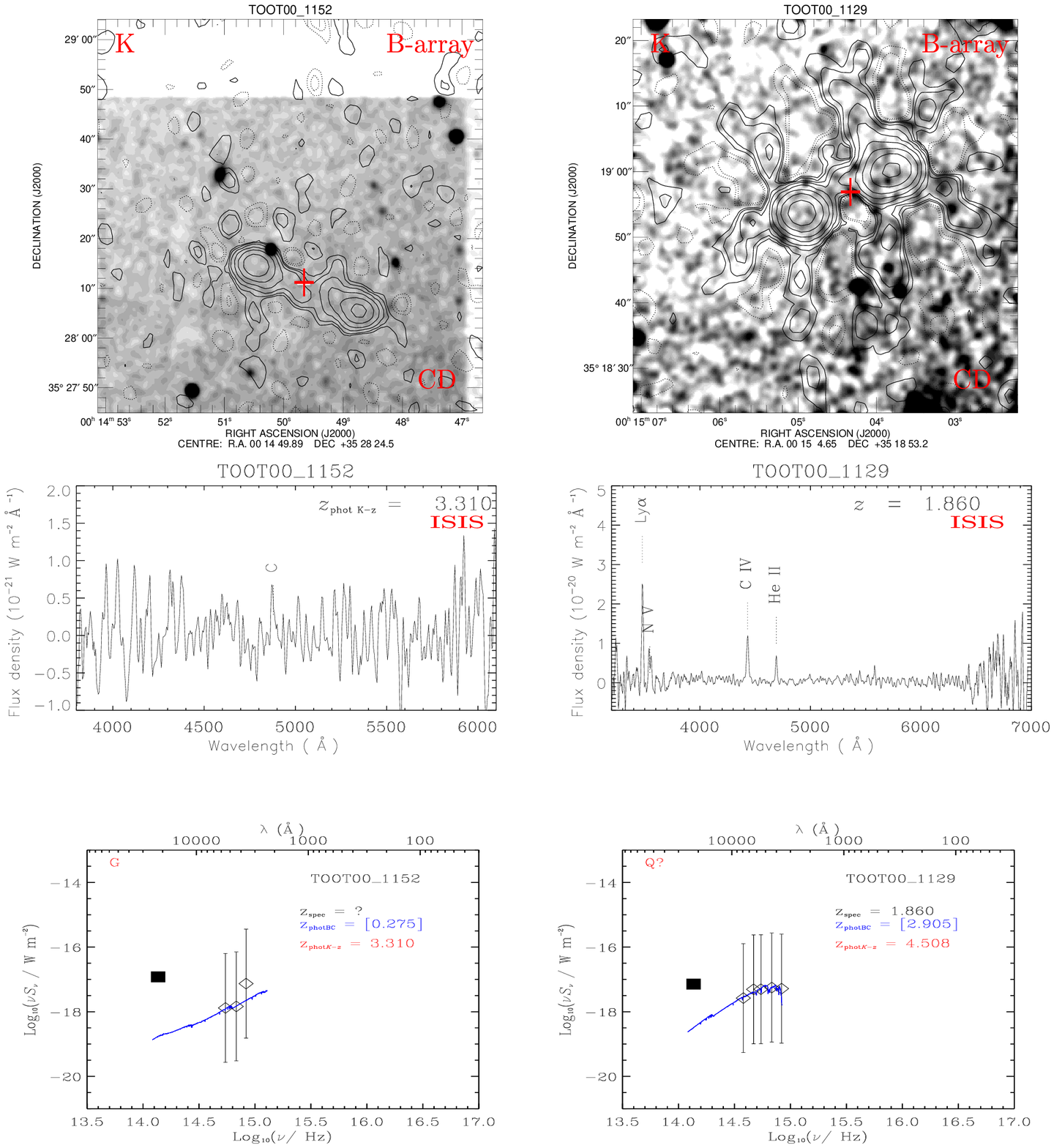}}
\end{picture}
\end{center}
\vspace{0.1in}
{\caption[junk]{(continued)
}}
\end{figure*}

\addtocounter{figure}{-1}

\clearpage

\begin{figure*}
\begin{center}
\setlength{\unitlength}{1mm}
\begin{picture}(150,220)
\put(-25,-40){\includegraphics{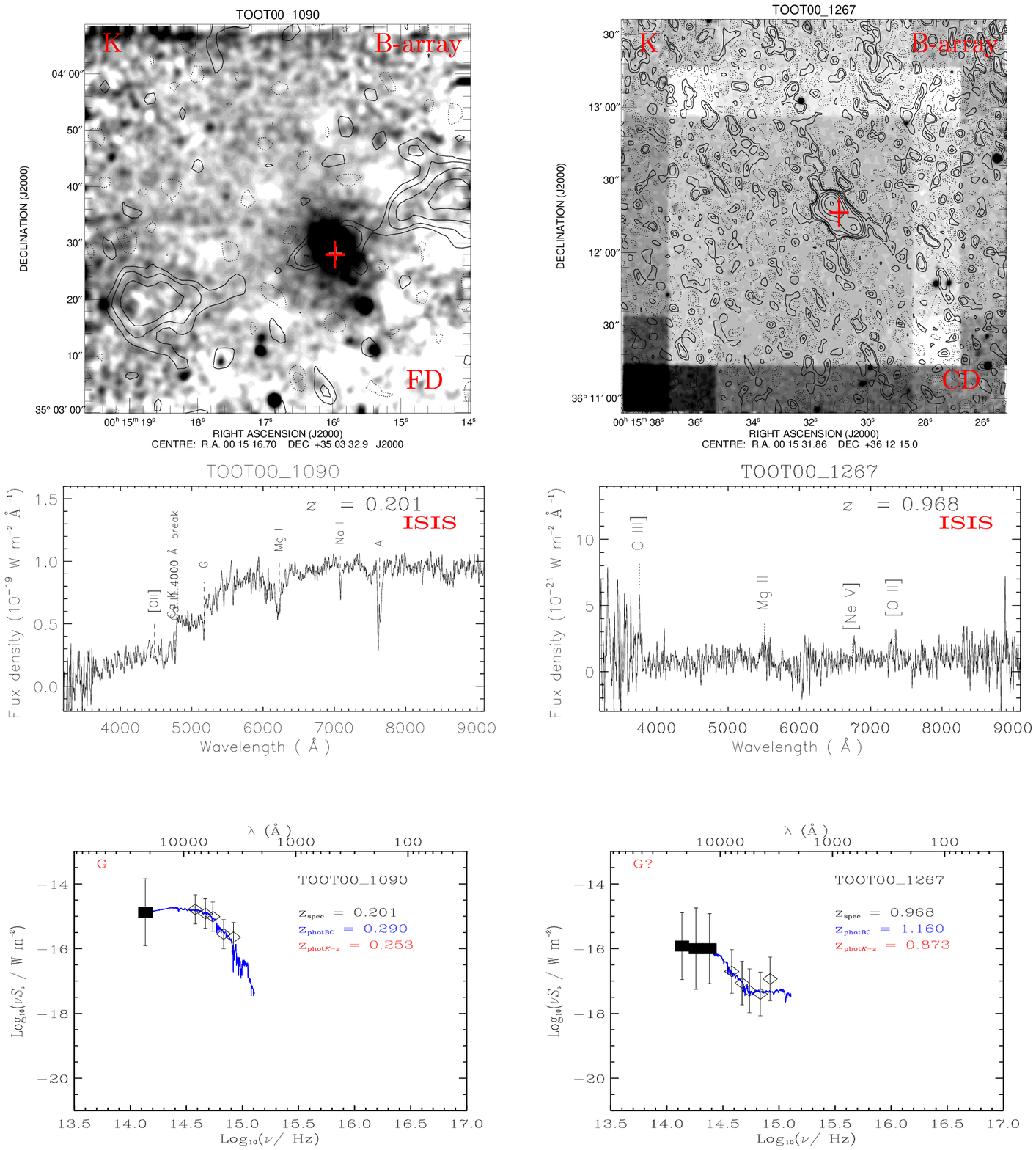}}
\end{picture}
\end{center}
\vspace{0.1in}
{\caption[junk]{(continued)
}}
\end{figure*}

\addtocounter{figure}{-1}

\clearpage
\begin{figure*}
\begin{center}
\setlength{\unitlength}{1mm}
\begin{picture}(150,220)
\put(-25,-40){\includegraphics{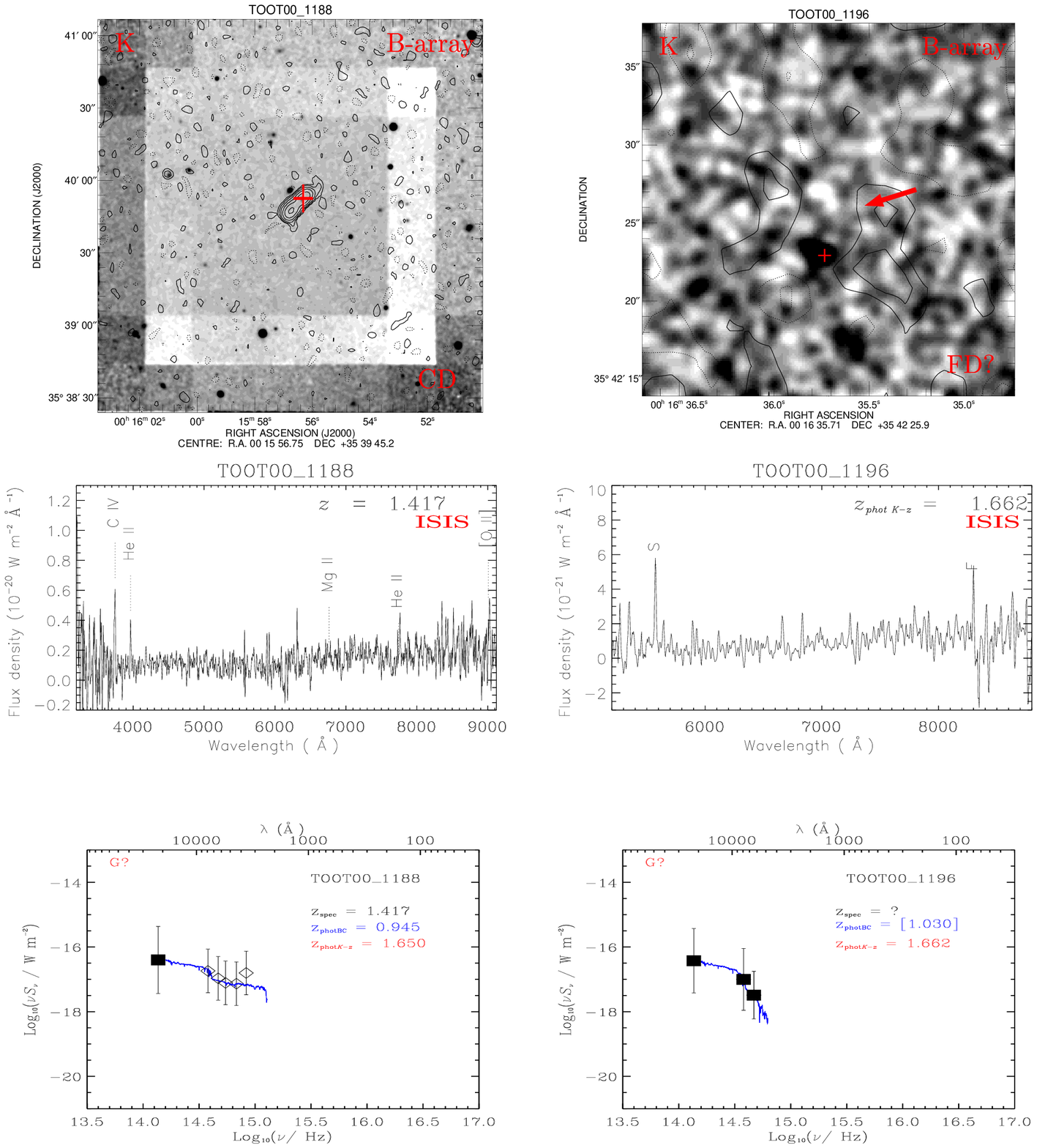}}
\end{picture}
\end{center}
\vspace{0.1in}
{\caption[junk]{(continued)
}}
\end{figure*}

\addtocounter{figure}{-1}

\clearpage

\begin{figure*}
\begin{center}
\setlength{\unitlength}{1mm}
\begin{picture}(150,220)
\put(-25,-40){\includegraphics{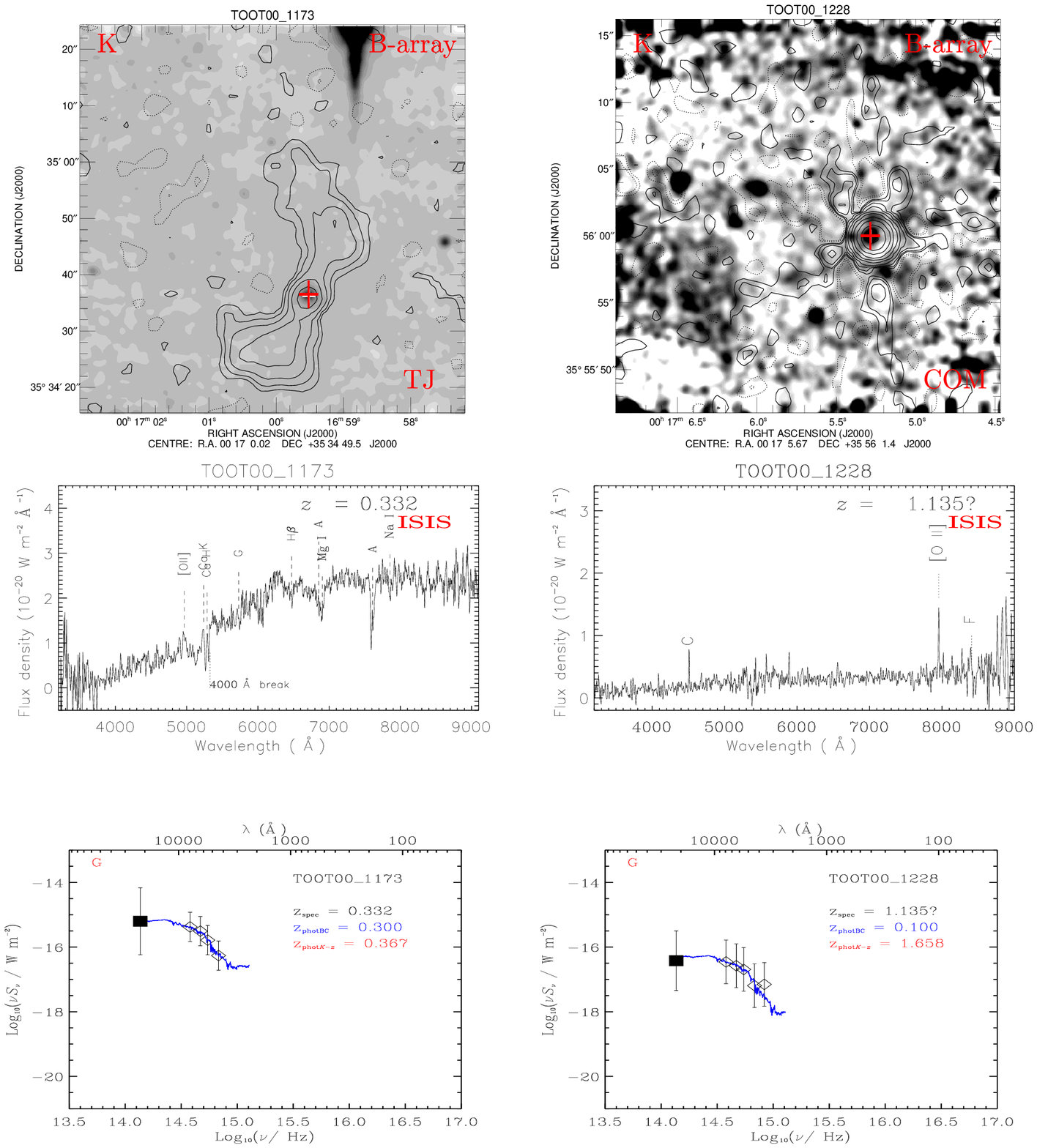}}
\end{picture}
\end{center}
\vspace{0.1in}
{\caption[junk]{(continued)
}}
\end{figure*}

\addtocounter{figure}{-1}

\clearpage

\begin{figure*}
\begin{center}
\setlength{\unitlength}{1mm}
\begin{picture}(150,220)
\put(-25,-40){\includegraphics{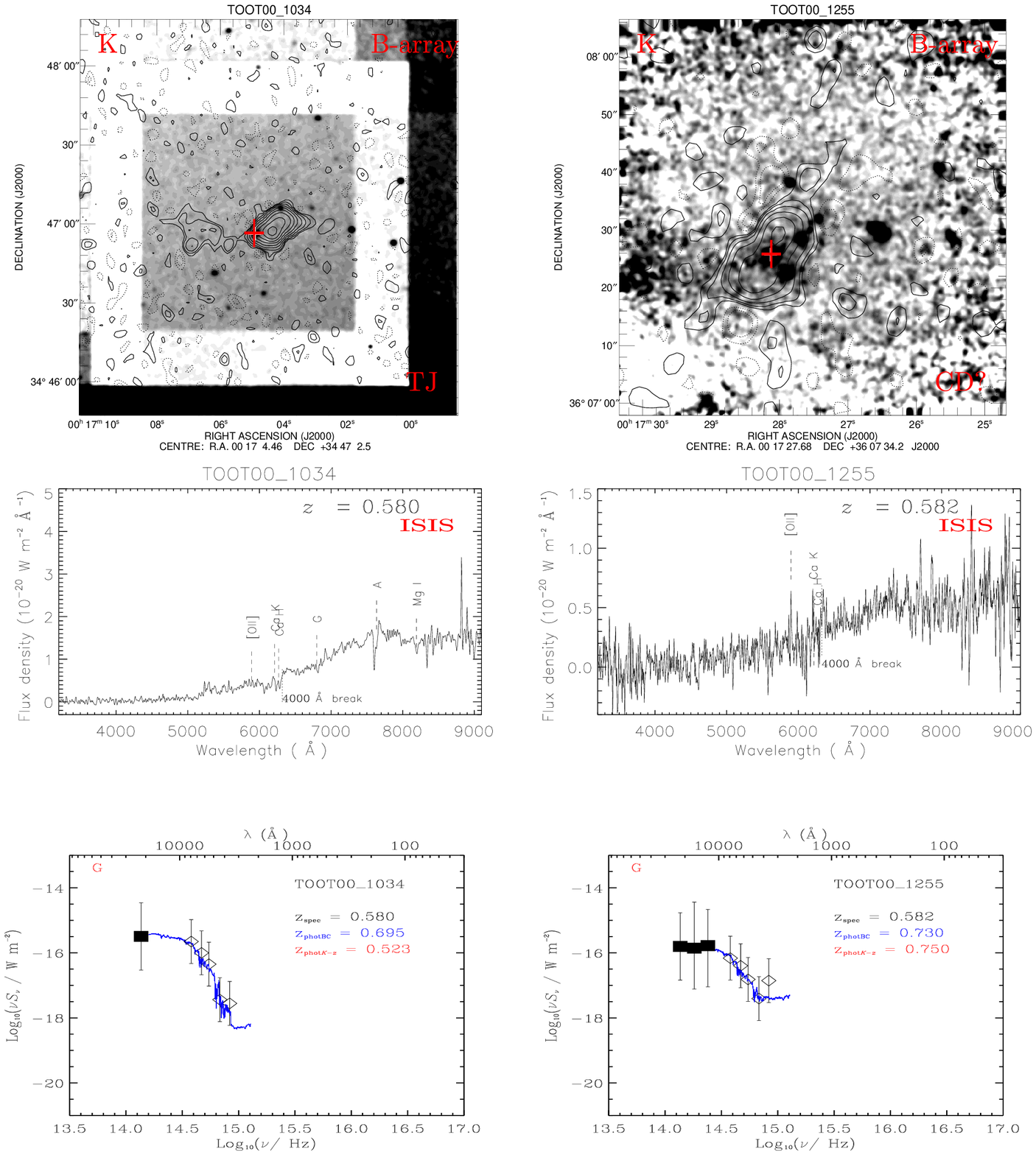}}
\end{picture}
\end{center}
\vspace{0.1in}
{\caption[junk]{(continued)
}}
\end{figure*}

\addtocounter{figure}{-1}

\clearpage

\begin{figure*}
\begin{center}
\setlength{\unitlength}{1mm}
\begin{picture}(150,220)
\put(-25,-40){\includegraphics{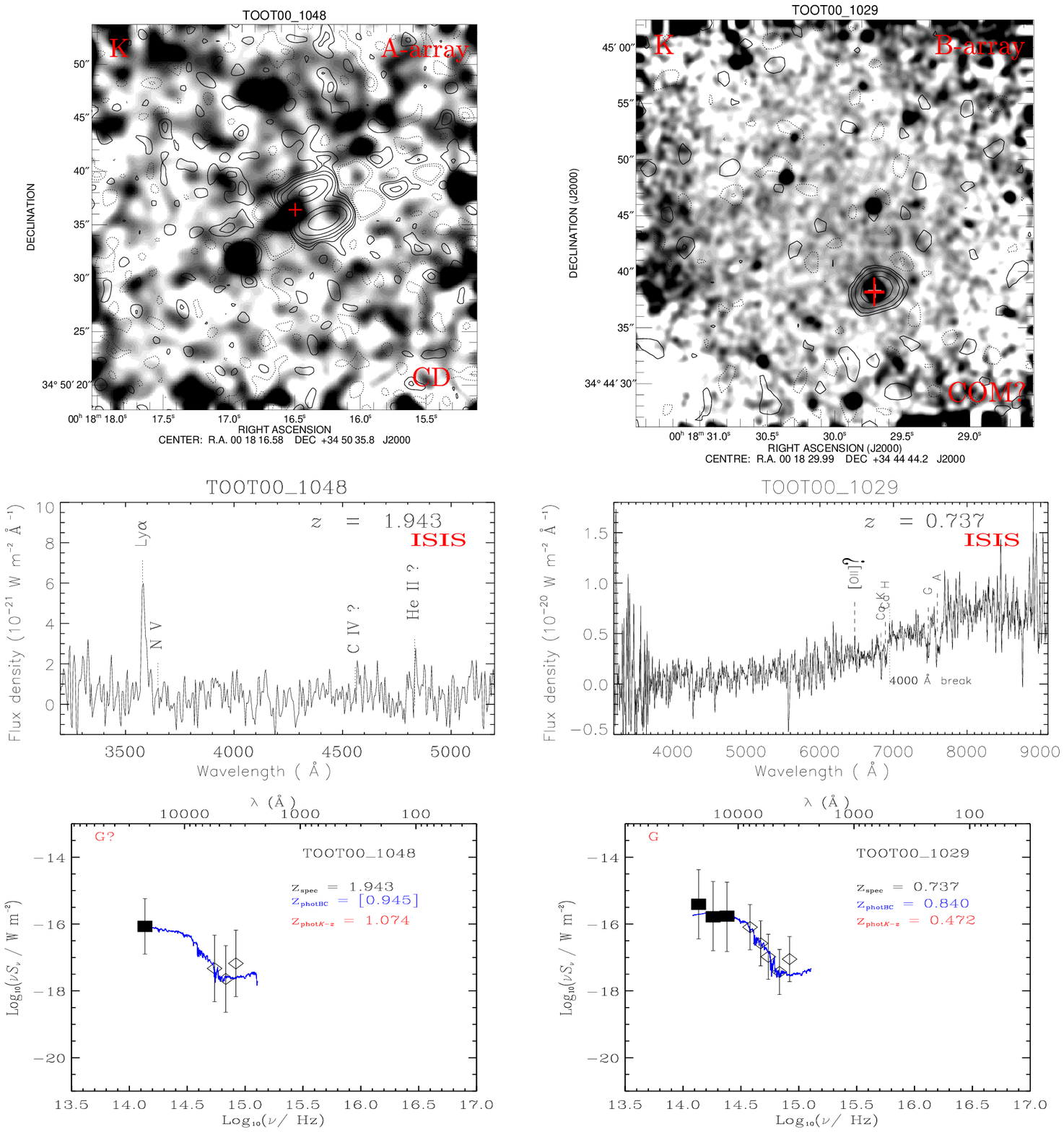}}
\end{picture}
\end{center}
\vspace{0.1in}
{\caption[junk]{(continued)
}}
\end{figure*}

\addtocounter{figure}{-1}

\clearpage
\begin{figure*}
\begin{center}
\setlength{\unitlength}{1mm}
\begin{picture}(150,220)
\put(-25,-40){\includegraphics{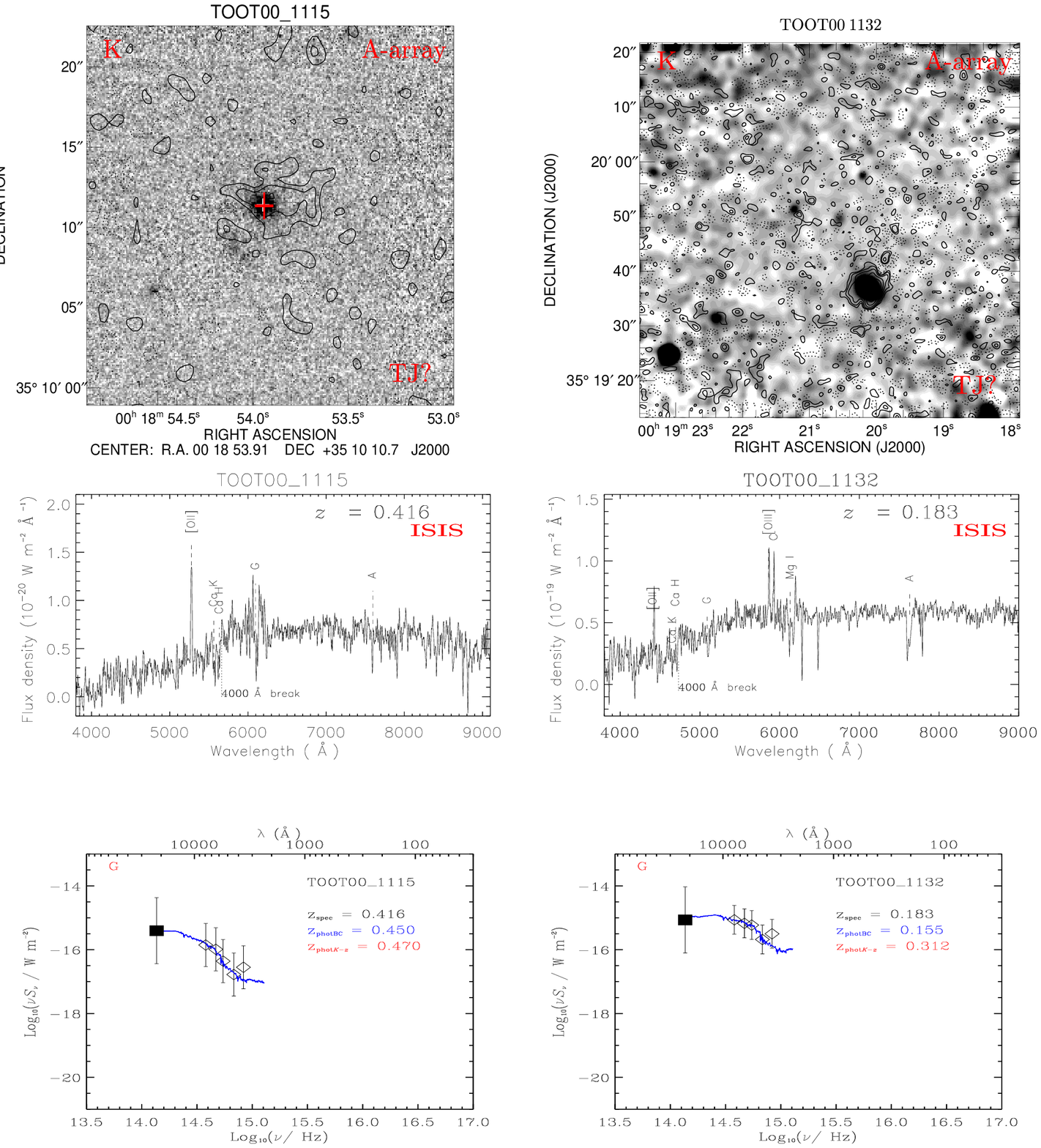}}
\end{picture}
\end{center}
\vspace{0.1in}
{\caption[junk]{(continued)
}}
\end{figure*}

\addtocounter{figure}{-1}

\clearpage
\begin{figure*}
\begin{center}
\setlength{\unitlength}{1mm}
\begin{picture}(150,220)
\put(-25,-40){\includegraphics{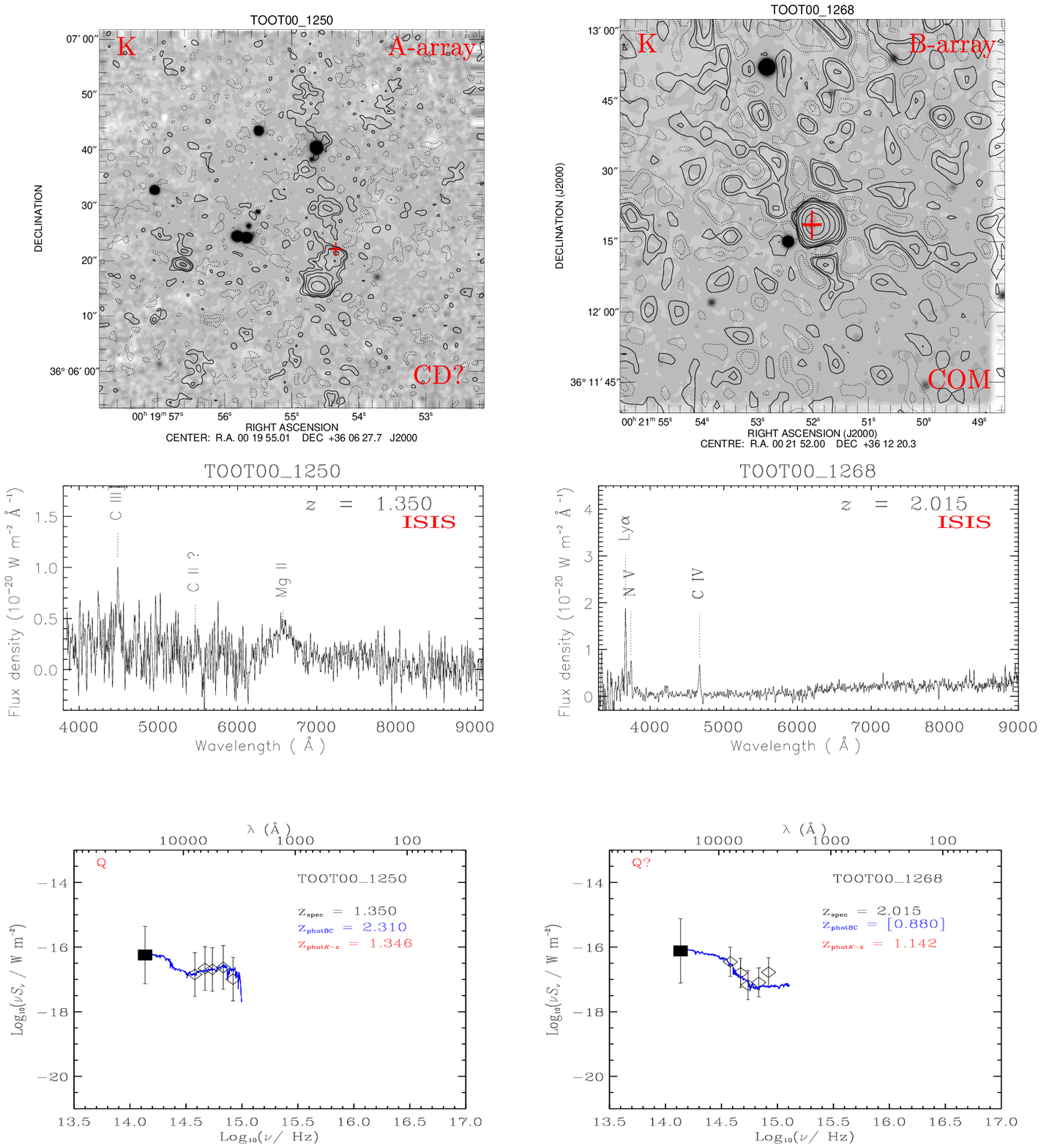}}
\end{picture}
\end{center}
\vspace{0.1in}
{\caption[junk]{(continued)
}}
\end{figure*}

\addtocounter{figure}{-1}

\clearpage

\begin{figure*}
\begin{center}
\setlength{\unitlength}{1mm}
\begin{picture}(150,220)
\put(-25,-40){\includegraphics{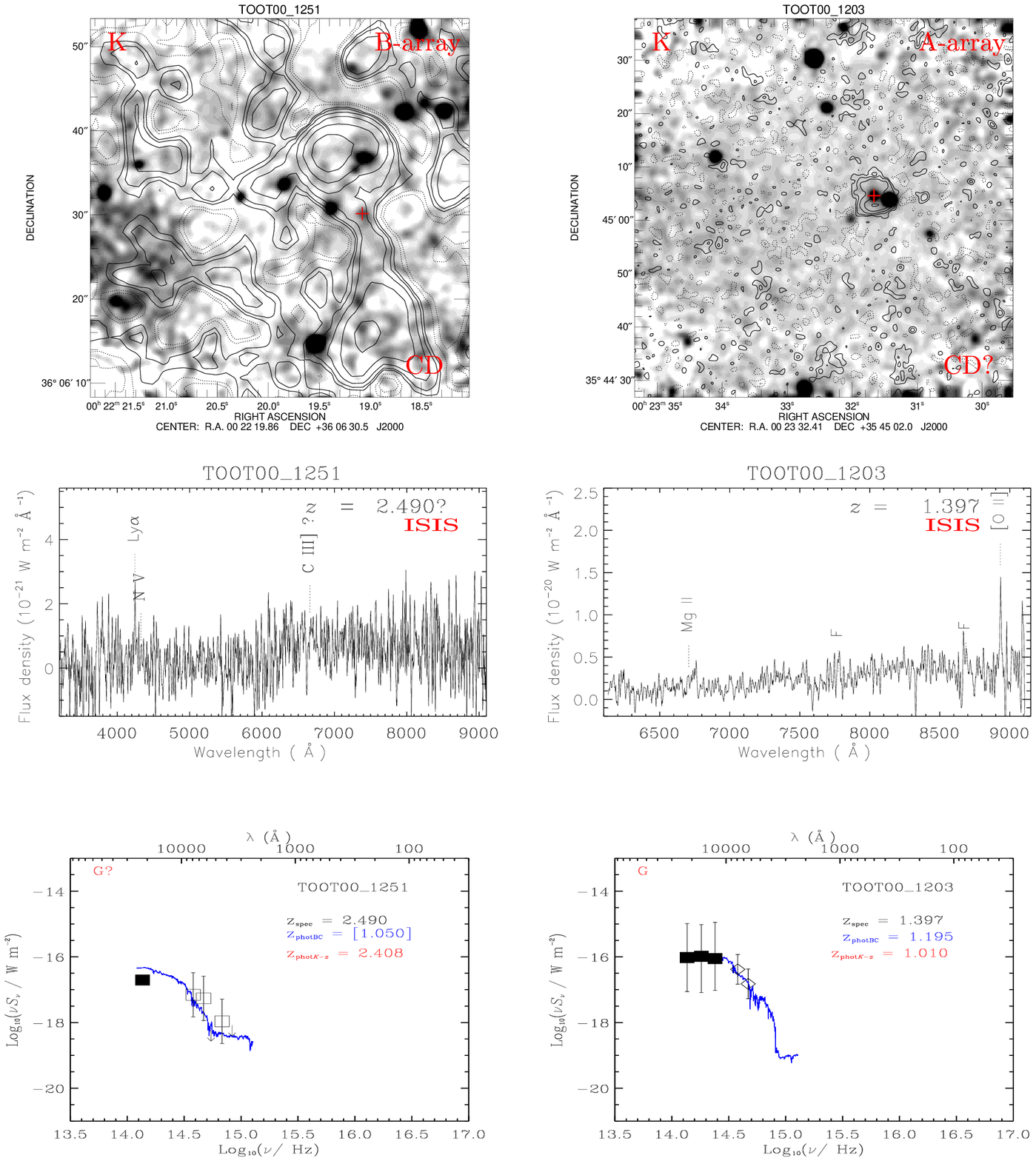}}
\end{picture}
\end{center}
\vspace{0.1in}
{\caption[junk]{(continued)
}}
\end{figure*}

\addtocounter{figure}{-1}

\end{document}